\documentclass[pre,aps,amssymb,onecolumn]{revtex4-1}
\usepackage{amsmath}
\usepackage{amssymb}
\usepackage{amsthm}
\usepackage{amsfonts}
\usepackage{listings}
\usepackage{longtable}
\usepackage{enumerate}
\usepackage{latexsym}
\usepackage{color}
\usepackage{dsfont}
\usepackage{setspace} 
\usepackage{blindtext}

\usepackage{array,etoolbox}
\preto\tabular{\setcounter{magicrownumbers}{0}}
\newcounter{magicrownumbers}

\newcommand{\irr}[2]{\overline{\textrm{#1}}_#2}
\newcommand{\irrg}[1]{\overline{\Gamma}_#1}

\newcommand{\irrl}[1]{\overline{\Lambda}_#1}
\newcommand{\irrr}[1]{\overline{\rho}_#1}
\newcommand{\beginsupplement}{%
        \setcounter{table}{0}
        \renewcommand{\thetable}{S\arabic{table}}%
        \setcounter{figure}{0}
        \renewcommand{\thefigure}{S\arabic{figure}}%
        \setcounter{section}{0}
        \renewcommand{\thesection}{S\arabic{section}}%
        \setcounter{equation}{0}
        \renewcommand{\theequation}{S\arabic{equation}}%
     }

\usepackage{bm}
\usepackage{hyperref}
\hypersetup{
 pdfnewwindow=true, colorlinks=true,
 linkcolor=blue, anchorcolor=blue,
 citecolor=blue, filecolor=blue,
 menucolor=blue, urlcolor=blue}

\usepackage{psfrag}

\usepackage{bm}
\usepackage{graphicx}
\usepackage{subfigure}

\begin{document}

\tolerance 10000

\newcommand{\vk}{{\bf k}}

\draft

\title{Application of the induction procedure and the Smith Decomposition in the calculation and topological classification of electronic band structures in the 230 space groups}

\author
{Luis Elcoro$^{1\dagger}$, Zhida Song$^{2}$, B. Andrei Bernevig$^{2,3,4}$\\
	\normalsize{$^1$Department of Condensed Matter Physics, University of the Basque Country UPV/EHU,} \\
	\normalsize{Apartado 644, 48080 Bilbao, Spain}\\
	\normalsize{$^{2}$Department of Physics, Princeton University, Princeton, New Jersey 08544, USA}\\
	\normalsize{$^3$Physics Department, Freie Universit\"at Berlin, Arnimallee 14, 14195 Berlin, Germany}\\
	\normalsize{$^4$Max Planck Institute of Microstructure Physics, 06120 Halle, Germany}\\
	\normalsize{$^\dagger$To whom correspondence should be addressed; E-mail:  luis.elcoro@ehu.eus}\\
\thanks{All authors contributed equally to this work}
}

\date{\today}

\begin{abstract}
The electronic properties in a solid depend on the specific form of the wave-functions that represent the electronic states in the Brillouin zone. Since the discovery of topological insulators, much attention has been paid to the restrictions that the symmetry imposes on the electronic band structures. In this work we apply two different approaches to characterize all types of bands in a solid by the analysis of the symmetry eingenvalues: the induction procedure and the Smith Decomposition method. The symmetry eigenvalues or irreps of any electronic band in a given space group can be expressed as the superposition of the eigenvalues of a relatively small number of building units (the \emph{basic} bands). These basic bands in all the space groups are obtained following a group-subgroup chain starting from P1. Once the whole set of basic bands are known in a space group, all other types of bands (trivial, strong topological or fragile topological) can be easily derived. In particular, we confirm previous calculations of the fragile root bands in all the space groups. Furthermore, we define an automorphism group of equivalences of the electronic bands which allows to define minimum subsets of, for instance, independent basic or fragile root bands.
\end{abstract}

\maketitle

\section{Introduction}
The theory of electronic band structure has underpinned our understanding of weakly interacting materials for the past century. It has been fundamental in areas from theoretical physics to material engineering, and has contributed to virtually each one of the technical advances in the past centuty.  At the basis of the theory rests Bloch's theorem - the fact that the Hamiltonian of any periodic lattice in real space has, when Fourier-transformed, a structure in momentum space which makes the energy levels in the first momentum space Brillouin Zone repeat at other electron momenta. While Bloch focused on the electron energies, it was realized early on, by Wigner, Von Neumman,  Herring, Harrison, and others, that eigenfunctions are fundamental to the electronic properties in crystals. 

The focus on the electron wavefunction properties experienced a fundamental breakthrough with the realization that topology plays an essential role in the physics of a given material. Thouless, Kohmoto, Nightingale, den Nijs (TKNN) \cite{thouless1982} proved, in the early 1980's, that the wavefunctions of a system under the influence of a magnetic field exhibit a topological invariant, the Chern number \cite{Hatsugai1993}, which - moreover - equals a physical observable, the Hall conductance. This is measured in the Quantum Hall Effect, and it is the first example of how topological protection can lead to experimental observables. The first topological classification was developed: wavefunctions described by different topological invariants - Chern numbers - cannot be adiabatically continued to one another. When two insulating materials described by  wavefunctions with different Chern number are placed next to one another,  gapless edge states develop at the interface between them. This is an example of the bulk-boundary correspondence: a topologically nontrivial gapped bulk gives rise to a gapless chiral edge -which is moving in one direction. The topological classification of Chern insulators is based on integers: every wavefunction can be described by a Chern number which is an integer. A nonzero Chern number requires the breaking of time-reversal \cite{Haldane1988}- usually realized by the application of a magnetic field. 

The field of electronic wavefunctions then experienced a lull until the early 2000's \cite{Kane2005_QSH,Kane2005_Z2,Bernevig2006,Konig2007,Fu2007,Zhang2009,Chen2009,Xia2009,kitaev2009}, when it was shown that adding symmetry provides new topological classification, which, for the first time, was different from a Chern number. The first papers of the new field of topological insulators showed that, by adding time-reversal to a system the integer Chern classification vanishes- nonzero Chern number  is not possible in the presence of time reversal. However, a new, $Z_2$ classification emerges, based on time-reversal pairs of edge states. A nontrivial time-reversal topological insulator, when put next to a trivial one, has a pair of helical edge states on one edge, whose crossing is protected by time reversal. 

The next 15 years have then produced many more topological classifications based, primarily of adding symmetries to the electron Hamiltonian \cite{teo2008,Fu2011,Hsieh2012,slager2013,shiozaki2014,Liu2014,Fang2015,Benalcazar2017,Benalcazar2017b,Langbehn2017,Song2017,Fang2017,Schindler2018,Ezawa2018}. Crystals are periodic arrangements of atoms, and as such, exhibit spatial symmetries classified by 230 non-magnetic space groups. When a symmetry is enforced on the system, new phases of matter become distinct. For example, when mirror symmetry is added to a system, we find that we can diagonalize the Hamiltonian in mirror symmetry sectors: each mirror symmetry sector has a Chern number, even if the full system - summed over all mirror sectors - did not have a Chern number. 

These realizations have pointed towards a fundamental refinement - based on topological features- of band theory. This refinement takes into account  the symmetry properties of a given space-group. In retrospect, the first re-writing of band theory was due to Zak \cite{Zak1982,Bacry1988}, although without emphasis or consideration  of  topological features. Zak mainly focused on atomic limit bands - electronic structures that are obtained from a set of localized orbitals. Zak's fundamental advances are multi-fold. First, Zak and co-authors realized that a Hamiltonian is not needed in order to describe the symmetry structure of bands that can be expressed from localized orbitals. Zak introduced the concept of band representations, induced representations from the lattice orbitals to the space group that fix the symmetry eigenvalues, in the Brillouin zone, of a band structure described by atomic orbitals. While Zak performed this program for a small set of orbitals, recent advances \cite{NaturePaper} completed his vision, tabulating all the 10398 elementary band representations (atomic limits) existent in the 230 space groups, with and without spin-orbit coupling. Even though it is supposed to only describe atomic limits, it was found \cite{NaturePaper} that Zak's theory intimately implies topological phases of matter, which cannot be described by atomic limits. This led to a re-writing of electronic band theory.

 In 2017 {\it Topological Quantum Chemistry} (TQC) classified all the elementary band representations that exist in Zak's theory. It also  put forward the thesis, following Soluyanov and Vanderbilt \cite{soluyanov2011}, that any band structure that is not an atomic limit has to be, by definition, topological.  It also provided \cite{NaturePaper} the entirety of the so-called compatibility relations, a set of linear algebraic constraints that determine whether bands starting from one high-symmetry point can connect to bands coming from a different high-symmetry point. This allows for a classification of the non-trivial (topological) band structures, as well as the enforced semimetals: any set of bands that do not satisfy the compatibility relations have to be metallic. All these constraints and data are tabulated at the Bilbao Crystallographic Server \cite{GroupTheoryPaper,progbandrep}. Several other similar formalisms were also proposed \cite{po2017,kruthoff2017}. Then, the topology criteria are explicitly mapped to topological invariants protected by TRS or crystalline symmetries \cite{Song2018a,Song2018b,khalaf2018}. These advances allowed high through-put work \cite{vergniory2018,zhang2018,tang2018} where we presented a catalogue with the topological classification of all the "high-quality" topological materials existent in the Inorganic Crystal Structure Database (ICSD: {\url {https://icsd.products.fiz-karlsruhe.de}}). The approach is based on classifying different types of bands that satisfy the compatibility relations. Elementary bands (representations) describe atomic limits. However, sometimes, they can be "split" into several branches, which are not EBRs, and are hence topological bands. In many symmetry groups, there also exist bands that are not EBRs - and hence are topological, but have nonetheless the same (or lower) dimension than EBRs. The question then becomes: what is a (smallest) basis for all the bands in every symmetry group? Clearly, and in general, these "basic" bands need to involve atomic limits, strong topological bands, and a new type of topological bands - fragile topological \cite{po2017b,cano2018b,bouhon2019,song2019b,Po2018,Ahn2018,po2019,paz2019,manes2019}.

In this paper, we present a different a bottom-up approach of obtaining all the topological and non-topological bands in any symmetry group. We start from SG 1 and build what we call the "basic" bands, by induction, successively, in supergroups. These basic bands can be characterized as atomic limits, topological, or trivial, as detailed in the main text. Upon building all sets of basic bands, we then show that we can find the same topological classification, both strong and fragile, as obtained in previous works. We derive several brute-force methods (Smith decomposition for strong and fragile bands) that obtain topological bands, and show that our basic band method matches the results obtained by brute-force. As such, a full classification of the basis of bands in a symmetry group is obtained.

\section{Types of electronic bands}
\label{sec:definitions}
In a solid, the electronic energy bands are continuous functions (wavefunctions) defined in the reciprocal space whose dependence on the wave-vector $\bf{k}$ is restricted by the space group of the structure. The existence of a translational periodicity forces the bands to be periodic in the reciprocal space, with the periodicity given by the reciprocal lattice. The symmetry operations whose rotational part is different from the identity impose additional restrictions on the band structure. In particular, at every $\bf{k}$-vector in the reciprocal space, the electronic states transform as a representation of the little group of $\bf{k}$. This representation is, in general, reducible, but it is identified by the multiplicities of the (tabulated) irreps at the $\mathbf{k}$-vector. In principle, the symmetry property of a band or a set of bands defined in the whole Brillouin Zone (BZ) are partially characterized by the multiplicities of the irreducible representations at every $\mathbf{k}$-vector. The set of irreps does not give complete information about the whole wave-function. However, it is enough to determine the connectivity properties of the band.

Although it is possible to determine the sets of irreps in all the whole BZ, it is not necessary to specify the multiplicities at every $\mathbf{k}$-vector to unambiguously identify a band. It is sufficient to know the multiplicities of the irreps of the little groups at a selected set of $\mathbf{k}$-vectors in the BZ: the $\mathbf{k}$-vectors of maximal symmetry or maximal $\mathbf{k}$-vectors \cite{GroupTheoryPaper,vergniory2017}. A vector $\mathbf{k}$ in the reciprocal space with little group $\mathcal{G}^{\mathbf{k}}$ is of maximal symmetry if it cannot be connected, following a continuous path in which all the $\mathbf{k}$-vectors have as little group $\mathcal{G}^{\mathbf{k}}$, to an end point $\mathbf{k}'$ whose little group is a supergroup of $\mathcal{G}^{\mathbf{k}}$. The $\mathbf{k}$-vectors of maximal symmetry in all the space groups form a small set and are listed in the program BANDREP \cite{progbandrep} (\emph{www.cryst.ehu.es/cryst/bandrep}). From the multiplicities of the irreps at the maximal $\mathbf{k}$-vectors, the multiplicities at the non-maximal $\mathbf{k}$-vectors can be derived making use of the \emph{compatibility relations}.

The knowledge of the set of maximal $\mathbf{k}$-vectors, the little groups of these $\mathbf{k}$-vectors, the irreps of these little groups and the sets of compatibility relations in the whole BZ allows us to define and classify the possible types of electronic bands that there can be in a solid. In this section we first define the types of bands from the point of view of their topology. 

\emph{\textbf{Definition 1:}} \textbf{Connected bands and Basic bands}. An electronic band is defined by the multiplicities of the irreps of the little groups of the maximal $\bf{k}$-vectors in the space group. These sets of irreps are interconnected along the intermediate paths (lines and planes) and satisfy the compatibility relations, i.e., for every pair of 
maximal $\bf{k}$-vectors, the set of irreps at both $\bf{k}$-vectors subduces into the same sets of irreps of the intermediate paths that connect the two maximal $\bf{k}$-vectors with identical multiplicities. If some subsets of irreps at each maximal $\bf{k}$-vector are fully connected and separated from the rest of subsets of irreps that form the whole set of electronic states in a structure, we say that it is a connected band. We define as \emph{basic band} a connected band that cannot be splitted into separate subsets of irreps at every maximal $\bf{k}$-vector satisfying internally the compatibility relations, even if we consider all the different possible arrangements of irreps at every $\mathbf{k}$-vector. As a consequence of this definition, any electronic band in a given space group is an integer linear combination (with non-negative coefficients) of basic bands. This linear combination is not necessarily unique, and different linear combinations of basic bands can give rise to the same whole set of symmetry eigenvalues at maximal $\bf{k}$-vectors. We can say that the basic bands constitute an, in general, overcomplete \emph{basis} of all the bands in a space group. In the next sections we will restrict the use of the term \emph{band} to refer only to connected bands.

\emph{\textbf{Definition 2:}} \textbf{Elementary band representation}. The concept of band representation was first introduced by Zak \cite{Zak1982} to refer to a set of energy bands spanned by a given collection of (exponentially) localized Wannier orbitals. Given the coordinates of a Wyckoff position in a space group ${\mathcal{G}}$, the symmetry operations that keep invariant the coordinates of this point belong to the so-called site-symmetry group of the Wyckoff position. This group is isomorphic to a point group and each irreducible representation of this group induces a representation onto ${\mathcal{G}}$ called band representation. A band representation that is equivalent to a direct sum of other band representations is called a composite band representation. A band representation that is not composite is called elementary. Therefore, if we identify a band representation through the multiplicities of the irreps at the maximal $\bf{k}$-vectors under subduction, the list of multiplicities of a composite band representation is a linear combination of the lists of multiplicities of the elementary band representations (EBRs) of the space group. All the EBRs of a space group are obtained from its maximal Wyckoff positions, although not all the band representations induced from the maximal Wyckoff positions are elementary. All the EBRs in all the space groups without spin-orbit coupling (SOC), which are induced from the so-called single-valued irreps, were deduced by Zak and Bacry \cite{Zak1982,Bacry1988} and the EBRs with SOC (induced from double-valued irreps) were calculated in ref.(\onlinecite{NaturePaper}) and implemented in the program BANDREP \cite{progbandrep}. The EBRs play a central role in the classification of the band structures into topological bands (strong or fragile) and trivial bands and, consequently, of the materials into topological or trivial. We can state that the EBRs form a basis for the band representations in a space group. Whereas the basic bands expand all the existing connected bands in a space group, the EBRs expand a particular type of connected bands: the band representations (or trivial bands). An EBR can coincide with one of the basic bands or can be a linear combination of basic bands with non-negative integer coefficients.

\emph{\textbf{Definition 3:}} \textbf{Fragile topological bands and fragile root bands}. A band that cannot be expressed as the direct sum (linear combination with non-negative integer coefficients) of EBRs but can be expressed as linear combinations of EBRs with positive and negative integer coefficients is a fragile (topological) band. A fragile band that cannot be expressed as a linear combination of EBRs and another fragile bands with non-negative integer coefficients is a fragile root. Then, any fragile band that is not a fragile root band can be written as linear combination of EBRs and fragile root bands. The EBRs and the fragile root bands form a basis for all the fragile bands in a space group.

\emph{\textbf{Definition 4:}} \textbf{Strong topological bands}. A band that cannot be expressed as an integer linear combination of EBRs is a strong (topological) band. Two strong bands are EBR-equivalent if its difference can be written as a linear combination of EBRs, with positive or negative integer coefficients.

According to the above definitions, the direct sums of different kinds of bands give rise to the following results:
\begin{itemize}
	\item The direct sum of two trivial bands (as a particular case the sum of two EBRs) gives a trivial band.
	\item The direct sum of two fragile bands (being roots or not) can give rise to another fragile band or to a trivial band. In some cases the trivial band is a single EBR. This EBR is thus identified as decomposable \cite{NaturePaper}.
	\item The direct sum of two strong bands can give rise to another strong band, to a trivial band (elementary or not) or to a fragile band (root or not). When the result is another strong band, this is not equivalent to any of the two previous strong bands. If the result is an EBR, the EBR is thus decomposable \cite{NaturePaper}.
	\item The direct sum of a fragile band and a trivial band can be a trivial band or a fragile band.
	\item The direct sum of a strong band and a not strong band is always another strong band equivalent to the first one.
\end{itemize}
Once all the basic bands in a space group have been identified, through combinations of these basic bands, it is possible to determine all kinds of topological bands that can be realized in a space group and, in particular, the fragile roots. In the next section, we describe a method to derive all the basic bands in a space group based in a standard group theory technique: the induction procedure.

\section{Calculation of the basic bands through the induction procedure}
\label{sec:induction}
The induction procedure has been the standard way to obtain the irreps of the point and space groups and programs to apply the induction have been developed \cite{hovestreydt,repres,GroupTheoryPaper}.  If ${\mathcal{H}}$ is a normal subgroup of ${\mathcal{G}}$, ${\mathcal{H}}\lhd{\mathcal{G}}$, the irreps of ${\mathcal{G}}$ can be obtained from the irreps of ${\mathcal{H}}$ through induction. The general procedure is simplified in the case of crystallographic point and space groups because, for any point or space group ${\mathcal{G}}$, there exists a normal subgroup ${\mathcal{H}}$ of index 2 or 3 in ${\mathcal{G}}$. In Section (\ref{sup:induction}) of the supplementary material (SM) we summarize the main steps of the induction procedure applied to crystallographic groups with index 2 or 3 \cite{repres}.

As an extension of the induction procedure, it can also be applied to induce the basic bands in a space group from the basic bands of one of its maximal subgroups. This purpose requires the systematic application of the induction to the little group of every $\mathbf{k}$-vector in the subgroup to derive the irreps of the corresponding little group of the $\mathbf{k}$-vector in the supergroup. In the following, we summarize the main steps followed in the systematic identification of the basic bands in a space group.

Given a $\mathbf{k}$-vector, its little co-group is the set of point-group operations $R$ that keep the $\mathbf{k}$-vector invariant, mod reciprocal lattice translations, i.e., those rotational operations such that the relation
\begin{equation}
\label{eq:cogroup}
\mathbf{k}\cdot R=\mathbf{k}+\mathbf{K}
\end{equation}
is fulfilled for some $\mathbf{K}$ in the reciprocal lattice. The little group of $\mathbf{k}$, $\mathcal{G}^{\mathbf{k}}$, is the subset of symmetry operations of $\mathcal{G}$ whose rotational part satisfies Eq. (\ref{eq:cogroup}). Therefore, $\mathcal{G}^{\mathbf{k}}$ is a subgroup of $\mathcal{G}$.

When the little groups of a given $\mathbf{k}$-vector in a space group $\mathcal{G}$ and in one of its subgroups $\mathcal{H}$ are compared, there can be two different possibilities depending on the $\mathbf{k}$-vector:
\begin{enumerate}
	\item[\textbf{T1}] The little group is the same in both groups, $\mathcal{G}^{\mathbf{k}}=\mathcal{H}^{\mathbf{k}}$. This means that no symmetry operation $g\in\mathcal{G}$ and $g\notin\mathcal{H}$ belongs to the little group of $\mathbf{k}$. For these $\mathbf{k}$-vectors, there is a 1 to 1 mapping between the irreps of the little groups, which have the same dimensions, i.e., an irrep $\rho_{\mathcal{H}^{\mathbf{k}}}\in \mathcal{H}^{\mathbf{k}}$ induces a single irrep $\rho_{\mathcal{G}^{\mathbf{k}}}\in \mathcal{G}^{\mathbf{k}}$.
	\item[\textbf{T2}] The little group in $\mathcal{G}$ is a supergroup of the little group in $\mathcal{H}$. The irreps of $\mathcal{G}^{\mathbf{k}}$ can be induced from the irreps in $\mathcal{H}^{\mathbf{k}}$ following the procedure summarized in Section (\ref{sup:induction}) of the SM. In general, there are two different types of results when the induction of the whole set of irreps $\rho_{\mathcal{H}^{\mathbf{k}}}^i\in \mathcal{H}^{\mathbf{k}}$ is considered (see the details in Section (\ref{sup:induction}) of the SM): \textbf{T2(a)} a given irrep $\rho_{\mathcal{H}^{\mathbf{k}}}$ can induce 2 or 3 irreps into $\mathcal{G}^{\mathbf{k}}$ (depending on the index between the two little groups: 2 or 3 for crystallographic space groups if the group-subgroup pair $\mathcal{H}\lhd\mathcal{G}$ is appropriately chosen) or \textbf{T2(b)} 2 or 3 irreps of $\mathcal{H}^{\mathbf{k}}$ combine to induce a single irrep in $\mathcal{G}^{\mathbf{k}}$.
\end{enumerate} 

Once the mapping between the irreps in the subgroup and the irreps in the supergroup has been performed, the multiplicities that define the different possible induced bands in the supergroup are obtained from the multiplicities of the given basic band in the subgroup and the restrictions imposed by the compatibility relations. In Section (\ref{compatibilityfirst}) of the SM we summarize the theoretical background of the compatibility relations and their application in the analysis of the connectivity of the bands, but we here outline some general results:
\begin{itemize}
	\item If the list of irreps at $\mathbf{k}$-vectors of maximal symmetry that define the bands in a group contains irreps of type \textbf{T2(a)}, every basic band induces, in principle, a band in the supergroup that, in general, is not basic.
	In principle, to identify all the basic bands in the supergroup we could first calculate all the induced bands from basic bands in the subgroup, and then find all the possible ways of decomposition of every induced band into sets of irreps at every $\mathbf{k}$-vector of maximal symmetry that form a fully connected band. However, computationally it is more efficient to proceed in a different way. Let $\rho_{\mathcal{H}^{\mathbf{k}}}$ an irrep of the little group $\mathbf{k}$  and let this irrep induce a reducible representation in the little group $\mathcal{G}^{\mathbf{k}}$,
	\begin{equation}
	   \rho_{\mathcal{H}^{\mathbf{k}}}\uparrow\mathcal{G}=\rho_{\mathcal{G}^{\mathbf{k}}}^1\oplus\rho_{\mathcal{G}^{\mathbf{k}}}^2
	\end{equation}
	If the multiplicity of  $\rho_{\mathcal{H}^{\mathbf{k}}}$ in a basic band of $\mathcal{H}$ is $n(\rho_{\mathcal{H}^{\mathbf{k}}})$, the multiplicities of the two irreps $\rho_{\mathcal{G}^{\mathbf{k}}}^1$ and $\rho_{\mathcal{G}^{\mathbf{k}}}^2$ in the induced band in $\mathcal{G}$ are $n(\rho_{\mathcal{G}^{\mathbf{k}}}^1)=n(\rho_{\mathcal{G}^{\mathbf{k}}}^2)=n(\rho_{\mathcal{H}^{\mathbf{k}}})$. Now, if we consider all the possible splits of this induced band into basic bands of $\mathcal{G}$, in any basic band the multiplicities (not necessarily identical) of these two irreps must be an integer between 0 and $n(\rho_{\mathcal{H}^{\mathbf{k}}})$. In the systematic search of all the possible basic bands, we can consider that every irrep $\rho_{\mathcal{H}^{\mathbf{k}}}$ \emph{can induce} in a basic band of $\mathcal{G}$ a single irrep $\rho_{\mathcal{G}^{\mathbf{k}}}^1$ \textbf{or} a single irrep $\rho_{\mathcal{G}^{\mathbf{k}}}^2$. This slightly different procedure, which can be called \emph{partial induction}, reduces the number of combinations to be checked. Once we have determined all the possible alternative ways to perform a partial induction at every $\mathbf{k}$-vector, we choose a particular result in every $\mathbf{k}$-vector to construct a set of irreps that can potentially form a basic band. If the compatibility relations are fulfilled and the sets of irreps at any pair of maximal $\mathbf{k}$-vectors are fully connected, they form a basic band.
	\item The list of irreps contains irreps of type \textbf{T2(b)}, i.e., at a given maximal $\mathbf{k}$-vector some irreps (2 or 3 in crystallographic groups) combine to induce a single irrep in the supergroup. This means that a basic band in the subgroup with different multiplicities of these 2 (or 3) irreps cannot induce by itself a band into the supergroup. In these cases, several basic bands in the subgroup must be combined to get a band with the appropriate multiplicities to induce a basic band into the supergroup.
\end{itemize}

More details of the whole process can be found in Sections (\ref{sup:induction}-\ref{compatibilityfirst}) of the SM.

However, as the single and double-valued irreps of the little group of all the $\mathbf{k}$-vectors in the space groups have been tabulated \cite{miller1967,stokes2013,GroupTheoryPaper}, the work to be done can be simplified using the opposite procedure to the induction (or the above described partial induction): the subduction, much easier and faster to compute. Therefore, in practice, we will use the subduction to determine the required relations between the two sets of irreps in the group-subgroup pair. We can summarize the subduction process in the following way: let $\rho_{\mathcal{G}^{\mathbf{k}}}$ be an irrep of the little group $\mathcal{G}^{\mathbf{k}}$ and $D_{\rho_{\mathcal{G}^{\mathbf{k}}}}(g)$ the matrix of the irrep of the symmetry operation $g\in\mathcal{G}^{\mathbf{k}}$. Let $\mathcal{H}$ a subgroup of $\mathcal{G}$ and $\mathcal{H}^{\mathbf{k}}$ the little group of $\mathbf{k}$ in $\mathcal{H}$. The matrices $D\rho_{\mathcal{G}^{\mathbf{k}}}(h)$ restricted to the symmetry operations  $h\in\mathcal{H}^{\mathbf{k}}$ form a representation, in general reducible, of $\mathcal{H}^{\mathbf{k}}$. The correlations between $\rho_{\mathcal{G}^{\mathbf{k}}}$ and the irreps $\rho^i_{\mathcal{H}^{\mathbf{k}}}$ of $\mathcal{H}^{\mathbf{k}}$ are easily calculated through the reduction formula (Eq. \ref{eq:multiplicities} in the SM). In Section \ref{sup:subduction} of the SM we give the details of the application of the subduction process in a group-subgroup pair of crystallographic groups. We have also implemented the program DCORREL (\cite{dcorrel}) in the BCS which gives all the correlations between the irreps in any group-subgroup pair. The program uses the subduction process and requires as input just the number of the group, the number of the subgroup and the transformation matrix \cite{ita} of the group-subgroup pair (see the details in Section \ref{sup:subduction} of the SM).

In principle, the systematic application of the induction-subduction process in the whole BZ making use of DCORREL allows us to calculate all the possible induced sets of irreps in $\mathcal{G}$ from the basic bands of $\mathcal{H}$. In every $\mathbf{k}$-vector, all the possible sets of induced irreps from $\mathcal{H}^{\mathbf{k}}$ to $\mathcal{G}^{\mathbf{k}}$ must be considered, along with the compatibility relations in all the intermediate paths (lines and planes) between all pairs of maximal $\mathbf{k}$-vectors in the space group $\mathcal{G}$. These conditions impose restrictions on the multiplicities of the irreps at different maximal $\mathbf{k}$-vectors to form a band. As an example of the application of the induction procedure to the determination of the basic bands in a space group, in Section (\ref{sup:inducedbands}) of the SM we derive the basic bands of the space group I$2_12_12_1$ with time-reversal (TR) symmetry from the basic bands of its maximal subgroup C2.

All the different kinds of bands (basic, strong topological, fragile, trivial) in a space group can be derived from the bands in one of its subgroups, ideally from one of its maximal subgroups, following the above outlined induction-subduction procedure and described in detail in Sections (\ref{sup:induction}-\ref{sup:inducedbands}) of the SM. Therefore, considering different subduction chains, all the basic bands in all the space groups can be ultimately derived from the unique basic band of the space group P1 (N. 1). 

In Section (\ref{sup:bandsall}) of the SM, starting from the single basic band in the space group P1, we describe the derivation of all the double-valued basic bands in the first steps of the different subduction chains, P1$\rightarrow$P$\overline{1}$ (N. 2), P1$\rightarrow$P2 (N. 3), P1$\rightarrow$P3 (N. 143), P1$\rightarrow$R3 (N. 146), P2$\rightarrow$P4 (N. 75), P2$\rightarrow$P$\overline{4}$ (N. 81) and the more elaborated cases P$\overline{1}\rightarrow$C$2/c$ (N. 15)$\rightarrow$F$ddd$ (N. 70) and P1$\rightarrow$C2 (N. 5)$\rightarrow$I$2_22_12_1$ (N. 24)$\rightarrow$I$2_13$ (N. 199).

Following different group-subgroup chains, we have identified all the double-valued basic bands in all the 230 space groups with TR symmetry. These are the relevant bands when the hamiltonian of the system depends on the spin (for instance, when SOC is considered). Table \ref{basicbands} shows the number of basic bands in each space group divided into elementary, strong and fragile  (columns e, s and f, respectively). The table also shows the numbers of independent bands  of each type (see section \ref{equivalences}). At the end of the SM, Table \ref{allbasicbands} lists all the basic bands for each space group through the set of multiplicities of the irreps of the little groups of the maximal $\mathbf{k}$-vectors. The order of irreps to which the multiplicities are referred is given in Table \ref{allmaximalirreps}. In Table \ref{allbasicbands} we also indicate the type of band (elementary, strong or fragile) of each basic band, and we identify a subset of independent basic bands. A similar process can also be applied to derive the single-valued basic bands, but they are not considered in this work.

\begin{table}[!ht]
	\caption{List of number of basic bands for each space group divided as (e) number of elementaty bands, (s) number of strong topological bands and (f) number of fragile topological bands. The columns (ie), (is) and (if) show the number of independent (see section \ref{equivalences}) elementary, strong topological and fragile topological basic bands, respectively.\label{basicbands}}
	\begin{tabular}{|c|cccccc|c|cccccc|c|cccccc|c|cccccc|c|cccccc|}\hline SG&e&ie&s&is&f&if&SG&e&ie&s&is&f&if&SG&e&ie&s&is&f&if&SG&e&ie&s&is&f&if&SG&e&ie&s&is&f&if\\
		\hline
		1&1&1&0&0&0&0&47&16&1&240&4&0&0&93&1&1&0&0&0&0&139&8&1&56&5&0&0&185&2&2&0&0&2&2\\
		\hline
		2&16&1&240&2&0&0&48&1&1&8&2&0&0&94&1&1&0&0&0&0&140&10&4&92&18&0&0&186&4&4&0&0&2&2\\
		\hline
		3&1&1&0&0&0&0&49&9&2&72&5&0&0&95&1&1&0&0&0&0&141&4&1&20&3&0&0&187&15&2&60&4&0&0\\
		\hline
		4&1&1&0&0&0&0&50&1&1&8&2&0&0&96&1&1&0&0&0&0&142&3&2&6&2&0&0&188&12&3&30&5&6&2\\
		\hline
		5&1&1&0&0&0&0&51&9&2&72&6&0&0&97&2&1&0&0&0&0&143&6&2&0&0&6&2&189&5&2&20&4&0&0\\
		\hline
		6&1&1&0&0&0&0&52&5&2&10&4&0&0&98&1&1&0&0&0&0&144&1&1&0&0&0&0&190&9&5&22&8&1&1\\
		\hline
		7&1&1&0&0&0&0&53&8&1&28&3&0&0&99&4&1&0&0&0&0&145&1&1&0&0&0&0&191&12&2&228&20&0&0\\
		\hline
		8&1&1&0&0&0&0&54&5&2&20&4&0&0&100&3&2&0&0&0&0&146&2&2&0&0&0&0&192&9&4&94&27&5&3\\
		\hline
		9&1&1&0&0&0&0&55&8&1&28&3&0&0&101&1&1&0&0&0&0&147&8&2&48&8&8&2&193&7&4&72&25&5&3\\
		\hline
		10&16&1&240&4&0&0&56&5&2&10&4&0&0&102&1&1&0&0&0&0&148&8&2&24&4&0&0&194&12&6&152&36&4&2\\
		\hline
		11&9&2&72&4&0&0&57&5&2&20&4&0&0&103&4&1&0&0&0&0&149&6&2&0&0&6&2&195&2&2&0&0&0&0\\
		\hline
		12&8&1&56&4&0&0&58&8&1&16&2&0&0&104&2&1&0&0&0&0&150&2&2&0&0&2&2&196&2&2&0&0&0&0\\
		\hline
		13&9&2&72&6&0&0&59&1&1&8&2&0&0&105&1&1&0&0&0&0&151&1&1&0&0&0&0&197&2&2&0&0&0&0\\
		\hline
		14&8&1&28&3&0&0&60&5&2&10&4&0&0&106&1&1&0&0&0&0&152&1&1&0&0&0&0&198&2&2&0&0&0&0\\
		\hline
		15&9&2&72&8&0&0&61&4&1&6&2&0&0&107&2&1&0&0&0&0&153&1&1&0&0&0&0&199&1&1&0&0&1&1\\
		\hline
		16&1&1&0&0&0&0&62&5&2&10&4&0&0&108&3&2&0&0&0&0&154&1&1&0&0&0&0&200&8&2&40&6&4&1\\
		\hline
		17&1&1&0&0&0&0&63&5&2&22&6&0&0&109&1&1&0&0&0&0&155&2&2&0&0&0&0&201&1&1&12&3&0&0\\
		\hline
		18&1&1&0&0&0&0&64&4&1&14&3&0&0&110&1&1&0&0&0&0&156&6&2&0&0&6&2&202&8&2&20&5&4&1\\
		\hline
		19&1&1&0&0&0&0&65&8&1&56&4&0&0&111&8&1&8&1&0&0&157&2&2&0&0&2&2&203&1&1&16&5&0&0\\
		\hline
		20&1&1&0&0&0&0&66&9&3&72&13&0&0&112&5&2&4&1&0&0&158&6&2&0&0&6&2&204&4&2&14&5&2&1\\
		\hline
		21&1&1&0&0&0&0&67&9&2&72&6&0&0&113&5&2&4&1&0&0&159&4&4&0&0&2&2&205&8&2&10&3&0&0\\
		\hline
		22&1&1&0&0&0&0&68&1&1&8&2&0&0&114&4&1&2&1&0&0&160&2&2&0&0&0&0&206&8&2&32&5&5&2\\
		\hline
		23&1&1&0&0&0&0&69&4&1&28&2&0&0&115&8&1&8&1&0&0&161&2&2&0&0&0&0&207&5&2&0&0&0&0\\
		\hline
		24&1&1&0&0&0&0&70&1&1&8&3&0&0&116&5&2&4&1&0&0&162&8&2&48&8&8&2&208&1&1&0&0&1&1\\
		\hline
		25&1&1&0&0&0&0&71&8&1&24&2&0&0&117&5&2&4&1&0&0&163&8&6&36&18&10&6&209&3&2&0&0&0&0\\
		\hline
		26&1&1&0&0&0&0&72&5&2&22&6&0&0&118&5&2&4&1&0&0&164&8&2&48&8&8&2&210&1&1&0&0&1&1\\
		\hline
		27&1&1&0&0&0&0&73&5&2&20&4&0&0&119&8&1&8&1&0&0&165&6&4&24&12&6&4&211&3&2&0&0&0&0\\
		\hline
		28&1&1&0&0&0&0&74&9&2&72&8&0&0&120&5&2&4&1&0&0&166&8&2&24&4&0&0&212&2&2&0&0&0&0\\
		\hline
		29&1&1&0&0&0&0&75&4&1&0&0&0&0&121&4&1&4&1&0&0&167&6&4&12&6&0&0&213&2&2&0&0&0&0\\
		\hline
		30&1&1&0&0&0&0&76&1&1&0&0&0&0&122&4&1&2&1&0&0&168&3&2&0&0&3&2&214&1&1&0&0&1&1\\
		\hline
		31&1&1&0&0&0&0&77&1&1&0&0&0&0&123&16&1&240&6&0&0&169&1&1&0&0&0&0&215&5&2&12&3&8&3\\
		\hline
		32&1&1&0&0&0&0&78&1&1&0&0&0&0&124&12&2&96&8&0&0&170&1&1&0&0&0&0&216&9&2&22&4&12&2\\
		\hline
		33&1&1&0&0&0&0&79&2&1&0&0&0&0&125&6&2&28&4&0&0&171&1&1&0&0&0&0&217&3&2&4&1&2&1\\
		\hline
		34&1&1&0&0&0&0&80&1&1&0&0&0&0&126&2&1&4&1&0&0&172&1&1&0&0&0&0&218&2&2&6&3&6&2\\
		\hline
		35&1&1&0&0&0&0&81&8&1&8&1&0&0&127&12&2&124&11&0&0&173&4&4&0&0&2&2&219&2&2&6&3&6&2\\
		\hline
		36&1&1&0&0&0&0&82&8&1&8&1&0&0&128&8&1&40&5&0&0&174&15&2&60&4&0&0&220&5&2&4&2&5&3\\
		\hline
		37&1&1&0&0&0&0&83&16&1&240&6&0&0&129&6&2&28&4&0&0&175&12&2&228&20&0&0&221&12&2&116&11&0&0\\
		\hline
		38&1&1&0&0&0&0&84&12&3&132&13&0&0&130&5&3&14&4&0&0&176&12&6&152&36&4&2&222&3&2&6&2&0&0\\
		\hline
		39&1&1&0&0&0&0&85&6&2&28&4&0&0&131&12&3&132&13&0&0&177&3&2&0&0&3&2&223&4&2&30&9&9&4\\
		\hline
		40&1&1&0&0&0&0&86&4&1&12&2&0&0&132&8&2&40&6&0&0&178&1&1&0&0&0&0&224&2&1&16&3&2&1\\
		\hline
		41&1&1&0&0&0&0&87&8&1&56&5&0&0&133&3&2&6&2&0&0&179&1&1&0&0&0&0&225&12&2&108&15&8&1\\
		\hline
		42&1&1&0&0&0&0&88&4&1&20&3&0&0&134&4&1&12&2&0&0&180&1&1&0&0&0&0&226&7&4&52&16&2&1\\
		\hline
		43&1&1&0&0&0&0&89&4&1&0&0&0&0&135&7&4&20&6&0&0&181&1&1&0&0&0&0&227&4&1&36&5&0&0\\
		\hline
		44&1&1&0&0&0&0&90&3&2&0&0&0&0&136&4&1&12&2&0&0&182&4&4&0&0&2&2&228&1&1&8&3&3&2\\
		\hline
		45&1&1&0&0&0&0&91&1&1&0&0&0&0&137&4&1&4&1&0&0&183&3&2&0&0&3&2&229&6&2&38&9&0&0\\
		\hline
		46&1&1&0&0&0&0&92&1&1&0&0&0&0&138&7&3&20&5&0&0&184&3&2&0&0&3&2&230&3&2&22&7&5&3\\
		\hline
	\end{tabular}
\end{table}
As stressed before, there are different kinds of basic bands: elementary, strong topological and fragile topological. In our complete analysis of the 230 space groups with TR symmetry, we have found four different types of space groups according to the kinds of basic bands found:
\begin{itemize}
	\item All the basic bands are elementary band representations. In these space groups there are neither strong nor fragile topological bands.
	\item Some basic bands are elementary band representations and the rest are fragile bands. In these space groups there are not strong bands.
	\item Some basic bands are elementary band representations and the rest are strong topological bands. These groups can have fragile bands, but they are combinations of basic strong topological bands.
	\item There are basic bands of the three types: elementary, strong and fragile. 
\end{itemize}

The classification of the 230 space groups into these four categories is shown in Table \ref{tb:spacegroups}.

\begin{table}[!h]	
	\caption{Classification of the 136 space groups that have strong and/or fragile topological bands with spin-orbit coupling and TR. In the remaining 94 space groups all the basic bands are elementary.\label{tb:spacegroups}}
	\begin{tabular}{l|l}
		\hline
		&2,10,11,12,13,14,15,47,49,51,53,55,58,63,64,65,66,67,69,70,71,72,74,81,82,83,84,85,86,87,88,\\
		space groups with strong&111,115,119,121,123,124,125,126,127,128,129,130,131,132,134,135,136,137,138,139,140,\\
		and fragile bands&141,147,148,162,163,164,165,166,167,174,175,176,187,188,189,190,191,192,193,194,200,201,\\
		&202,203,204,205,206,215,216,217,218,219,220,221,222,223,224,225,226,227,228,229,230\\
		\hline
		space groups with strong&48,50,52,54,56,57,59,60,61,62,68,73,112,113,114,116,117,118,120,122,133,142\\
		but without fragile bands&\\
		\hline
		space groups with fragile&143,149,150,156,157,158,159,168,173,177,182,183,184,185,186,199,208,210,214\\
		but without strong bands&\\
		\hline
	\end{tabular}
\end{table}

\subsection{Determination of the fragile root bands}
Once all the basic bands in a space group have been identified, using also the tabulated EBRs of the space group \cite{progbandrep}, it is possible to construct all the kinds of bands in the space group (trivial, strong topological, fragile root bands, fragile but not root bands...), though in this work we will focus on the derivation, in particular, of the fragile roots. Looking at the four kinds of space groups found, it is possible to come to some preliminary conclusions:
\begin{itemize}
	\item In space groups whose basic bands are elementary or fragile, by definition these basic fragile bands are the only fragile roots in the space group.
	\item In those groups whose basic bands are elementary or strong topological, the fragile bands are combinations of basic strong bands and, therefore, all the fragile roots (if any) are direct sums of strong basic bands. 
	\item In the space groups with basic bands of the three types, the basic fragile bands are also root fragile bands, but there can be additional roots as combinations of strong bands.
\end{itemize}

The determination of the fragile roots in a space group is thus immediate in those groups with no strong basic bands. All the fragile basic bands (if any) form the complete set of fragile roots. In the space groups with strong basic bands the determination of the fragile roots has been performed in a step-like process. (1) The starting subset of fragile roots are the fragile basic bands. At this point we define \emph{basis} as the union of the elementary basic bands and the fragile basic bands. (2) In the next step, we construct all the combinations of two strong basic bands and remove from the set those than can be expressed as linear combination with non-negative integers of at least a band of \emph{basis} and strong basic bands. The fragile bands in the remaining set are root bands and are added to \emph{basis}. (3) The \emph{rest} of bands (those direct sums of two strong bands that result in another strong band) are considered in the next step: we combine them with all the strong basic bands to get all the relevant combinations of three strong basic bands. We repeat the process explained in step (2): remove the bands that are combinations of bands in \emph{basic} and another strong band, keep the remaining fragile bands as fragile root bands, add these new roots to \emph{basis}, and consider the combinations that result in another strong bands for the next step (\emph{rest} of bands). We repeat the process combining in each step the remaining set of bands with the strong basic bands.
We have checked that, in all the 117 space groups that have basic strong bands (Table \ref{tb:spacegroups}), at the beginning, the \emph{rest} of bands (to be considered in subsequent steps) increases with the number of strong basic bands combined, until it reaches a maximum. Then, the number of bands in \emph{rest} decreases in all the space groups, until it goes to 0 when a given number of strong basic bands are combined. This number depends on the space group (it goes from 2 in space groups with low symmetry, to a maximum value 18 in space groups P$6/m$ (N. 175) and P$6/mmm$ (N. 191)). Note that, in principle, as the number of strong basic bands in these two extreme space groups is 228 (see Table \ref{basicbands}) the number of combinations of 18 strong basic bands is $228^{18}\simeq 10^{42}$. However, as most of the combinations have been removed in previous steps, the maximum number of bands to be considered is never higher than half a million, what makes the problem tractable. The results obtained through the induction method confirm the results previously obtained using the polyhedron method \cite{song2019} based on the computation of the normalizations of affine monoids, which can be represented by linear diophantine equations and inequalities \cite{bruns2010}. An example of application of this step-like process can be found in section \ref{bands:fragile83} of the SM, where we give details about the determination of the fragile roots in the space group P$4/m$ (N. 83).

\subsection{Relations between the types of bands in the group-subgroup pair.}
The results of the induction process establishes features between the basic bands in the subgroup and the induced bands. We can state that:

\begin{itemize}
	\item EBRs are always induced from EBRs in the subgroup, independently of the group-subgroup pair. Alternatively, given an EBR in a space group, it always subduces into a direct sum of EBRs in any of its subgroups.
	\item Strong topological bands are induced from strong topological bands, fragile bands or from EBRs. In the last two cases, it is only possible to induce a strong topological band when one of the added symmetry operations to the subgroup has as rotational part an improper operation, except a mirror plane, i.e., there must be in the supergroup at least one symmetry operation (not present in the subgroup) that keeps invariant a single point. 
	\item Fragile bands are induced from fragile bands or from EBRs. The second case is only possible when one of the added symmetry operations to the subgroup has as rotational part a 3-fold axis and/or an improper operation different from a mirror plane. 
\end{itemize}

\section{Calculation of the strong bands and strong topological indices through the Smith Normal Form}
\label{sec:smith}
In the preceding sections we have described an extension of the standard induction-subduction procedure to the determination of the types of bands in a solid. The calculation of the basic bands in a space group is based on the knowledge of the basic bands in one of its subgroups and the classification of the basic bands and combinations of basic bands into different types, trivial, fragile topological or strong topological, relies on the known EBRs in the space group. In this section, we introduce an alternative but equivalent formulation of the problem based just on the knowledge of the EBRs in a space group and on the fact that the set of EBRs contains all the necessary information to derive all kinds of bands in the space group.

Let $\rho_{\mathbf{k}_j}^{i}$, $i=1,\ldots,N_{\mathbf{k}_j}$ be the list of the $N_{\mathbf{k}_j}$ irreps (single-valued irreps if spin-orbit coupling is not considered or double-valued irreps when it is considered) at the $N_{\mathbf{k}}$ maximal $\mathbf{k}_j$-vectors ($j=1,\ldots,N_{\mathbf{k}}$) in a space group and $m_{i,\mathbf{k}_j}\ge0$ the multiplicity of the irrep $\rho_{\mathbf{k}_j}^{i}$ in the decomposition of a band B into the irreps at maximal $\mathbf{k}$-vectors. We address the question: can the band given by the $m_i$ multiplicities be expressed as a linear combination of EBRs?

The band $B$ can be represented as a $N$-dimensional "symmetry data vector" whose components are the integers $m_{i,\mathbf{k}_j}$,
\begin{equation}
\label{Bvector}
B=\left(m_{1,\mathbf{k}_1},\ldots,m_{N_{\mathbf{k}_1},\mathbf{k}_1},\ldots\right)^T
\end{equation}
and the EBRs of the space group can also be described in the same form. $N=N_{\mathbf{k}_1}+N_{\mathbf{k}_2}+\ldots$ is the total number of irreps at maximal $\mathbf{k}$-vectors. Given an EBR, we denote as $EBR_i$ the $N$-dimensional column vector whose $j$-th component $EBR_{j,i}$ represents the multiplicity of the $j^{\textrm{th}}$ irrep in the decomposition of the $i^{\textrm{th}}$ EBR into irreps at the maximal $\mathbf{k}$-vectors. Then, the band in Eq. (\ref{Bvector}) can be expressed as a linear combination of the $N_{EBR}$ elementary  band representations if there exist an $N_{EBR}$-dimensional vector $X=(x_1,\ldots,x_{N_{EBR}})$ with integer components such that the set of linear diophantine equations given in matrix form,
\begin{equation}
\label{eqsystem}
EBR\cdot X=B
\end{equation}
is fulfilled. Note that the rank of the matrix $EBR$ in some groups is rank$(EBR)<N_{EBR}$ because some EBRs can be linear combinations of other EBRs with integer coefficients. The equation system (\ref{eqsystem}) can be simplified through the Smith Decomposition of $EBR$ which can be stated as,

\textbf{Theorem 1}  (Smith Normal Form). If $EBR$ is any $m \times n$ integer matrix, then there is a unimodular integer invertible $m \times m$ matrix $L$ and a unimodular integer invertible $n \times n$ matrix $R$ such that
\begin{equation}
\label{schmidtequation} 
\Delta = L\cdot EBR \cdot R, 
\end{equation}
where $\Delta$ is a diagonal $m \times n$ matrix, not necessarily square, known as the Smith Normal Form of EBR. The elements of $\Delta$ are $\Delta_{i,j} = 0$ if $i\ne  j$ and the number of elements in the diagonal different from 0 is the rank of the matrix $EBR$. $L$ and $R$ can be chosen such that $0<\Delta_{1,1}\le\Delta_{2,2},\ldots,\le\Delta_{r,r}$ with $r$=rank$(EBR)$.

Using the Smith Decomposition, the equation (\ref{eqsystem}) can be written as,
\begin{equation}
\label{schmidtequationtransf} 
\Delta \cdot Y = C\hspace{1cm}\textrm{with }Y=R^{-1}\cdot X\textrm{ and }C=L\cdot B
\end{equation}
Since $L$ and $R$ are unimodular matrices, Eq. (\ref{schmidtequationtransf}) is also an equation over the integers.

Due to the diagonal form of $\Delta$, the set of equations (\ref{schmidtequationtransf}) has an integer solution if and only if 
\begin{equation}
\label{comprel}
c_i=0,\,i>r
\end{equation}
and
\begin{equation}
\label{strongindices}
c_i/\Delta_{i,i}\in\mathbb{Z},\,i=1,\ldots,r
\end{equation}
If a solution exists, the $x_i$ integers in Eq. (\ref{eqsystem}) are,
\begin{equation}
\label{eq:Xvec}
X=R\cdot Y\hspace{1cm}\textrm{with}\hspace{1cm}Y=\left(c_1/\Delta_{11},\ldots,c_r/\Delta_{rr},y_1,\ldots,y_{N_{EBR}-r}\right)^T
\end{equation}
where $\left(y_1,\ldots,y_{N_{EBR}-r}\right)$ are free variable integers. Setting all $y_i=0$ we obtain a particular solution of Eq. (\ref{eqsystem}) in terms of a set of linearly independent EBRs.

\subsection{Strong indices, topological classes and compatibility relations}

The existence of, at least, a solution of Eq. (\ref{eqsystem}) requires that the components of the $B$ vector fulfill two types of conditions given by eqs.(\ref{comprel}) and (\ref{strongindices}). The condition (\ref{comprel}) can be written as,
\begin{equation}
\label{Lmatcomp}
\widetilde{C}=\widetilde{L}\cdot B=0
\end{equation}
where the matrix $\widetilde{L}$ is the matrix $L$ once the first $r$ rows have been removed and the $\widetilde{C}$-vector has as components the last $N-r$ components of the $C$-vector in Eq. (\ref{schmidtequationtransf}) \cite{po2017,tang2019}. Its components are, thus, $\widetilde{L}_{i,j}=L_{i+r,j}$ with $i=1,\ldots,N-r$ and $j=1,\ldots,N_{EBR}$.

These conditions are equivalent to the conditions imposed by the compatibility relations along any path between any pair of maximal $\mathbf{k}$-vectors in the space group. As mentioned before, if we consider two maximal $\mathbf{k}$-vectors $\mathbf{k}_1$, $\mathbf{k}_2$, an intermediate $\mathbf{k}$-vector $\mathbf{k}_l$ (a line or a plane) that connects the two maximal vectors $\mathbf{k}_1$ and $\mathbf{k}_2$, and an irrep $\rho_l$ of the little group of $\mathbf{k}_l$, the total multiplicity of the irrep $\rho_l$ upon subduction of the irreps of the little group of $\mathbf{k}_1$ and $\mathbf{k}_2$ must be the same at both sides. These conditions can be expressed as a set of linear equations on the components of $B$,
\begin{equation}
\label{comprelred}
C_{\textrm{comp}}\cdot B=0
\end{equation}
where the compatibility matrix $C_{omp}$ contains as many rows as irreps at all the possible intermediate paths between any pair of maximal $\mathbf{k}$-vectors in the space group. For a detailed explanation of the construction of the compatibility matrix see Section (\ref{sup:compasmith}) of the SM. We have checked that, rank($\widetilde{L}$)=rank($C_{omp}$)=rank($\widetilde{L}\cup C_{omp}$) in the 230 space groups with and without TR for single-valued and double-valued irreps in agreement with the results in ref. [\onlinecite{tang2019}]. Therefore, the set of restrictions given by the matrix $\widetilde{L}$ is equivalent to the restrictions imposed by the compatibility relations. In other words, a set of multiplicities that define a $B$-vector (\ref{Bvector}) that do not fulfill Eq. (\ref{Lmatcomp}) do not form a band as defined in Section (\ref{sec:definitions}).

The second condition (\ref{strongindices}) implies extra restrictions when $\Delta_{i,i}>1$. In these cases, the restrictions can be written as,
\begin{equation}
\label{restmod}
c_i=0\,\,\textrm{mod }\Delta_{i,i}
\end{equation}
If the $B$-vector satisfies the compatibility relations (\ref{Lmatcomp}) but do not fulfill all the conditions (\ref{restmod}), the band given by Eq. (\ref{Bvector}) is strong topological and the $c_i$ components in Eq. (\ref{restmod}) can be considered a set of strong topological indices of the space group, with $c_i=L_i\cdot B$ and $L_i$ the $i$-th row of the $L$ matrix in the Smith Decomposition of $EBR$ \cite{tang2019}.

For instance, the Smith Normal Form $\Delta$ of the $EBR$ matrix in the space group P$\overline{1}$ (N. 2) with TR has three diagonal elements $\Delta_{i,i}=2$ and one diagonal element $\Delta_{i,i}=4$. Therefore, there are 4 indices $c_i$ than can take $2\times2\times2\times4=32$ different values (32 topological classes), being $=c_{2,1}=c_{2,1}=c_{2,3}=c_4=0$ the only set of values of the coefficients that fulfills Eq. (\ref{strongindices}) or (\ref{restmod}) and, therefore, it corresponds to a trivial or fragile band. A detailed analysis of the determination of the strong topological indices of this space group (first in 2-D and then in 3-D) is explained in Section (\ref{sup:strongtopo}) of the SM.
\section{Determination of the fragile phases and fragile roots through the Smith Decomposition}
\label{fragileroots:main}
The determination of the topological strong phases and the definition of the topological indices in a space group is almost immediate once the multiplicities of the EBRs are known, as it has been shown in the previous section. However, the determination of the fragile root is much more difficult. We can state the problem in the following terms: which conditions must satisfy a set of multiplicities given by the symmetry data $B$-vector in Eq. (\ref{Bvector}) to form a fragile band and, among the fragile bands, which are fragile roots?

First, the multiplicities must fulfill the compatibility relations. From the Smith decomposition of the matrix $EBR$ the condition for the symmetry data $B$ vector to fulfill the compatibility relations is Eq. (\ref{Lmatcomp}). We now have obtained an integer equation which would give us the general solution for the band vector $B$ in Eq. (\ref{Bvector}). This is easily solved as follows: first, we do a Smith decomposition 
\begin{equation}
\Delta_{\text{comp}} = L _{\text{comp}} \cdot C_{\textrm{comp}} \cdot R_{\text{comp}} 
\end{equation}
We then look for the nonzero components in the matrix $\Delta_{\text{comp}}$. This is a diagonal matrix and has the first $p$ components equal to 1: $(\Delta_{\text{comp}})_{1,1}=\ldots=(\Delta_{\text{comp}})_{p,p}=1$ and $(\Delta_{\text{comp}})_{p+1,p+1} =\ldots = (\Delta_{\text{comp}})_{m,m}  = 0 $. The condition (\ref{Lmatcomp}) can be written as,
\begin{equation}
L _{\text{comp}}^{-1}\cdot\Delta_{\text{comp}}\cdot R_{\text{comp}}^{-1}\cdot B=0\hspace{0.5cm}\rightarrow\hspace{0.5cm}\Delta_{\text{comp}}\cdot R_{\text{comp}}^{-1}\cdot B=0
\end{equation}
and due to the special values of the diagonal matrix $\Delta_{\text{comp}}$,
\begin{equation}
R_{\text{comp}}^{-1}\cdot B=Y_p=(0,\ldots,0,y_{p+1},\ldots,y_m)^T
\end{equation}
$Y_p$ is a $m$-dimensional vector whose first $p$ components are 0 and the remaining $m-p$ components are, for now, integers that must fulfill some conditions. The general $B$-vector that satisfies the compatibility relations is thus,
\begin{equation}
\label{eq:finalBvec}
B=R_{\text{comp}}\cdot Y_p
\end{equation}
As the components of the $B$-vector must be non-negative integers, Eq. (\ref{eq:finalBvec}) restricts the possible sets of allowed components of $Y_p$ through the matrix $R_{\text{comp}}$.

Once the general form of the $B$-vector that fulfills the compatibility relations has been obtained, we now have to ensure that the band is not strong topological. We then build the $C=L\cdot B$ matrix (\ref{schmidtequationtransf}) and define the diagonal $n\times N_{EBR}$ matrix $\Delta^{-1}$ as: $\Delta^{-1}_{i,i}=1/\Delta_{i,i}$ for $i\le r$, being $r$ the rank of $\Delta$ in Eq. (\ref{schmidtequation}). So constructed, as in Eqs. (\ref{schmidtequationtransf}) and (\ref{comprel}), the $C$ matrix has coefficients $c_i=0$ for $i>r$, provided that $B$ has been forced to fulfill the compatibility relations. Finally, we build the EBR vector $V_{EBR}$,
\begin{equation}
\label{eq:vEBRvector}
V_{EBR}=R\cdot\Delta^{-1}\cdot L\cdot B=R\cdot\Delta^{-1}\cdot L\cdot R_{\text{comp}}\cdot Y_p
\end{equation}
whose components are the coefficients of the EBRs in the linear equations that give the parametrization of any band $B$ that fulfills the compatibility relations. Note that $V_{EBR}$ is $X$ in Eq. (\ref{eq:Xvec}), but we have changed the notation here to stress explicitly that this vector is a linear combination of EBRs. If at least one component is not an integer number, the band represented by $B$ is strong topological. If all the coefficients are integer numbers it is fragile or trivial. If all the coefficients are non-negative, the band is trivial. However, as the set of EBRs is, in general, an overcomplete basis, a non-negative component $V_{EBR}$ does not ensure that the band is not trivial. It should be checked that there is no other $V_{EBR}$-vector with non-negative coefficients that gives the same $B$. If there is such $V_{EBR}$-vector, the band is trivial.

The whole procedure to identify the fragile phases thus relies on the Smith decomposition of the $EBR$ matrix. It is clear that the number of fragile phases, i.e., number of solutions of Eq. (\ref{eq:vEBRvector}) with at least a negative integer will depend on the rank of $EBR$. For instance, in those groups where rank($EBR$)=1 there cannot exist fragile phases. The vector (of non-negative components) that represents any EBR is a multiple of the unique basis-vector which, in principle, could be or not an EBR. In any case, the components of this vector are non-negative integers and, ultimately, any band in this group can be expressed as a multiple (positive) of this vector basis. This does not mean that no fragile phases exist. It just means that, if they exist, they cannot be identified by symmetry indices. In fact, in several groups analyzed in section \ref{fragileroots:Smith} in the SM, decomposable EBRs do exist. However, each branch of a decomposable EBR has characters at high symmetry points that can be expressed entirely as sums of other EBRs. The non-wannierizable character of their bands has to be proved by other methods which make use of Berry phases \cite{bradlyn2019}. We call these phases \emph{Berry Fragile Phases}, to differentiate them from the Eigenvalue Fragile Phases which can only be written in terms of a sum and (necessarily) a difference of EBRs.

 It is convenient, thus, to do the analysis starting from the groups whose $EBR$ matrix has the lowest rank. The ranks of the double-valued $EBR$ matrices in the space groups with TR covers all the integer numbers from 1 to 14, except the value 12. If single-valued $EBR$s are considered (no SOC), the highest rank is 27. Tables \ref{tb:ranksdoubles} and \ref{tb:rankssingles} give the rank of the $EBR$ matrix for each space group with and without SOC, respectively. It is interesting to remark that, according to Table (\ref{tb:ranksdoubles}), when spin-orbit coupling is considered, the space groups with highest ranks (from 11 to 14) are symmorphic space groups that contain two or more symmetry operations whose rotational part keeps a single point fixed. Moreover, symmorphic space groups with primitive unit cell in the standard setting have higher rank than symmorphic groups with the same point group but non-primitive centering. In section \ref{fragileroots:Smith} of the SM the above procedure is applied to space groups of rank 2 and 3.

\begin{table}
\caption{List of ranks of the double-valued $EBR$ matrices (with SOC) in the space groups with TR.\label{tb:ranksdoubles}}
\begin{tabular}{l|l}
	rank&space groups\\
\hline
1&1,3,4,5,6,7,8,9,16,17,18,19,20,21,22,23,24,25,26,27,28,29,30,31,32,33,34,35,\\ &36,37,38,39,40,41,42,43,44,45,46,76,77,78,80,91,92,93,94,95,96,98,101,102,\\
&105,106,109,110,144,145,151,152,153,154,169,170,171,172,178,179,180,181\\
\hline
2&79,90,97,100,104,107,108,146,155,160,161,195,196,197,198,199,208,210,212,213,214\\
\hline3&48,50,52,54,56,57,59,60,61,62,68,70,73,75,89,99,103,112,113,114,116,\\
&117,118,120,122,133,142,150,157,159,173,182,185,186,209,211\\
\hline
4&63,64,72,121,126,130,135,137,138,143,149,156,158,168,177,183,184,207,218,219,220\\
\hline
5&11,13,14,15,49,51,53,55,58,66,67,74,81,82,86,88,111,115,119,134,136,141,167,217,228,230\\
\hline
6&69,71,85,125,129,132,163,165,190,201,203,205,206,215,216,222\\
\hline
7&12,65,84,128,131,140,188,189,202,204,223\\
\hline
8&124,127,148,166,193,200,224,226,227\\
\hline
9&2,10,47,87,139,147,162,164,176,192,194\\
\hline
10&174,187\\
\hline
11&225,229\\
\hline
13&83,123\\
\hline
14&175,191,221\\
\hline
\end{tabular}
\end{table}

\begin{table}
	\caption{List of ranks of the double-valued $EBR$ matrices (without SOC) in the space groups with TR.\label{tb:rankssingles}}
	\begin{tabular}{l|l}
		rank&space groups\\
		\hline
		1&1,4,7,9,19,29,33,76,78,144,145,169,170\\
		\hline
		2&8,31,36,41,43,80,92,96,110,146,161,198\\
		\hline
		3&5,6,18,20,26,30,32,34,40,45,46,61,106,109,151,152,\\
		&153,154,159,160,171,172,173,178,179,199,212,213\\
		\hline
		4&24,28,37,39,60,62,77,79,91,95,102,104,\\
		&143,155,157,158,185,186,196,197,210\\
		\hline
		5&3,14,17,27,42,44,52,56,57,94,98,100,\\
		&101,108,114,122,150,156,182,214,220\\
		\hline
		6&11,15,35,38,54,70,73,75,88,90,103,105,107,\\
		&113,142,149,167,168,184,195,205,219\\
		\hline
		7&13,22,23,59,64,68,82,86,117,118,120,130,163,\\
		&165,180,181,203,206,208,209,211,218,228,230\\
		\hline
		8&21,58,63,81,85,97,116,133,135,137,148,183,190,201,217\\
		\hline
		9&2,25,48,50,53,55,72,99,121,126,138,141,147,188,207,216,222\\
		\hline
		10&12,74,93,112,119,176,177,202,204,215\\
		\hline
		11&66,84,128,136,166,227\\
		\hline
		12&51,87,89,115,129,134,162,164,174,189,193,223,226\\
		\hline
		13&16,67,111,125,194,224\\
		\hline
		14&49,140,192,200\\
		\hline
		15&10,69,71,124,127,132,187\\
		\hline
		17&225,229\\
		\hline
		18&65,83,131,139,175\\
		\hline
		22&221\\
		\hline
		24&191\\
		\hline
		27&47,123\\
		\hline
	\end{tabular}
\end{table}

\section{Determination of a minimum set of independent basic bands}
\label{equivalences}
In the previous sections we have described two ways to calculate the basic bands of a space group and, from these basic bands, the identification of the fragile roots. In general, there are correlations between the multiplicities of different basic bands. In particular, we can define two different kinds of isomorphism in the whole set of basic bands:  (a) through conjugation of operations that belong to the affine normalizer (or affine stabilizer) \cite{koch2016} of the space group and (b) through a special case of the Kronecker product of irreducible representations of the space group, when one of the irreps is 1-dimensional. These two kinds of isomorphisms allows to reduce the number of independent basic bands to a minimum set of bands. Each subset of isomorphisms form an automorphism group, and the join of both automorphism groups form the automorphism group of the set of bands in a space group. In the next two sections we analyze these automorphism groups.
\subsection{Reduction on the number of independent basic bands through elements of the affine normalizer of the group}
\label{main:normalizers}
Given a space group ${\mathcal{G}}$ and one of its supergroups ${\mathcal{S}}$, there is a unique intermediated group ${\mathcal{N}_S(\mathcal{G})}$ called the normalizer of ${\mathcal{G}}$ with respect to ${\mathcal{S}}$ \cite{ita}. A symmetry operation $\{N|\mathbf{n}\}$ of ${\mathcal{S}}$ belongs to ${\mathcal{N}_S(\mathcal{G})}$ if it maps the group ${\mathcal{G}}$ into itself through conjugation, i.e., 
\begin{equation}
	\mathcal{N}_S(\mathcal{G}):=\{\{N|\mathbf{n}\}\in\mathcal{S}|\{N|\mathbf{n}\}^{-1}\mathcal{G}\{N|\mathbf{n}\}\in\mathcal{G} \}.
\end{equation}

But in general the elements of ${\mathcal{N}_S(\mathcal{G})}$ do not map a subgroup of ${\mathcal{G}}$ into itself. In general, it maps a subgroup of ${\mathcal{G}}$ into another (conjugated) subgroup of ${\mathcal{G}}$. In particular, let $\{R|\mathbf{t}\}$ a symmetry operation that belongs to the little group of a given $\mathbf{k}$ vector, i.e. $\mathbf{k}=\mathbf{k}\cdot R$ mod  translations of the reciprocal lattice. The symmetry operation,
\begin{equation}
	\{N|\mathbf{n}\}^{-1}\{R|\mathbf{t}\}\{N|\mathbf{n}\}=\{N^{-1}\cdot R\cdot N|N^{-1}\cdot\left(-\mathbf{n}+\mathbf{t}+R\cdot\mathbf{n}\right)\}
\end{equation}
belongs to the little group of $\mathbf{k}'=\mathbf{k}\cdot N$,
\begin{equation}
	\mathbf{k}'\cdot N^{-1}\cdot R\cdot N=\mathbf{k}\cdot N\cdot N^{-1}\cdot R\cdot N=\mathbf{k}\cdot R\cdot N=\mathbf{k}\cdot N=\mathbf{k}'
\end{equation}
mod translations of the reciprocal lattice. The little groups of $\mathbf{k}$ and $\mathbf{k}'$ are thus conjugated and it is possible to establish a 1 to 1 relation between the irreps of both groups. If $\{\mathbf{k}_1,\ldots,\mathbf{k}_r\}$ is the set of $\mathbf{k}$-vectors of maximal symmetry of ${\mathcal{G}}$, $\{\mathbf{k}_1\cdot N,\ldots,\mathbf{k}_r\cdot N\}$ is also a set of $\mathbf{k}$-vectors of maximal symmetry, being $N$ the rotational part of an element of the normalizer. As the subset of $\mathbf{k}$-vectors of maximal symmetry is unique in each space group, the second list must contain all the $\mathbf{k}$-vectors of the first list. Therefore, the operation $\{N|\mathbf{n}\}$ of the normalizer maps the set of maximal irreps into itself,
\begin{equation}
	\label{eq:conjirreps}
	\left(\rho_1^{\mathbf{k}_1},\ldots,\rho_{n_{{\mathbf{k}}_1}}^{\mathbf{k}_1},\rho_1^{\mathbf{k}_2},\ldots,\rho_{n_{{\mathbf{k}}_2}}^{\mathbf{k}_2},\ldots,\rho_1^{\mathbf{k}_r},\ldots,\rho_{n_{{\mathbf{k}}_r}}^{\mathbf{k}_r}\right)\xrightarrow{N}\left(\rho_1^{\mathbf{k}_1\cdot N},\ldots,\rho_{n_{{\mathbf{k}}_1\cdot N}}^{\mathbf{k}_1\cdot N},\rho_1^{\mathbf{k}_2\cdot N},\ldots,\rho_{n_{{\mathbf{k}}_2\cdot N}}^{\mathbf{k}_2\cdot N},\ldots,\rho_1^{\mathbf{k}_r\cdot N},\ldots,\rho_{n_{{\mathbf{k}}_r\cdot N}}^{\mathbf{k}_r\cdot N}\right)
\end{equation}  
The set of maximal irreps on the left and on the right are the same but, in general, re-ordered (see the example in section \ref{sup:equivnormal} of the SM). We can define a $N_{irr}\times N_{irr}$ matrix $M_{\{N|\mathbf{n}\}}$, with $N_{irr}=n_{{k}_1}+\ldots+n_{{k}_r}$ being the number of maximal irreps, such that the $ij$ element is zero unless the $i$-th irrep on the left in Eq. (\ref{eq:conjirreps}) and the $j$-th irrep on the right are related through conjugation. In this case, the element is 1. Therefore $M_{\{N|\mathbf{n}\}}$ is a permutation matrix. i.e., a square matrix that has exactly one entry of 1 in each row and each column and 0 elsewhere.

For each member $\{N|\mathbf{n}\}$ of the affine normalizer, we can define a unique $M_{\{N|\mathbf{n}\}}$ matrix and these matrices form a finite group (it is an automorphism group) which we can denote as $\mathcal{N}_G$. Although the number of operations of the normalizer is, in general, infinite, the order of $\mathcal{N}_G$ is finite. This group is a subgroup of the group of all the permutation matrices of dimension $N_{irr}\times N_{irr}$ whose order is $N_{irr}!$.

Note that some operations of the normalizer correspond just to an origin shift (when $N=E$). In these special cases, as $\mathbf{k}R=\mathbf{k}$, this particular automorphism maps the irreps of the little group of  $\mathbf{k}$ into themselves, for any $\mathbf{k}$-vector.

In section \ref{sup:equivnormal} we present an example of the calculation of $\mathcal{N}_G$ in the space group P$\overline{1}$. Table \ref{normalizers:data} in the SM gives, for each space group, the order of $\mathcal{N}_G$.
\subsection{Reduction on the number of independent basic bands through Kronecker products of irreps}
\label{kronecker}
The Kronecker (on inner) product of irreps of a group $\mathcal{G}$ is a representation of $\mathcal{G}$. In general, it is equivalent to the direct sum of irreps of $\mathcal{G}$. In a physical system, if we have two set of states (in two different subspaces) that transform under two representations, the tensor product of the two sets of states transforms under the Kronecker product of the representations.

Let $\rho^{\mathbf{k}_1}$ and $\rho^{\mathbf{k}_2}$ two irreps of the little groups $\mathcal{G}^{\mathbf{k_1}}$ and $\mathcal{G}^{\mathbf{k_2}}$ of $\mathbf{k}_1$ and $\mathbf{k}_2$, respectively. We denote as $^{*}\rho^{\mathbf{k}_1}$ and $^{*}\rho^{\mathbf{k}_2}$ the induced irreps (known as \emph{full} irreps) from $\rho^{\mathbf{k}_1}$ and $\rho^{\mathbf{k}_2}$ into $\mathcal{G}$, i.e.,
\begin{equation}
^{*}\rho^{\mathbf{k}_i}=\rho^{\mathbf{k}_i} \uparrow\mathcal{G},\hspace{1cm} \rho^{\mathbf{k}_i}\in \mathcal{G}^{\mathbf{k}_i},\,^{*}\rho^{\mathbf{k}_i}\in \mathcal{G}
\end{equation}			

Let $\{\mathbf{k}_1^{i}\}$ with $i=1,\ldots,n_1$ be the $n_1$ $\mathbf{k}$-vectors of the star of $\mathbf{k}_1\equiv \mathbf{k}_1^1$ and $\{\mathbf{k}_2^{i}\}$ with $i=1,\ldots,n_2$ be the $n_2$ $\mathbf{k}$-vectors of the star of $\mathbf{k}_2\equiv \mathbf{k}_2^1$. In general, the set of $n_1*n_2$ vectors $\{\mathbf{k}_1^i+\mathbf{k}_2^j\}$ with $i=1,\ldots,n_1$ and $j=1,\ldots,n_2$ is the direct sum of different stars of $\mathcal{G}$
\begin{equation}
	\label{kroneckersum}
\{\mathbf{k}_1^i+\mathbf{k}_2^j\}=\cup\{\mathbf{k}_m^i\}\hspace{1cm} m=3,\ldots 
\end{equation}			

The Kronecker product of the two full irreps $^{*}\rho^{\mathbf{k}_1}$
 and $^{*}\rho^{\mathbf{k}_2}$ is equivalent to the direct sum of full irreps induced from the irreps of the little groups of $\mathbf{k}_m\equiv\mathbf{k}_m^1$, $m=3,\ldots$ in the right side of Eq. (\ref{kroneckersum}),
 
 \begin{equation}
 	\label{kroneckerprod}
 	^{*}\rho^{\mathbf{k}_1}\otimes\, ^{*}\rho^{\mathbf{k}_2}=\sum_{m=3}\sum_{i=1}^{n_m}c_{m,i,1,2}\,^{*}\rho_i^{\mathbf{k}_m}
 \end{equation}			
where $^{*}\rho_i^{\mathbf{k}_m}$ is the full irrep induced from the $i$-th irrep $\rho_i^{\mathbf{k}_m}$ of the little group $\mathcal{G}^{\mathbf{k_m}}$ of $\mathbf{k}_m$ into $\mathcal{G}$, $n_m$ is the number of irreps of $\mathcal{G}^{\mathbf{k_m}}$ and $c_{m,i,1,2}$ are the Clebsch-Gordan coefficients. The summation in $m$ extends to the number of stars in the decomposition in Eq. (\ref{kroneckersum}).

For our purposes, we consider a particular case of the general Kronecker product of irreps: we take as $^{*}\rho^{\mathbf{k}_1}$ in Eq. (\ref{kroneckerprod}) a 1-dimensional full irrep for a given $\mathbf{k}_1$-vector. This means that, on the one hand, the star of $\mathbf{k}_1$ contains just $\mathbf{k}_1$ (the little group of $\mathbf{k}_1$ is then the whole space group $\mathcal{G}$) and, on the other hand, the irrep $\rho^{\mathbf{k}_1}$ of the little group of $\mathbf{k}_1$ is 1-dimensional. In this particular case, the full irrep $^{*}\rho^{\mathbf{k}_1}$ coincides with the irrep $\rho^{\mathbf{k}_1}$ and the decomposition in Eq. (\ref{kroneckersum})
reduces to a single star: the star of $\mathbf{k}_3=\mathbf{k}_1+\mathbf{k}_2$. Note also that, as $\mathbf{k}_1.R\equiv\mathbf{k}_1$ for any operation $R$ of the point group of $\mathcal{G}$, the little group of $\mathbf{k}_2$ and $\mathbf{k}_3$ is the same, $\mathcal{G}^{\mathbf{k_2}}=\mathcal{G}^{\mathbf{k_3}}$ and there is a 1:1 correspondence between the full irreps induced from these two little groups. As $^{*}\rho^{\mathbf{k}_2}$ is an irrep of $\mathcal{G}$ and the dimension of $^{*}\rho^{\mathbf{k}_1}\otimes ^{*}\rho^{\mathbf{k}_2}$ in Eq. (\ref{kroneckerprod}) is the same as the dimension of $^{*}\rho^{\mathbf{k}_2}$, then $^{*}\rho^{\mathbf{k}_1}\otimes ^{*}\rho^{\mathbf{k}_2}$ is also an irrep. Therefore, in the decomposition given by Eq. (\ref{kroneckerprod})  one coefficient is 1 and the rest ones are 0. This equation establish thus an homomorphism between the full irreps $^{*}\rho_i^{\mathbf{k}_2}$ and $^{*}\rho_i^{\mathbf{k}_3}$. This  also implies an homomorphism  between the irreps $\rho_i^{\mathbf{k}_2}$ and $\rho_i^{\mathbf{k}_3}$ of the little groups $\mathcal{G}^{\mathbf{k_2}}$ and $\mathcal{G}^{\mathbf{k_3}}$.
If we calculate the Kronecker product of $^{*}\rho^{\mathbf{k}_1}$ and all the irreps of the little groups of all the maximal $\mathbf{k}$-vectors, the Kronecker product maps the set of maximal irreps into itself, as in Eq. (\ref{eq:conjirreps}).
 
It is important to stress that the above automorphism is only established when the full irrep $^{*}\rho^{\mathbf{k}_1}$ is  1-dimensional. If TR is not considered, the 1-dimensional   irrep $^{*}\rho^{\mathbf{k}_1}$ can be real or not, but with TR symmetry the irrep must be real and thus, it must be a single-valued irrep. Therefore, when TR is considered, for each 1-dimensional irrep, an automorphism is established between the single-valued irreps on one hand (without SOC), and another automorphism between double-valued irreps on the other hand (with SOC). The automorphism under the Kronecker product by $^{*}\rho^{\mathbf{k}_1}=\rho^{\mathbf{k}_1}$ can also be described by a $N_{irr}\times N_{irr}$ permutation matrix $M_{\rho^{\mathbf{k}_1}}$. 

The existence of the homomorphism between the irreps at two different maximal $\mathbf{k}$-vectors under the Kronecker product by a 1-d irrep can be easily checked for the matrices of the translation operators. The matrices of the translations $\{E|\mathbf{T}\}$ of any full irrep of the little group $\mathcal{G}^{\mathbf{k}}$ of $\mathbf{k}$ are diagonal matrices,  $D_{\rho^{\mathbf{k}}}(\mathbf{T})=e^{i\mathbf{k}\cdot\mathbf{T}}\mathds{1}_d$, where $d$ is the dimension of the irrep. We can write this matrix indicating only the diagonal elements as,

\begin{equation}
	\label{eq:mattras}
D_{\rho^{\mathbf{k}}}(\mathbf{T})=\{e^{i\mathbf{k}\cdot\mathbf{T}},\ldots,e^{i\mathbf{k}\cdot\mathbf{T}}\}
\end{equation}
The matrices of the full representations are 
\begin{equation}
	\label{eq:mattrasfull}
^{*}D_{^{*}\rho^{\mathbf{k}}}(\mathbf{T})=\{e^{i\mathbf{k}\cdot\mathbf{T}},\ldots,e^{i\mathbf{k}\cdot\mathbf{T}},e^{i\mathbf{k}^2\cdot\mathbf{T}},\ldots,e^{i\mathbf{k}^2\cdot\mathbf{T}},\ldots,e^{i\mathbf{k}^n\cdot\mathbf{T}},\ldots,e^{i\mathbf{k}^n\cdot\mathbf{T}}\}
\end{equation}
where $\{\mathbf{k}\equiv\mathbf{k}^1,\mathbf{k}^2,\ldots,\mathbf{k}^n\}$ is the star of $\mathbf{k}$ and the symbol $^{*}$ stands for a matrix of the full irrep.

If $\mathbf{k}_1$ is a vector whose little group is the whole space group $\mathcal{G}^{\mathbf{k}_1}=\mathcal{G}$ (therefore its star contains just the vector $\mathbf{k}_1$ itself) and $\rho^{\mathbf{k}_1}$ is a 1-d irrep of $\mathcal{G}^{\mathbf{k}_1}$, the matrices of the translations of $\mathcal{G}^{\mathbf{k}_1}$ and $\mathcal{G}$ for this irrep are,

\begin{equation}
D_{\rho^{\mathbf{k}_1}}(\mathbf{T})=\, ^{*}D_{^{*}\rho^{\mathbf{k}_1}}(\mathbf{T})=\{e^{i\mathbf{k}_1\cdot\mathbf{T}}\}
\end{equation}

Now we consider a $\mathbf{k}$-vector with maximal symmetry, $\mathbf{k}_2$. The matrices for the translations of an irrep $\rho^{\mathbf{k}_2}$ of the little group $\mathcal{G}^{\mathbf{k}_2}$ of $\mathbf{k}_2$ and the matrices of the full irrep are given by Eqs.  (\ref{eq:mattras}) and (\ref{eq:mattrasfull}), respectively, with $\mathbf{k}=\mathbf{k}_2$. 

The Kronecker product of the matrices of the full irreps are,

\begin{equation}
^{*}D_{^{*}\rho^{\mathbf{k}_1}}(\mathbf{T})^{*}D_{^{*}\rho^{\mathbf{k}_2}}(\mathbf{T})=\{e^{i\left(\mathbf{k}_1+\mathbf{k}_2\right)\cdot\mathbf{T}},\ldots,e^{i\left(\mathbf{k}_1+\mathbf{k}_2\right)\cdot\mathbf{T}},e^{i\left(\mathbf{k}_1+\mathbf{k}_2^2\right)\cdot\mathbf{T}},\ldots,e^{i\left(\mathbf{k}_1+\mathbf{k}_2^2\right)\cdot\mathbf{T}},\ldots,e^{i\left(\mathbf{k}_1+\mathbf{k}_2^n\right)\cdot\mathbf{T}},\ldots,e^{i\left(\mathbf{k}_1+\mathbf{k}_2^n\right)\cdot\mathbf{T}}\}
\end{equation}

They correspond to the matrices of the translations of a full irrep $^{*}\rho^{\mathbf{k}_3}$ induced by an irrep $\rho^{\mathbf{k}_3}$ of the little group of $\mathbf{k}_3=\mathbf{k}_1+\mathbf{k}_2$.

Once we have identified  all the 1-dimensional single-valued irreps in all the single-vector stars in a space group, we can define one automorphism (and then a permutation matrix) for each Kronecker product. This set of matrices form an automorphism group which we denote as $\mathcal{K}_G$,
and which is also a subgroup of the group of all the permutation matrices of dimension $N_{irr}\times N_{irr}$. Table \ref{normalizers:data} in the SM gives, for each space group, the order of $\mathcal{K}_G$. 

In section \ref{sec:kronecker}we present an example of the calculation of $\mathcal{K}_G$ in the space group P$\overline{1}$.

Finally, the join of the two previous automorphism groups $\mathcal{J}_G=\langle\mathcal{N}_G,\mathcal{K}_G\rangle$ gives the complete automorphism group of the space group $\mathcal{G}$ that maps the irreps of the little groups of maximal $\mathbf{k}$-vectors into itself. This group is also a subgroup of the permutation group of matrices of dimension $N_{irr}\times N_{irr}$ and, in general, is not the direct product of $\mathcal{N}_G$ and $\mathcal{K}_G$. Table \ref{normalizers:data} in the SM also gives the order of $\mathcal{J}_G$. It can be checked that, in general, $|\mathcal{J}_G|\le|\mathcal{N}_G|*|\mathcal{K}_G|$.

The automorphism group can be applied to the set of basic bands, to the set of EBRs of to the set of fragile roots of each space group to identify a minimal subset of independent basic, EBRs and fragile roots, respectively. Table \ref{frag:ind:roots} in the SM gives the number of fragile roots and the number of independent fragile roots by space group.

\section{Conclusions}
The set of electronic bands in a solid can be described as a linear combination of a limited number of basic units (basic bands), identified by the irreducible representations at every $\mathbf{k}$-vector of maximal symmetry. These basic bands are the smallest units that fulfill the compatibility relations, in the sense that they cannot be split into subsets of bands that internally satisfy the compatibility relations. These basic bands are thus the building blocks of any electronic band structure. On the other hand, the elementary band representations are the basis of any set of bands that can be induced from localized wannier functions in the unit cell of the solid. Both subsets, the basic bands and the elementary band representations, allows to characterize all the possible types of bands (topologically trivial, strong or fragile) that can appear in a material.

The induction procedure is a powerful technique to derive all the basic bands in all the space groups starting from the unique trivial basic band in the space group P1 following group-subgroup chains. On the other hand, we have demonstrated that the Smith Decomposition of the matrix constructed from the multiplicities of the elementary bands representations on the irreps of the little groups at the $\mathbf{k}$-vector of maximal symmetry contains all the information about the all kinds of bands in all the space groups. Both methods give rise to the same result and can be considered thus as equivalent.

Once all the basic bands have been calculated through the induction procedure, we have identified all the fragile root bands in all the space groups. The result confirms the previous derivation through the polyhedron method \cite{song2019}. We have also calculated the automorphism group of all the space groups, whose elements map the set of basic bands into itself. The elements of this group are deduced from the operations of the affine normalizer of the space group and from the Kronecker products of the irreps at $\mathbf{k}$-vectors of maximal symmetry and the 1-dimensional single-valued irreps in the space group. These mappings enable us to reduce the number of independent basic bands. These equivalences have also been applied to reduce the number of independent fragile root bands.

\acknowledgments{
	Z.S. and B.B. are supported by the Department of Energy Grant No. desc-0016239, the National Science Foundation EAGER Grant No. DMR 1643312, Simons Investigator Grants No. 404513, No. ONR N00014-14-1-0330, No. NSF-MRSECDMR DMR 1420541, the Packard Foundation No. 2016-65128, the Schmidt Fund for Development of Majorama Fermions funded by the Eric and Wendy Schmidt Transformative Technology Fund.
	L.E. is supported by the Government of the Basque Country (project IT1301-19). 
	This research used resources of the National Energy Research Scientific Computing Center (NERSC), a U.S. Department of Energy Office of Science User Facility operated under Contract No. DE-AC02-05CH11231.}
\bibliography{references.bib}

\newpage

\setcounter{page}{1}
\begin{center}
\textbf{\center{\large{Supplementary Material}}}
\end{center}

\beginsupplement

\section{Summary of the induction procedure in the calculation of the irreps in a space group from the irreps of one of its maximal subgroups.}
\label{sup:induction}
In this section we summarize the standard induction procedure for the calculation of the irreducible representations (irreps) of a space group ${\mathcal{G}}$ induced from the irreps of one of its normal subgroups, ${\mathcal{H}}\lhd{\mathcal{G}}$. The first step consists in the decomposition of ${\mathcal{G}}$ into coset representatives with respect to ${\mathcal{H}}$,
\begin{equation}
\label{eq:cosets}
\mathcal{G}=\mathcal{H}g_1 \cup  \mathcal{H} g_2 \cup  \ldots \cup \mathcal{H} g_n
\end{equation}
where the first representative $g_1\in\mathcal{H}$ is usually taken as the identity, and for $i\neq1$, $g_i\in\mathcal{G}$ but $g_i\notin\mathcal{H}$. In crystallographic groups, it is always possible for every space group $\mathcal{G}$ to choose as $\mathcal{H}$ one of its maximal subgroups such that $n=2,3$. When $n=3$, we will choose $g_3=g_2^2$.

In the next step the irreps of ${\mathcal{H}}$ are distributed into orbits under $G$. Let be $\rho$ an irrep of $\mathcal{H}$ and $D_{\rho}(h)$ the matrix of this irrep of the symmetry operation $h\in\mathcal{H}$. Being $\mathcal{H}$ a normal subgroup, the symmetry operation obtained through conjugation of $h\in\mathcal{H}$ by the coset representatives $g_i$ in Eq. (\ref{eq:cosets}) belongs to $\in\mathcal{H}$, i.e., $g_i^{-1}hg_i\in\mathcal{H}$. Therefore, the set of matrices $D_{\rho}(g_i^{-1}hg_i)$ form an irrep of ${\mathcal{H}}$. It is said that this irrep belongs to the orbit of $\rho$ in ${\mathcal{G}}$ and the irreps that belong to the same orbit are thus mutually conjugated in ${\mathcal{G}}$. Given an irrep $\rho$, two different kinds of orbits can result:
\begin{itemize}
	\item The irrep $\rho$ is self-conjugated in ${\mathcal{G}}$, i.e., the set of matrices $D_{\rho}(g_i^{-1}hg_i)$ correspond to the same irrep as $D_{\rho}(h)$. In crystallographic groups, where the index $|G|/|H|=2,3$, the irrep $\rho$ of ${\mathcal{H}}$ induces a reducible representation in ${\mathcal{G}}$ which is reduced into 2 (when $n=2$) or 3 (when $n=3$) irreps in ${\mathcal{G}}$ of the same dimension of $\rho$.
	\item The orbit contains 2 (for $n=2$) or 3 (for $n=3$) different irreps. These 2 or 3 irreps of $\mathcal{H}$ \emph{combine} to give a single irrep of ${\mathcal{G}}$ whose dimension is 2 (for $n=2$) or 3 (for $n=3$) times the dimension of $\rho$.
\end{itemize}
Once the orbits of each irrep of $\mathcal{H}$ have been identified, the matrices of the induced irreps in $\mathcal{G}$ can be easily obtained. For a full description of the procedure to calculate the matrices see ref. (\onlinecite{aroyo2006}). For any of the 230 crystallographic space groups $\mathcal{G}$, it is possible to construct a composition series,
\begin{equation}
\mathcal{T}=\mathcal{H}_1\lhd\mathcal{H}_2\lhd\ldots\lhd\mathcal{H}_n=\mathcal{G}
\end{equation}
such that $\mathcal{H}_{i-1}$ is a normal subgroup of $\mathcal{H}_i$ of index 2 or 3, and $\mathcal{T}$ is the translation group. Then, we can apply the induction procedure to successive pairs of group-subgroup \emph{climbing up} from the irreps of $\mathcal{T}$ to the irreps of $\mathcal{G}$. The irreps of the 230 space groups have been calculated using the above induction procedure and have been tabulated \cite{miller1967}, and on-line versions of the tables have been implemented \cite{stokes2013}. The irreps of the double space groups have also been recently implemented \cite{GroupTheoryPaper}.

 \section{The Compatibility Relations}
 \label{compatibilityfirst}
 The compatibility relations give the correlations between the irreps of the little groups of a point in the BZ and a line where the point sits. The little group of the line is a subgroup of the little group of the point. They also give the correlations between the irreps in a point-plane or line-plane pairs. The compatibility relations are of particular interest in the analysis of the connectivity of electronic bands.
 
 Any number of bands described as band representations of the space group are \emph{partially} identified by their decomposition into irreps of the little group of $\bf{k}$ upon subduction, for all the $\bf{k}$-vectors in the BZ. However, for gapped bands, it is not necessary to consider the multiplicities of the irreps in the whole BZ due to the compatibility relations between the irreps of two connected subsets of $\bf{k}$-vectors. In fact, what makes the problem of decomposing bands into irreps of the little group of the continuous variable $\bf{k}$ in the BZ feasible is that the irrep decomposition in the whole BZ is fully fixed by the irrep decomposition of only a finite, very small number of $\bf k$ points, called maximal $\bf k$ vectors \cite{NaturePaper,GroupTheoryPaper,vergniory2017,cano2018}. Consider a $\bf{k}$ point in the momentum space that belongs to a line ($l$) $\bf{k}_l$  and let's assume that the little group of $\bf{k}$, $\mathcal{G}^{\bf{k}} $ is a supergroup of the little group of $\bf{k}_l$, $\mathcal{G}^{\bf{k}_l}< \mathcal{G}^{\bf{k}}$. For example, $\bf k$ can be a high-symmetry point and $\bf{k}_l$ can be the line of momenta going through the high symmetry point $\bf k$. Then, the matrices of an irreducible representation $\rho^i_{\mathcal{G}^{\bf{k}}}$ of the little group $\mathcal{G}^{\bf{k}}$ associated with the symmetry operations that belong to $\mathcal{G}^{\bf{k}_l}$ form a representation, not necessarily irreducible, of  $\mathcal{G}^{\bf{k}_l}$. This representation, in general, is the direct sum of irreducible representations $\rho_{\mathcal{G}^{\bf{k}_l}}^j$ of $\mathcal{G}^{\bf{k}_l}$,
 \begin{equation}
 \rho_{\mathcal{G}^{\bf{k}}}^i\downarrow \mathcal{G}^{\bf{k}_l}=\bigoplus_{i=1}^sm_{ij}^{\bf{k},\bf{k}_l}\rho_{\mathcal{G}^{\bf{k}_l}}^j \label{representationreduction}
 \end{equation}
 being $s$ the number of irreducible representations of $\mathcal{G}^{\bf{k}_l}$ and $m_{ij}^{\bf{k},\bf{k}_l}$ the (integer) multiplicity of $\rho_{\mathcal{G}^{\bf{k}_l}}^j$ in the decomposition of $\rho^i_{\mathcal{G}^{\bf{k}}}$. The same arguments are used to calculate the correlations of irreps in a point and a plane where the point sits, or between a line and a plane where the line sits. The program DCOMPREL  \cite{dcomprel} on the Bilbao Crystallographic Server (BCS), gives the multiplicities for each pair of connected $\bf{k}$-vectors in the momentum space (point-line, line-plane, etc...) for any double space group.
 
 If $\bf{k}_1$ and $\bf{k}_2$ are two points of maximal symmetry in the momentum space and  $\bf{k}_l$ is a line or a plane that connects the two points, a set of compatibility relations are defined between $\bf{k}_1$ and $\bf{k}_l$ on one hand and between $\bf{k}_2$ and $\bf{k}_l$ on the other hand. As the electronic bands are continuous functions of $\bf{k}$, the compatibility relations fix the possible ways to connect two sets of irreducible representations at $\bf{k}_1$ and $\bf{k}_2$. Consequently, the two sets of irreps $\bf{k}_1$ and $\bf{k}_2$ can form a disconnected subset of electronic bands if and only if the sum of multiplicities of each irrep of the little group of $\bf{k}_l$ in the decomposition of the irreps at $\bf{k}_1$ and at $\bf{k}_2$ is the same \cite{NaturePaper,GroupTheoryPaper,vergniory2017}.

\section{Subduction tables for group-subgroup pairs}
\label{sup:subduction}
The calculation of the basic bands of a space group from the basic bands of one of its subgroups requires the previous identification of the correlations between the irreps of the little groups of all the $\mathbf{k}$-vectors of the Brillouin zone (BZ). Instead of the induction process described in Section \ref{sup:induction}, it is easier and faster to calculate the correlations by the subduction process. In this section we explain the details of the application of the subduction to the crystallographic groups.

 The irreps at each $\mathbf{k}$-vector of the BZ in all the (double) space groups are tabulated. The program REPRESENTATIONS DSG (www.cryst.ehu.es/cryst/representationsDSG) in the Bilbao Crystallographic Server \cite{irrepsSG}) gives the irreps of the double space groups for any $\mathbf{k}$-vector. This program can be used to get all the irreps of the given space group ${\mathcal{G}}$ and the irreps of its subgroup ${\mathcal{H}}$ at all $\mathbf{k}$-vectors of the BZ. However, these wave-vectors in the database are given in the standard setting \cite {koch2016} of the group and, in general, in a group-subgroup relation, if the symmetry operations of ${\mathcal{G}}$ (and the $\mathbf{k}$-vectors) are given in its standard setting, the subset of symmetry operations that belong to ${\mathcal{H}}$ and the $\mathbf{k}$-vectors are not expressed, in general, in its standard setting. Therefore, it is necessary to know the transformation matrix between both settings.
A fully description of a ${\mathcal{G}}-{\mathcal{H}}$ group-subgroup pair must include the so-called transformation matrix, ($P,\mathbf{p}$), where $P$ is a $3\times3$ matrix and $\mathbf{p}$ is a vector of 3 components. 
Let $\{R|\mathbf{t}\}$ the Seitz symbol of a symmetry operation $h\in\mathcal{G},\mathcal{H}$, being $R$ the $3\times3$ matrix that represents the rotational part and $\mathbf{t}$ the translational part of the symmetry operation expressed in the standard setting of $\mathcal{G}$. If the transformation matrix associated to the ${\mathcal{G}}-{\mathcal{H}}$ group-subgroup pair is ($P,\mathbf{p}$), then,
\begin{equation}
\{R^s|\mathbf{t}^s\}=(P^{-1},-P^{-1}\mathbf{p})\{R|\mathbf{t}\}(P,\mathbf{p})
\end{equation}
i.e.
\begin{equation}
R^s=P^{-1}RP,\hspace{0.5 cm}\mathbf{t}^s=P^{-1}.\left(\mathbf{t}-\mathbf{p}+R\mathbf{p}\right)
\end{equation}
where $R^s$ and $\mathbf{t}^s$ are the rotational and translational part, respectively, of the symmetry operation $h$ but expressed in the standard setting of $\mathcal{H}$. 

In the reciprocal space, if $k_i$ are the coordinates of $\mathbf{k}$ in the standard setting of ${\mathcal{G}}$, its coordinates in the standard setting of $\mathcal{H}$ are $k_i^s=P_{ji}k_j$. Once these components have been calculated, the $\mathbf{k}$-vectors can be identified and the corresponding label can be assigned. Using this label, we can obtain from REPRESENTATIONS DSG the irreducible representations associated to $\mathbf{k}$ in ${\mathcal{H}}$.

The full list of subgroups of a given space group $\mathcal{G}$ under different constrains can be obtained by the program SUBGROUPS in the BCS \cite{subgroups}. This program gives also the transformation matrix of the group-subgroup pairs. In our case, we are interested in the subgroups that have the same lattice as the supergroup (translationgleiche subgroups). In the main menu of SUBGROUPS it is enough to introduce as basis vectors of the supercell three primitive basis vectors of the lattice of $\mathcal{G}$. The program lists all the translationgleiche subgroups and provides the transformation matrix to the standard setting of each subgroup.

For instance, the 4-fold axis in the space group P$4/m$ (N. 83) is parallel to the $\mathbf{c}$ axis in its standard setting. If the 4-fold axis is removed (but the 2-fold axis and the inversion center are kept), the resulting subgroup is P$2/m$ (N. 10), but the 2-fold axis is parallel to the $\mathbf{b}$ axis in the standard setting of this space group. The transformation matrix then interchanges the $\mathbf{b}$ and $\mathbf{c}$ axes, $P=((1,0,0),(0,0,-1),(0,1,0))$;  the translational part of the transformation matrix, or origin shift, is $\mathbf{p}=(0,0,0)$ in this particular case. The point X in the first Brillouin zone of the space group P$4/m$ has coordinates X:$(0,1/2,0)$, but in the standard setting of the subgroup, the coordinates are $(0,0,-1/2)$, which correspond to the point B. Therefore, the irreps at X in the space group P$4/m$ subduce into the irreps at B in the space group P$2/m$. In the opposite direction, the irreps at B in the space group P$2/m$ induce irreps into the space group P$4/m$ at X.

Once the irreps that correspond to a given $\mathbf{k}$-vector both in ${\mathcal{G}}$ and ${\mathcal{H}}$ have been identified, we can calculate the table of subductions. Let $\rho^i_{\mathcal{G}^{\mathbf{k}}}$ with $i=1,\ldots,n$ be the full list of irreps of the little group $\mathcal{G}^{\mathbf{k}}$ of $\mathbf{k}$  in ${\mathcal{G}}$ and 
$\rho^j_{\mathcal{H}^{\mathbf{k}}}$ with $j=1,\ldots,m$ the full list of irreps of the little group $\mathcal{H}^{\mathbf{k}}$ of $\mathbf{k}$ in ${\mathcal{H}}$. In general, the matrices of the irrep $\rho^i_{\mathcal{G}^{\mathbf{k}}}$ of the symmetry operations $h$ that belong to the subgroup $\mathcal{H}^{\mathbf{k}}$ form a representation of $\mathcal{H}^{\mathbf{k}}$ that is, in general, reducible. Then, this representation can be expressed as the direct sum of irreps $\rho^j_{\mathcal{H}^{\mathbf{k}}}$ with multiplicities $m_{ij}$,
\begin{equation}
\label{eq:subduction}
\rho^i_{\mathcal{G}^{\mathbf{k}}}\downarrow\mathcal{H}^{\mathbf{k}}=\bigoplus_{j=1}^mm_{ij}\rho^j_{\mathcal{H}^{\mathbf{k}}},\hspace{0.5 cm}i=1,\ldots,m
\end{equation}
The multiplicities $m_{ij}$ can be calculated by the reduction formula or Schur orthogonality relation,
\begin{equation}
\label{eq:multiplicities}
m_{ij}=\frac{1}{|{\mathcal{H}}_{\mathbf{k}}|}\sum_h\chi_i^G(h)\chi_j^H(h)
\end{equation}
where the summation is performed over symmetry operations $h$ that belong to the little group $\mathcal{H}^{\mathbf{k}}$. In the summation, we must consider only one element of the little group for each rotational part and $|\overline{{\mathcal{H}}}_{\mathbf{k}}|$ is the order of the little co-group of $\mathbf{k}$. $\chi^i_{\mathcal{G}^{\mathbf{k}}}(h)$ and $\chi^j_{\mathcal{H}^{\mathbf{k}}}(h)$ are the characters of the irreps $\rho^i_{\mathcal{G}^{\mathbf{k}}}$ and $\rho^j_{\mathcal{H}^{\mathbf{k}}}$, respectively, for the element $h$.
The program DCORREL \cite{dcorrel} gives the correlations between the irreps at all $\mathbf{k}$-vectors. These lists of correlations include both maximal and non-maximal $\mathbf{k}$-vectors. The bands in a space group are fully identified through the multiplicities of the irreps at maximal $\mathbf{k}$-vectors, and the information given at these points is enough to identify the bands. However, in some group-subgroup pairs, a maximal $\mathbf{k}$-vector in the supergroup $\mathcal{G}$ does not correspond to a maximal k-vector in the subgroup ${\mathcal{H}}$. Therefore, the full list of maximal and non-maximal $\mathbf{k}$-vectors are, in general, necessary to derive the bands of $\mathcal{G}$ from the bands in $\mathcal{H}$.

As stressed in the main text, if we compare the two little groups of a certain $\mathbf{k}$, ${\mathcal{G}}^{\mathbf{k}}$ and  $\mathcal{H}^{\mathbf{k}}$, we can distinguish two different cases:
\begin{itemize}
	\item The little groups ${\mathcal{G}}^{\mathbf{k}}$ and ${\mathcal{H}}^{\mathbf{k}}$ are in fact the same group. This happens when no one symmetry operation $g\in\mathcal{G}$ but $g\notin\mathcal{H}$ belongs to the little group ${\mathcal{G}}^{\mathbf{k}}$. This is the case of the point X:$(0,1/2,0)$ in the example above with ${\mathcal{G}}$: P$4/m$ and ${\mathcal{H}}$: P$2/m$ group-subgroup pair. The 4-fold axis $C_4$ does not belong to the little co-group of X. There is a 1:1 correspondence between the irreps in both little groups at $\mathbf{k}$: every irrep $\rho^i_{\mathcal{G}^{\mathbf{k}}}$ of ${\mathcal{G}}^{\mathbf{k}}$ subduces into a single irrep $\rho^j_{\mathcal{H}^{\mathbf{k}}}$ of ${\cal{H}}^{\mathbf{k}}$ (all the multiplicities in Eq. (\ref{eq:subduction}) are zero except one of them that is 1). In the opposite direction, the irrep $\rho^j_{\mathcal{H}^{\mathbf{k}}}$ of ${\mathcal{H}}^{\mathbf{k}}$ induces a single irrep onto ${\mathcal{G}}^{\mathbf{k}}$.
	\item The little group ${\mathcal{H}}^{\mathbf{k}}$ is a subgroup of ${\mathcal{G}}^{\mathbf{k}}$ different from ${\mathcal{G}}^{\mathbf{k}}$. In this case, in general, an irrep $\rho^i_{\mathcal{G}^{\mathbf{k}}}$ of ${\mathcal{G}}^{\mathbf{k}}$ subduces into the direct sum of several irreps with multiplicities that can be different from 1. For instance, the double TR invariant irrep $\overline{\Gamma}_{10}$ of the space group P$m\overline{3}m$ (N. 221) subduces into 2 times $\irrg{3}\irrg{3}$ in the space group P$\overline{1}$. Moreover, it can happen that several irreps $\rho^i_{\mathcal{G}^{\mathbf{k}}}$ with different $i$ subduce into the same $\rho^j_{\mathcal{H}^{\mathbf{k}}}$, i.e., some multiplicities $m_{ij}$ with different $i$ and same $j$ are different from 0. In our example of space group P$m\overline{3}m$, the irreps $\overline{\Gamma}_{6}$ and $\overline{\Gamma}_{7}$ subduce into $\overline{\Gamma}_3\overline{\Gamma}_3$. Then, when we consider the induction process from the space group P$\overline{1}$ onto P$m\overline{3}m$, a single irrep $\overline{\Gamma}_3\overline{\Gamma}_3$ can \emph{induce} a single irrep $\overline{\Gamma}_{6}$ or a single irrep $\overline{\Gamma}_{7}$ (in the sense given to the term \emph{partial induction} defined in Section \ref{sec:induction} of the main text. But the combination of 2 irreps can induce the following four different combinations of irreps:
	\begin{eqnarray}
	2\times\overline{\Gamma}_3\overline{\Gamma}_3&\rightarrow&2\,\overline{\Gamma}_{6}\nonumber\\
	2\times\overline{\Gamma}_3\overline{\Gamma}_3&\rightarrow&2\,\overline{\Gamma}_{7}\nonumber\\
	2\times\overline{\Gamma}_3\overline{\Gamma}_3&\rightarrow&\overline{\Gamma}_{6}\oplus\overline{\Gamma}_{7}\\
	2\times\overline{\Gamma}_3\overline{\Gamma}_3&\rightarrow&\overline{\Gamma}_{10}\nonumber
	\end{eqnarray}
	All the possible combinations must be considered in a induction procedure to get all the types of bands in a space group from the bands of one of its subgroups.
\end{itemize}

\section{Example of the calculation of the basic bands of the space group I$2_12_12_1$ from the basic bands of the space group C2}
\label{sup:inducedbands}
In this section we describe in detail the induction process applied to the calculation of the basic bands in a supergroup (I$2_12_12_1$ in our example) from the basic bands of one of its subgroups (C2 in our example) with TR. There is a group-subgroup relation between I$2_12_12_1$ (N. 24) and C2 (N. 5) with a transformation matrix \cite{subgroups},
\begin{equation}
\label{eq:trmat}
P=\left(
\begin{array}{rrr}
1&0&0\\0&1&0\\
1&0&1
\end{array}
\right),
\mathbf{p}=\left(
\begin{array}{r}
\frac{1}{4}\\
0\\
0
\end{array}
\right)
\end{equation}
In general, this means that, given a symmetry operation $\{R_G|\mathbf{t}_G\}$ of a space group $\mathcal{G}$ that also belongs to its subgroup $\mathcal{H}$ expressed in the standard setting of $\mathcal{G}$, in the standard setting of the subgroup $\mathcal{H}$ the operation is,
\begin{equation}
	\label{eq:changesetting}
	\{R_H|\mathbf{t}_H\}=\{P^{-1}R_GP|P^{-1}(\mathbf{t}_G-\mathbf{p}+R_G\mathbf{p})\}
\end{equation}
First, we order the double-valued irreps in the subgroup C2 at maximal $\mathbf{k}$-vectors as,
\begin{equation}
\left(\irrg{3}\irrg{4},\irr{A}{3}\irr{A}{4},\irr{L}{2}\irr{L}{2},\irr{M}{3}\irr{M}{4},\irr{V}{2}\irr{V}{2},\irr{Y}{3}\irr{Y}{4}\right)
\end{equation}
and at non-maximal $\mathbf{k}$-vectors as,
\begin{equation}
\left(\irr{B}{2},\irrl{3},\irrl{4},\irr{U}{3},\irr{U}{4},\irr{GP}{2}\right)
\end{equation}
There is a unique basic band in C2 and the multiplicity of all the irreps of each $\mathbf{k}$-vector in the previous lists is 1, except into $\irr{B}{2}$ and $\irr{GP}{2}$, where the multiplicity is 2. 

In the next step, we need the correlations between the maximal $\mathbf{k}$-vectors in the supergroup I$2_12_12_1$ and the $\mathbf{k}$-vectors in the subgroup C2. The correlations relevant for our example are shown in table (\ref{table:correlations}) and have been obtained using the program DCORREL.

\begin{table}[h]
	\caption{Correlation between the maximal $\mathbf{k}$-vectors $\mathbf{k}_{24}$ in the space group I$2_12_12_1$ (N. 24) and the $\mathbf{k}$-vectors $\mathbf{k}_5$ in C2 (N. 5) with transformation matrix given by \ref{eq:trmat}. The coordinates of the $\mathbf{k}$-vectors in both settings are related by, $\mathbf{k}_5=P^T\mathbf{k}_{24}$.}
	\label{table:correlations}
	\begin{tabular}{l|l}
		\multicolumn{1}{c|}{Maximal $\mathbf{k}$-vectors in}&\multicolumn{1}{c}{$\mathbf{k}$-vectors in the standard}\\
		\multicolumn{1}{c|}{SG I$2_12_12_1$}&\multicolumn{1}{c}{setting of SG C2}\\
		\hline
		$\Gamma:(0,0,0)$&$\Gamma:(0,0,0)$\\
		R:$\left(-\frac{1}{2},0,\frac{1}{2}\right)$&A:$\left(0,0,\frac{1}{2}\right)$\\
		R:$\left(\frac{1}{2},0,\frac{1}{2}\right)$&M:$\left(1,0,\frac{1}{2}\right)$\\
		S:$\left(0,\frac{1}{2},\frac{1}{2}\right)$&L:$\left(\frac{1}{2},\frac{1}{2},\frac{1}{2}\right)$\\
		T:$\left(\frac{1}{2},\frac{1}{2},0\right)$&V:$\left(\frac{1}{2},\frac{1}{2},0\right)$\\
		W:$\left(\frac{1}{2},\frac{1}{2},\frac{1}{2}\right)$&U:$\left(1,1/2,\frac{1}{2}\right)$\\
		WA:$\left(-\frac{1}{2},-\frac{1}{2},-\frac{1}{2}\right)$&U:$\left(-1,-1/2,-\frac{1}{2}\right)$\\
		X:$(1,1,1)$&Y:$(2,1,1)$\\
		Q$\left(\frac{1}{2},v,\frac{1}{2}\right)$&U:$\left(1,v,\frac{1}{2}\right)$\\
		D$\left(u,\frac{1}{2},\frac{1}{2}\right)$&GP:$\left(\frac{1}{2}+u,\frac{1}{2},\frac{1}{2}\right)$\\
		P$\left(\frac{1}{2},\frac{1}{2},w\right)$&GP:$\left(\frac{1}{2}+w,\frac{1}{2},w\right)$\\
	\end{tabular}
\end{table}
The symmetry operations of the space group C2 taken from the program \emph{Representations DSG} \cite{irrepsSG} are,
\begin{equation}
\label{eq:symop5}
\{1|t_1,t_2,t_3\},\{2_y|0,0,0\},\{\overline{E}|0,0,0\},\{ ^d2_y|0,0,0\}
\end{equation}
where the first operation represents a general lattice translation and the other three elements are the coset representatives of the double space group with respect to the translation group. The superscript $^d$ is used to distinguish the two symmetry operations of the space group that differ only in their action on the spinor space \cite{GroupTheoryPaper}.

The coset decomposition of the supergroup with respect to the subgroup can be expressed as,
\begin{equation}
\mathrm{I}2_12_12_1=\mathrm{C}2\cup\mathrm{C}2\,g
\end{equation}
with $g=\{2_z|-1/2,1/2,1/2\}$ and
\begin{equation}
2_z:\left(
\begin{array}{rrr}
-1&0&0\\
0&-1&0\\
2&0&1
\end{array}\right)
\end{equation}
in the setting of C2, given by \ref{eq:changesetting}.

First we consider the induction of the 1-dimensional double-valued irreps $\irrg{3}$ and $\irrg{4}$ into the supergroup. The little group of $\Gamma$ is the whole space group and the $1\times1$ matrices $D(h)$ of the symmetry operations of Eq. (\ref{eq:symop5}) are,
\begin{eqnarray}
\irrg{3}&\rightarrow& 1,-i,-1,i\nonumber\\
\irrg{4}&\rightarrow& 1,i,-1,-i
\end{eqnarray}
The little group of $\Gamma$ in I$2_12_12_1$ is also the whole space group and then the little group of $\Gamma$ in $\mathcal{G}$ is a supergroup of its little group in $\mathcal{H}$. The conjugated elements $g^{-1}hg$ of the symmetry operations in Eq. (\ref{eq:symop5}) are,
\begin{eqnarray}
\label{conjugation}
h&\rightarrow&g^{-1}hg\\
\{1|t_1,t_2,t_3\}&\rightarrow& \{1|-t_1,-t_2,2t_1+t_3\}\nonumber\\
\{2_y|0,0,0\}&\rightarrow&\{ ^d2_y|-1,0,1\}\nonumber\\
\{\overline{E}|0,0,0\}&\rightarrow&\{\overline{E}|0,0,0\}\nonumber\\
\{ ^d2_y|0,0,0\}&\rightarrow&\{2_y|-1,0,1\}
\end{eqnarray}
and the matrices D($g^{-1}hg$) of these operations are,
\begin{eqnarray}
\irrg{3}&\rightarrow& 1,i,-1,-i\nonumber\\
\irrg{4}&\rightarrow& 1,-i,-1,i
\end{eqnarray}
The list of characters of the irrep $\irrg{3}$ transform into the list of characters of the irrep $\irrg{4}$ and vice-versa by conjugation (Eq. \ref{conjugation}) through the coset representative $g=\{2_z|-1/2,1/2,1/2\}$. Therefore, $\irrg{3}$ and $\irrg{4}$ belong to the same orbit and the two irreps that in the space group C2 form the $\irrg{3}\irrg{4}$ irrep when TR is considered, induce a single TR-invariant irrep in I$2_12_12_1$, labeled as  $\irrg{5}$ (see Representations DSP \cite{irrepsPG}).

The symmetry operation $g=\{2_z|-1/2,1/2,1/2\}$ (or any other operation of $\mathcal{G}$ not included in $\mathcal{H}$) does not belong to the little group of S(L), T(T), R(A,M) and X(Y) $\mathbf{k}$-vectors in I$2_12_12_1$(C2). Therefore, there is a 1 to 1 equivalence between the irreps at S and L, T and T, R and A, R and M and X and Y. 

Now we apply the induction from the irreps at U:$(1,v,1/2)$. The little group of U is the whole space group in C2. Here we must distinguish between the general $\mathbf{k}$-vectors in the line U and the particular $\mathbf{k}$-vectors with $v=\pm 1/2$ or equivalents. In the first case, the symmetry operation $g=\{2_z|-1/2,1/2,1/2\}$ does not belong to the little group of U and there is a 1 to 1 relation between the irreps at U in C2 and the irreps at the corresponding line (labeled as Q) in I$2_12_12_1$. The relations are
\begin{eqnarray}
\label{eq:lineQ}
\irr{U}{3}&\rightarrow&\irr{Q}{3}\nonumber\\
\irr{U}{4}&\rightarrow&\irr{Q}{4}.
\end{eqnarray}
In the second case the special $\mathbf{k}$-vectors in the line U with $v=1/2$ and $v=-1/2$ correspond to the $\mathbf{k}$-vectors W and WA, respectively, in the supergroup. As the $g=\{2_z|-1/2,1/2,1/2\}$ symmetry operation belongs to the little groups of these two $\mathbf{k}$-vectors, first we calculate the orbit of each irrep at U with $v=1/2$. The matrices of the symmetry operations (\ref{eq:symop5}) of the two double-valued irreps are,
\begin{eqnarray}
\label{eq:irrepsU}
\irr{U}{3}&\rightarrow& e^{i\pi(2t_1+t_2+t_3)},-i,-1,i\nonumber\\
\irr{U}{4}&\rightarrow& e^{i\pi(2t_1+t_2+t_3)},i,-1,-i
\end{eqnarray}
and the matrices D($g^{-1}hg$) of the conjugated  operations (\ref{conjugation}) are,
\begin{eqnarray}
\label{eq:irrepsUconj}
\irr{U}{3}&\rightarrow& e^{i\pi(-t_2+t_3)},-i,-1,i\nonumber\\
\irr{U}{4}&\rightarrow& e^{i\pi(-t_2+t_3)},i,-1,-i
\end{eqnarray}
For any translation $\{1|t_1,t_2,t_3\}$ of C2, the irrep $\irr{U}{3}$ in Eq. (\ref{eq:irrepsUconj}) is equivalent to $\irr{U}{3}$ in Eq. (\ref{eq:irrepsU}) and $\irr{U}{4}$ is equivalent to $\irr{U}{4}$. Both irreps are self-conjugated. Each irrep in C2 can induce two different irreps in the supergroup,
\begin{eqnarray}
\label{eq:ind1}
\irr{U}{3}&\rightarrow&\irr{W}{3}|\irr{W}{4}\nonumber\\
\irr{U}{4}&\rightarrow&\irr{W}{2}|\irr{W}{5}
\end{eqnarray}
where we have used the symbol $|$ to stress that a single $\irr{U}{3}$ irrep, for instance, induces (in the sense given in the main text) a single $\irr{W}{3}$ irrep or a single $\irr{W}{4}$.
A similar calculation from U:$(-1,v,-1/2)$ to WA:$(-1/2,-1/2,-1/2)$ with $v=-1/2$ gives,
\begin{eqnarray}
\label{eq:ind2}
\irr{U}{3}&\rightarrow&\irr{WA}{2}|\irr{WA}{5}\nonumber\\
\irr{U}{4}&\rightarrow&\irr{WA}{3}|\irr{WA}{4}
\end{eqnarray}
As the multiplicity of $\irr{U}{3}$ and $\irr{U}{4}$ is the same in the basic band in the subgroup, the multiplicities of the irreps at W in the induced band fulfill the condition $n(\irr{W}{3})+n(\irr{W}{4})=n(\irr{W}{2})+n(\irr{W}{5})$ and at WA, $n(\irr{WA}{3})+n(\irr{WA}{4})=n(\irr{WA}{2})+n(\irr{WA}{5})$.

Another extra condition arises when the compatibility relations along the line U:$(1,v,1/2)$ is taken into consideration. This line of C2 corresponds to the line Q:$(1/2,v,1/2)$ in the supergroup I$2_12_12_1$. The compatibility relations (\ref{eq:lineQ}), (\ref{eq:ind1}) and (\ref{eq:ind2}) imply, $n(\irr{W}{3})+n(\irr{W}{4})=n(\irr{WA}{2})+n(\irr{WA}{5})$ for $\irr{Q}{3}$ and $n(\irr{W}{2})+n(\irr{W}{5})=n(\irr{WA}{3})+n(\irr{WA}{4})$ for $\irr{Q}{4}$. Moreover, as the multiplicities of $\irr{U}{3}$ and $\irr{U}{4}$ in the unique basic band in C2 are equal, the four quantities must satisfy, $n(\irr{W}{3})+n(\irr{W}{4})=n(\irr{WA}{2})+n(\irr{WA}{5})=n(\irr{W}{2})+n(\irr{W}{5})=n(\irr{WA}{3})+n(\irr{WA}{4})$.

If we consider now the induced irreps from the irreps of the particular line of the general position GP:$(1/2+u,1/2,1/2)$ which corresponds to the line D:$(u,1/2,1/2)$ in the supergroup, we get extra conditions. Note that for the particular value $u=1/2$ the $\mathbf{k}$-vector in the supergroup is W:$(1/2,1/2,1/2)$ and for $u=-1/2$ the resulting $\mathbf{k}$-vector is equivalent to WA:(-1/2,-1/2,-1/2) by a translation of the reciprocal lattice. The compatibility relations along the W$\leftrightarrow$D$\leftrightarrow$WA line imply the extra relations 
$n(\irr{W}{2})+n(\irr{W}{3})=n(\irr{WA}{4})+n(\irr{WA}{5})$ for $\irr{D}{3}$ and $n(\irr{W}{4})+n(\irr{W}{5})=n(\irr{WA}{2})+n(\irr{WA}{3})$ for $\irr{D}{4}$. Moreover, as the S point is also connected to W and WA through the line D, we have the extra relations, $n(\irr{W}{2})+n(\irr{W}{3})=n(\irr{W}{4})+n(\irr{W}{5})$ and $n(\irr{WA}{2})+n(\irr{WA}{3})=n(\irr{WA}{4})+n(\irr{WA}{5})$.

Finally, the particular line of the general position in C2, GP:$(1/2+w,1/2,w)$ corresponds to the line P:$(1/2,1/2,w)$ in the supergroup and, for the particular cases $w=1/2$ and $w=-1/2$, the resulting $\mathbf{k}$-vectors in the supergroup are again W and an equivalent $\mathbf{k}$-vector to WA, respectively.  The compatibility relations along the P line give $n(\irr{W}{3})+n(\irr{W}{4})=n(\irr{WA}{2})+n(\irr{WA}{5})$ for $\irr{P}{3}$ and $n(\irr{W}{2})+n(\irr{W}{5})=n(\irr{WA}{3})+n(\irr{WA}{4})$ for $\irr{P}{4}$. Moreover, as the T point is also connected to W and WA through the line P, we have the extra relations, $n(\irr{W}{2})+n(\irr{W}{4})=n(\irr{W}{3})+n(\irr{W}{5})$ and $n(\irr{WA}{2})+n(\irr{WA}{4})=n(\irr{WA}{3})+n(\irr{WA}{5})$.

The fulfillment of all the restrictions requires an even number of basic bands in C2 to combine to induce a band in  the supergroup I$2_12_12_1$. Therefore, ordering the double-valued irreps at maximal $\mathbf{k}$-vectors with TR in this way,
\begin{equation}
\label{eq:labelsSG24}
\left(\irrg{5},\irr{W}{2},\irr{W}{3},\irr{W}{4},\irr{W}{5},\irr{WA}{2},\irr{WA}{3},\irr{WA}{4},\irr{WA}{5},\irr{X}{5},\irr{R}{3}\irr{R}{4},\irr{S}{3}\irr{S}{4},\irr{T}{3}\irr{T}{4}\right).
\end{equation}
the induced single basic band into I$2_12_12_1$ has multiplicities
\begin{equation}
\label{eq:multSG24}
\left(2,1,1,1,1,1,1,1,1,2,2,2,2\right).
\end{equation}
 
 \section{Examples of systematic calculation of the basic bands in a space group from the basic bands of one of its maximal subgroups}
 \label{sup:bandsall}
 In this section we will outline the derivation of the basic bands in some space groups with low symmetry. These simple examples, however, show the different general issues that can be found in the general induction procedure. The results obtained for these space groups can be further used to derive the types of bands in space groups of higher symmetry. It should be noted that, considering the double space groups, there are two types of irreducible representations and, therefore, two types of bands: the single valued irreps and the double-valued irreps. Single valued irreps always induce single-valued irreps and double-valued irreps induce double-valued irreps. Therefore, in the subduction chains of bands, single and double valued bands follow two separate chains that do  not intersect. Then, they can be derived separately. In this section we will consider only double-valued irreps, being the procedure to construct the single-valued bands completely equivalent. The labels of the irreps, the set of EBRs and the compatibility relations of each group together with other crystallographic data used in this section can de obtained from the programs BANDREP (\cite{progbandrep}), DCOMPREL (\cite{dcomprel}), REPRESENTATIONS DPG (\cite{irrepsPG}) and REPRESENTATIONS DSG (\cite{irrepsSG}). 
\subsection{Bands in the space group P1 (N. 1)}
This is the starting point of the whole induction process for all the space groups. In the space group P1 there is only one Wyckoff position with site-symmetry group isomorphic to the trivial point group and, therefore, there exist only one EBR induced from the unique irrep of the site-symmetry group of the Wyckoff position. In the reciprocal space there are 8 maximal $\bf{k}$-vectors when TR is considered, $\Gamma$:(0,0,0), X:(1/2,0,0), Y:(0,1/2,0), V:(1/2,1/2,0), Z:(0,0,1/2), U:(1/2,0,1/2), T:(0,1/2,1/2) and R:(1/2,1/2,1/2), whose common little group contains the lattice translations and the lattice translations combined with TR. At every maximal $\bf{k}$-vector there is only one double-valued irrep of dimension 2 (\cite{irrepsSG}). If we consider the set of irreps at the maximal $\bf{k}$-vectors in this specific order, 
\begin{equation}
\label{eq:labelsSG1}
\left(\irrg{2}\irrg{2},\irr{X}{2}\irr{X}{2},\irr{Y}{2}\irr{Y}{2},\irr{V}{2}\irr{V}{2},\irr{Z}{2}\irr{Z}{2},\irr{U}{2}\irr{U}{2},\irr{T}{2}\irr{T}{2},\irr{R}{2}\irr{R}{2}\right),
\end{equation}
any band in P1 can be denoted by a 8-component vector, whose $i$-th component is the multiplicity of the $i$-th irrep in the previous list upon subduction of the band.

Every pair of maximal $\bf{k}$-vectors are connected through the general position (GP) and thus, the only condition imposed by the compatibility relations on the multiplicities of the irreps at the maximal $\bf{k}$-vectors to form a band is to be equal. The general expression for a band in this space group is $(n,n,n,n,n,n,n,n)^T$ with $n>0$. There is only one basic band ($n=1$) which is also the unique EBR in this SG, of dimension 2. Any other band of dimension $2n$ is the direct sum of $n$ EBRs and thus there exist no fragile or strong bands in this group. 

The only non-maximal $\bf{k}$-vector in this space group is the general position. The multiplicity of the double irrep $\irr{GP}{2}$ in the basic band is 2. This fact will be used in the derivation of the basic bands in those groups (as the space group P3 below) in which some maximal $\bf{k}$-vectors do not correspond to maximal $\bf{k}$-vectors of P1.

\subsection{Bands in the space group P$\overline{1}$ (N. 2)}
\label{subsec:sp2}
In this section we will derive all the basic bands in the space group P$\overline{1}$ from the basic bands of its (unique) translationgleiche subgroup P1. The transformation matrix of the group-subgroup pair can be taken as the identity. The set of maximal $\mathbf{k}$-vectors in this space group is exactly the same as in P1, with the same labels. The little group of all the maximal $\mathbf{k}$-vectors is the whole space group and, then, at all the maximal $\mathbf{k}$-vectors, the little group in $\mathcal{G}:$P$\overline{1}$ is a supergroup of the little group in $\mathcal{H}:$P1. At all maximal $\mathbf{k}$-vectors, the little group $\mathcal{G}^{\bf{k}}$ has two double-valued irreps: $\irrg{2}\irrg{2}$ and $\irrg{3}\irrg{3}$ at $\Gamma$, $\irr{X}{2}\irr{X}{2}$ and $\irr{X}{3}\irr{X}{3}$ at X, etc\ldots \cite{irrepsSG}. The two irreps at each maximal $\mathbf{k}$-vector subduce into the unique double-valued of the little group of the corresponding $\mathbf{k}$-vector in P1, i-e., both $\irrg{2}\irrg{2}$ and $\irrg{3}\irrg{3}$ in P$\overline{1}$ subduce into $\irrg{2}\irrg{2}$ in P1, $\irr{X}{2}\irr{X}{2}$ and $\irr{X}{3}\irr{X}{3}$ into $\irr{X}{2}\irr{X}{2}$, etc\ldots. Therefore, in the opposite direction, an irrep  $\irrg{2}\irrg{2}$ in P1 can induce two different irreps in P$\overline{1}$, $\irrg{2}\irrg{2}$ or $\irrg{3}\irrg{3}$. The same is true for the other 7 irreps at maximal $\mathbf{k}$-vectors in P1. If we consider all the possible combinations at each of the 8 maximal $\mathbf{k}$-vectors, there are $2^8=256$ possible different bands induced from the single basic irrep in P1. As the intermediate path between any pair of maximal $\mathbf{k}$-vectors is the GP, the compatibility relations are trivially fulfilled by all the induced bands, and therefore all the 256 bands are basic bands. They are the only basic bands in this space group. 

If we list the 16 irreps at maximal $\mathbf{k}$-vectors in this order:
\begin{equation}
\label{eq:labelsSG2}
\left(\irrg{2}\irrg{2},\irrg{3}\irrg{3},\irr{X}{2}\irr{X}{2},\irr{X}{3}\irr{X}{3},\irr{Y}{2}\irr{Y}{2},\irr{Y}{3}\irr{Y}{3},\irr{V}{2}\irr{V}{2},\irr{V}{3}\irr{V}{3},\irr{Z}{2}\irr{Z}{2},\irr{Z}{3}\irr{Z}{3},\irr{U}{2}\irr{U}{2},\irr{U}{3}\irr{U}{3},\irr{T}{2}\irr{T}{2},\irr{T}{3}\irr{T}{3},\irr{R}{2}\irr{R}{2},\irr{R}{3}\irr{R}{3}\right)
\end{equation}
the 256 basic bands can be written as a column vector,
\begin{equation}
\label{eq:multSG2}
\left(n(\irrg{2}\irrg{2}),n(\irrg{3}\irrg{3}),\ldots,n(\irr{R}{2}\irr{R}{2}),n(\irr{R}{3}\irr{R}{3})\right)^T
\end{equation}
with $n(\rho_2\rho_2)+n(\rho_3\rho_3)=1$ for all $\rho=\Gamma,$ X, \ldots. The 16 non-negative integer components of the vectors in Eq. (\ref{eq:multSG2}) give the multiplicities the irreps in Eq. (\ref{eq:labelsSG2}) in the band.

In order to identify which type of band every basic band in (\ref{eq:multSG2}) belongs to, it is necessary to know the EBRs, which can be obtained from the program BANDREP (\cite{progbandrep}). In the space group P$\overline{1}$ there are 16 EBRs, also described through the multiplicities of the irreps in Eq. (\ref{eq:labelsSG2}),

\begin{equation}
\label{eq:EBRsSG2}
EBR=\left(
\begin{array}{cccccccccccccccc}
1 & 1 & 1 & 1 & 1 & 1 & 1 & 1 & 0 & 0 & 0 & 0 & 0 & 0 & 0 & 0 \\
0 & 0 & 0 & 0 & 0 & 0 & 0 & 0 & 1 & 1 & 1 & 1 & 1 & 1 & 1 & 1 \\
1 & 1 & 0 & 0 & 0 & 0 & 1 & 1 & 0 & 0 & 1 & 1 & 1 & 1 & 0 & 0 \\
0 & 0 & 1 & 1 & 1 & 1 & 0 & 0 & 1 & 1 & 0 & 0 & 0 & 0 & 1 & 1 \\
1 & 0 & 1 & 0 & 0 & 1 & 0 & 1 & 0 & 1 & 0 & 1 & 1 & 0 & 1 & 0 \\
0 & 1 & 0 & 1 & 1 & 0 & 1 & 0 & 1 & 0 & 1 & 0 & 0 & 1 & 0 & 1 \\
1 & 1 & 1 & 1 & 0 & 0 & 0 & 0 & 0 & 0 & 0 & 0 & 1 & 1 & 1 & 1 \\
0 & 0 & 0 & 0 & 1 & 1 & 1 & 1 & 1 & 1 & 1 & 1 & 0 & 0 & 0 & 0 \\
1 & 0 & 0 & 1 & 0 & 1 & 1 & 0 & 0 & 1 & 1 & 0 & 1 & 0 & 0 & 1 \\
0 & 1 & 1 & 0 & 1 & 0 & 0 & 1 & 1 & 0 & 0 & 1 & 0 & 1 & 1 & 0 \\
1 & 0 & 0 & 1 & 1 & 0 & 0 & 1 & 0 & 1 & 1 & 0 & 0 & 1 & 1 & 0 \\
0 & 1 & 1 & 0 & 0 & 1 & 1 & 0 & 1 & 0 & 0 & 1 & 1 & 0 & 0 & 1 \\
1 & 1 & 0 & 0 & 1 & 1 & 0 & 0 & 0 & 0 & 1 & 1 & 0 & 0 & 1 & 1 \\
0 & 0 & 1 & 1 & 0 & 0 & 1 & 1 & 1 & 1 & 0 & 0 & 1 & 1 & 0 & 0 \\
1 & 0 & 1 & 0 & 1 & 0 & 1 & 0 & 0 & 1 & 0 & 1 & 0 & 1 & 0 & 1 \\
0 & 1 & 0 & 1 & 0 & 1 & 0 & 1 & 1 & 0 & 1 & 0 & 1 & 0 & 1 & 0 \\
\end{array}
\right)
\end{equation}
Each column gives the multiplicities of an EBR. These 16 bands belong to the full set of basic bands (\ref{eq:multSG2}). None of the other 240 bands in Eq. (\ref{eq:multSG2}) can be expressed as integer (positive or negative) linear combination of the EBRs (the set of diophantine equations that can be established to express the bands as linear combinations of EBRs has no solution). As a consequence, they are all strong topological bands. It can be checked that they can be divided into 30 subsets of 8 bands. In each subset, the 8 basic bands are EBR-equivalent, i.e, the difference between the multiplicities of any pair of bands in a subset is a linear integer (positive and/or negative) combination of EBRs. Basic bands that belong to different subsets are not EBR-equivalent.

The space group P$\overline{1}$ has 4 topological indices or indicators with SOC, $z_{2w,1}$, $z_{2w,2}$, $z_{2w,3}$ and $z_4$,  (\cite{Song2018a}) with $2\times2\times2\times4=32$ possible sets of values, if we include the trivial or non-topological case, i.e., all $z=0$. The 8 EBR-equivalent basic bands in each subset have the same topological indices, and bands that belong to different subsets have, at least, one index different. The 30 different subsets cover 30 out of the 31 possible non-trivial combinations of indices. There is no subset with indices $z_{2w,1}=z_{2w,2}=z_{2w,3}=0$ and $z_4=2$, but these  values can be obtained by any combination of two basic bands with indices ($z_{2w,1},z_{2w,2},z_{2w,3}$,$z_4$) and ($z_{2w,1},z_{2w,2},z_{2w,3}$,2-$z_4$).

Therefore, there are no basic fragile bands in this space group. The fragile bands must be obtained through the direct sum of two or more strong topological bands. Any combination of two basic strong topological bands with complementary indices (we can define as complementary indices of $z_{2w,1}$, $z_{2w,2}$, $z_{2w,3}$ and $z_4$ the indices $z_{2w,1}$, $z_{2w,2}$, $z_{2w,3}$ and $4-z_4$) gives rise to a trivial band (linear combination of EBRs with positive indices) or to a fragile band. At this point, in this particular space group, the search of the fragile roots becomes a problem of combinatory and \emph{brute force}. Out of 1016 different possible combinations of two complementary basic bands, 112 correspond to fragile bands of dimension 4 which are fragile root bands (there are no fragile bands of lower dimension, so they are necessarily fragile root bands). In the next step we identify the combinations of three basic bands that give rise to trivial or fragile bands. We find that some of the resulting bands can be expressed as combinations of 3 EBRs and all the other bands are the sum of one EBR and one fragile root of dimension 4. Therefore there is no fragile root bands of dimension 6. Finally, we check which combinations of 4 basic bands fulfill the requirements. There are 1024 different 8-dimensional fragile bands that are no linear combinations with positive integers of EBRs and fragile bands of dimension 4. Moreover, we have checked that all the ($240^5$) combinations of 5 strong basic bands can be expressed as the linear combination of an EBR or a fragile root and several strong bands. This implies that no fragile root bands can exist as combination of 5 or more strong basic bands. Therefore in this space group there are 1136 fragile root bands.

\subsection{Bands in the space group P2 (N. 3)}
\label{subsec:P2}
The basic bands in space group P2 can be easily derived from the basic band of space group P1. The set of maximal $\bf{k}$-vectors is exactly the same but the site-symmetry groups are different (note that also the standard labels change). For all the maximal $\mathbf{k}$-vectors, the little group is the whole space group and all the irreps are self-conjugated. For instance, at $\Gamma$ point the irrep $\irrg{2}$ without TR can induce $\irrg{3}$ or $\irrg{4}$ into P2 without TR. However, if TR is added, the double-valued $\irrg{2}\irrg{2}$ induces the TR-invariant $\irrg{3}\irrg{4}$ irrep in P2. The other possible combinations do not fulfill TR. The same arguments apply at all the other 7 maximal $\mathbf{k}$-vectors. Then, if we order the irreps in P2 as, 
\begin{equation}
\label{eq:labelsSG3}
(\irrg{3}\irrg{4},\irr{Y}{3}\irr{Y}{4},\irr{Z}{3}\irr{Z}{4},\irr{C}{3}\irr{C}{4},\irr{B}{3}\irr{B}{4},\irr{A}{3}\irr{A}{4},\irr{D}{3}\irr{D}{4},\irr{E}{3}\irr{E}{4})
\end{equation}
there is a 1:1 correspondence between the $i^{\textrm{th}}$ irrep of P2 in Eq. (\ref{eq:labelsSG3}) and the $i^{\textrm{th}}$ irrep of P1 in (\ref{eq:labelsSG1}) upon induction-subduction. Then, the unique basic band in P1 induces a unique basic band in P2. The multiplicities of this band are identical to the multiplicities of the four EBRs induced from the four maximal Wyckoff positions. Therefore, there are no fragile and no strong topological bands in the space group P2. An equivalent result is obtained in all the space groups whose point group contains just a 2-fold axis, a 2-fold screw axis, a mirror plane or a glide plane.

\subsection{Bands in the space group P3 (N. 143)}
\label{subsec:P3}
The basic bands in P3 are also obtained from the basic bands of P1. The space group contains 8 maximal $\bf{k}$-vectors, A, $\Gamma$, H, HA, K, KA, L and M. The little groups of A and $\Gamma$ coincide with the space group (included TR). At each of these two points, there are two double-valued irreps which subduce into the same irrep. For example, the two irreps at A, $\irr{A}{4}\irr{A}{4}$ and $\irr{A}{5}\irr{A}{6}$ (that correspond to the pairs of irreps TR-related with eigenvalues of the 3-fold axis 1 ($\irr{A}{4}$), $e^{-i\pi/3}$ ($\irr{A}{5}$) and $e^{i\pi/3}$ ($\irr{A}{6}$)), subduce into $\irr{Z}{2}\irr{Z}{2}$ in P1 (see \cite{dcorrel}). In the opposite direction, this means that the irrep $\irr{Z}{2}\irr{Z}{2}$ of P1 can induce two different irreps into P3, $\irr{A}{4}\irr{A}{4}$ or $\irr{A}{5}\irr{A}{6}$. The same is true for the two irreps at $\Gamma$ in P3, $\irrg{4}\irrg{4}$ and $\irrg{5}\irrg{6}$, induced from $\irrg{2}\irrg{2}$ in P1. In principle, just considering the A and $\Gamma$ points, it could be inferred that four combinations are possible as a result of the induction from the basic band in P1. However, the compatibility relations along the line  $\Delta$ that connects both points, forces the multiplicities of $\irr{A}{i}\irr{A}{j}$ and $\irrg{i}\irrg{j}$ to be the same. Then, only two combinations are allowed by the compatibility relations.

The little groups of H, HA, K and KA contain the symmetry operations of P3, but not these operations combined with TR. There are three irreps at each $\bf{k}$-vector: $\irr{H}{i}$, $\irr{HA}{i}$, $\irr{K}{i}$ and $\irr{KA}{i}$ at H, HA, K and KA, respectively, with $i=1,2,3$, that correspond to the three eigenvalues $1,e^{\pm i\pi/3}$ of the 3-fold axis. The coordinates of these four points correspond, in all the cases, to particular points of the general position in P1. Then the three irreps at each $\bf{k}$-vector subduce into the same irrep of the general position of P1. Note that, in P1, the multiplicity of the irrep at the general position of the unique basic band is 2. Then, in the induction process from P1, there are six possible induced pairs of irreps at H from the basic band in P1 (three pairs that contain two different irreps and three pairs of irreps with the same label). The same argument applies for HA, K and KA. In principle, there are $6^4=1296$ possible combinations of irreps at these four points in the induced bands. However, on the one hand, the compatibility relations along the line P that connects H and K forces the multiplicities of $\irr{H}{i}$ and $\irr{K}{i}$ to be the same. The same is true for $\irr{HA}{i}$ and $\irr{KA}{i}$ along the line PA. On the other hand, TR symmetry forces the multiplicities of $\irr{H}{i}$ and $\irr{HA}{i}$ (and the multiplicities of $\irr{K}{i}$ and $\irr{KA}{i}$) to be the same. As a consequence, out of the 1296 combinations, only 6 combinations of irreps at these 4 points fulfill all the requirements to form a band.

Finally, the little groups of the other two maximal $\bf{k}$-vectors L and M contain only the identity and TR and there is a unique irrep at each point. The remaining paths that connect the four sets of $\bf{k}$-vectors, (A,$\Gamma$), (H,HA,K,KA),(L) and (M) are in all cases the general position, and then no additional restrictions are imposed by the compatibility relations. 

Ordering the irreps in this way, 
\begin{equation}
\left(\irr{A}{4}\irr{A}{4},\,\irr{A}{5}\irr{A}{6},\,\irrg{4}\irrg{4},\,\irrg{5}\irrg{6},\,\irr{H}{4},\,\irr{H}{5},\,
\irr{H}{6},\,\irr{HA}{4},\,\irr{HA}{5},\,\irr{HA}{6},\,\irr{K}{4},\,\irr{K}{5},\,\irr{K}{6},\,\irr{KA}{4},\,\irr{KA}{5},\,\irr{KA}{6},\,
\irr{L}{2}\irr{L}{2},\,\irr{M}{2}\irr{M}{2}\right)
\end{equation}
the resulting 12 allowed basic bands are,
\begin{equation}
\left(n,m,n,m,p,q,r,p,q,r,p,q,r,p,q,r,1,1\right)^T
\end{equation}
with
\begin{equation}
n=1,m=0\hspace{0.1cm}\textrm{or}\hspace{0.1cm}n=0,m=1\hspace{2cm}\textrm{or}\hspace{2cm}n+m=1
\end{equation}
and
\begin{equation}
p=2,q=r=0\hspace{0.1cm}\textbf{or}\hspace{0.1cm}q=2,p=r=0\hspace{0.1cm}\hspace{0.1cm}\textbf{or}\hspace{0.1cm}r=2,p=q=0\hspace{0.1cm}\textbf{or}\hspace{0.1cm}p=0,q=r=1\hspace{0.1cm}\textbf{or}\hspace{0.1cm}q=0,p=r=1\hspace{0.1cm}\textbf{or}\hspace{0.1cm}r=0,p=q=1\nonumber
\end{equation}
or $p+q+r=2$. Six of these 12 combinations correspond to the 6 EBRs of the space group: ($n=1,p=2$), ($n=1,q=2$), ($n=1,r=2$),($n=0,p=q=1$), ($n=0,q=r=1$), ($n=0,r=p=1$) (\cite{progbandrep}). The other 6 bands can be expressed as linear combinations of the EBRs with at least one negative component and then they are fragile (and root) bands. As any other band in this space group is a combination of the 12 basic bands (6 EBRs and 6 fragile roots), there are no more fragile roots in P3 and no strong topological bands are allowed in this space group.

\subsection{Bands in the space group R3 (N. 146)}
\label{subsec:R3}
Although the point group of this space group is the same as the point group of P3, the results obtained are very different. In the standard setting, the space group R3 can be expressed as the coset decomposition  R3=P3+P3$\mathbf{t}_1$+P3$\mathbf{t}_2$ with $\mathbf{t}_2=2\mathbf{t}_1=(4/3,2/3,2/3)$. We calculate the basic bands of R3 from the basic band in P1. The transformation matrix between these two groups (which share the same lattice) is $(2\mathbf{a}/3+\mathbf{b}/3+\mathbf{c}/3,\mathbf{a}/3+2\mathbf{b}/3-\mathbf{c}/3,-1\mathbf{a}/3+\mathbf{b}/3+\mathbf{c}/3;0,0,0)$. The maximal $\mathbf{k}$-vectors are $\Gamma$, T, F and L and these $\mathbf{k}$-vectors correspond to $\Gamma$, R, U and Z, respectively, in the subgroup. With the help of DCORREL, we see that the irreps $\irr{U}{2}\irr{U}{2}$ at U and $\irr{Z}{2}\irr{Z}{2}$ at Z induce the irreps $\irr{F}{2}\irr{F}{2}$ at F and $\irr{L}{2}\irr{L}{2}$ at L in the supergroup. There is thus a 1 to 1 relationship between the irreps at these points. The irreps $\irrg{2}\irrg{2}$ at $\Gamma$ and $\irr{R}{2}\irr{R}{2}$ at R can induce two different irreps ($\irrg{4}\irrg{4}|\irrg{5}\irrg{6}$) at $\Gamma$ and ($\irr{T}{4}\irr{T}{4}|\irr{T}{5}\irr{T}{6}$) at T, respectively. In principle, there are 4 possible combinations to get basic bands from the unique band in P1, but the compatibility relations along the path $\Gamma\leftrightarrow\Lambda\leftrightarrow T$, imply these restrictions: $n(\irrg{4}\irrg{4})=n(\irr{T}{4}\irr{T}{4})$ and $n(\irrg{5}\irrg{6})=n(\irr{T}{5}\irr{T}{6})$. Therefore, ordering the irreps at the maximal $\mathbf{k}$-vectors in this way,
\begin{equation}
	\label{irreps:R3}
	\left(\irrg{4}\irrg{4},\irrg{5}\irrg{6},\irr{T}{4}\irr{T}{4},\irr{T}{5}\irr{T}{6},\irr{F}{2}\irr{F}{2},\irr{L}{2}\irr{L}{2}\right)
\end{equation}
the multiplicities of the two possible basic bands are,
\begin{equation}
	\begin{array}{l}
	\label{multi:R3}
	\left(1,0,1,0,1,1\right)\\
	\left(0,1,0,1,1,1\right)
	\end{array}
\end{equation}
Both basic bands are EBRs. Therefore, there are neither strong nor fragile topological bands in R3.

\subsection{Bands in the space group P4 (N. 75)}
\label{subsec:P4}
The space group P4 has only one maximal translationgleiche subgroup, P2. The basic bands in P2, identified in Section \ref{subsec:P2}, can be used to obtain the basic bands of P4 through the induction-subduction process. In the space group P4 there are six maximal $\mathbf{k}$-vectors: $\Gamma:(0,0,0)$, A:(1/2,1/2,1/2), M:(1/2,1/2,0), Z:(0,0,1/2), R:(0,1/2,1/2),(1/2,0,1/2) and X:(0,1/2,0),(1/2,0,0), which correspond to the $\mathbf{k}$-vectors ($\Gamma$), (E), (A), (Z), (D,C) and (B,Y), respectively, in the space group P2. Note that the 2-fold axis in the group P4 is parallel to the $z$ direction, whereas in the standard setting of the subgroup P2 the 2-fold axis is parallel to the $y$ direction. Thus, the identification of the $\mathbf{k}$-vectors in the subgroup (and the irreps) needs a previous change of setting given by the transformation matrix P=((1,0,0),(0,0,-1),(0,1,0)) and $\mathbf{p}=(0,0,0)$. The $\mathbf{k}$-vectors R and X in P4 have two branches in the star, and these points correspond to two different no symmetry-related pairs of maximal $\mathbf{k}$-vectors in the subgroup P2, D and C in the first case and B and Y in the second one.

The little group of the point R:(0,1/2,1/2) in P4 and the little group of the point D:(0,1/2,1/2) P2 are the same. The same is true for R:(1/2,0,1/2) in P4 and C:(1/2,1/2,0) in P2, for X:(0,1/2,0) in P4 and B:(0,0,1/2) in P2, and X:(1/2,0,0) in P4 and Y:(1/2,0,0) in P2. Then, there is a 1 to 1 relationship between the irreps at these pairs of $\mathbf{k}$-vectors. However, the little group of points  $\Gamma$, A, M and Z in the space group P4 is a supergroup of the little group of the corresponding $\mathbf{k}$-vector in P2, $\Gamma$, E, A and Z, respectively. The unique irrep with TR at these points in the subgroup is self-conjugated in all the cases. Then, any irrep can induce two different irreps into P4. With the help of the program DCORREL \cite{dcorrel}, we can state the relations, $\irrg{3}\irrg{4}\rightarrow\irrg{5}\irrg{7}|\irrg{6}\irrg{8}$, $\irr{E}{3}\irr{E}{4}\rightarrow\irr{A}{5}\irr{A}{7}|\irr{A}{6}\irr{A}{8}$, $\irr{A}{3}\irr{A}{4}\rightarrow\irr{M}{5}\irr{M}{7}|\irr{M}{6}\irr{M}{8}$ and $\irr{Z}{3}\irr{Z}{4}\rightarrow\irr{Z}{5}\irr{Z}{7}|\irr{Z}{6}\irr{Z}{8}$.

In principle, the unique basic band in P2 can induce $1\times1\times2\times2\times2\times2=16$ sets of irreps. However, only 4 of these combinations fulfill the compatibility relations. Then, there are only 4 basic bands in this space group and the four bands are EBRs. Putting the irreps in this order,
\begin{equation}
	\label{eq:irr75}
	\left(\irr{A}{5}\irr{A}{7},\irr{A}{6}\irr{A}{8},\irrg{5}\irrg{7},\irrg{6}\irrg{8},\irr{M}{5}\irr{M}{7},\irr{M}{6}\irr{M}{8},\irr{Z}{5}\irr{Z}{7},\irr{Z}{6}\irr{Z}{8},\irr{R}{3}\irr{R}{4},\irr{X}{3}\irr{X}{4}\right)
\end{equation}
the multiplicities of the four basic bands are,
\begin{equation}
	\label{eq:multi75}
	\begin{array}{l}
	\left(1, 0, 0, 1, 1, 0, 0, 1, 1, 1\right)\\
	\left(0, 1, 1, 0, 0, 1, 1, 0, 1, 1\right)\\
	\left(0, 1, 0, 1, 0, 1, 0, 1, 1, 1\right)\\
	\left(1, 0, 1, 0, 1, 0, 1, 0, 1, 1\right).
	\end{array}
\end{equation}
As a consequence, there are neither fragile nor strong topological bands in this space group.

\subsection{Bands in the space group P$\overline{4}$ (N. 81)}
As in the previous example, the basic bands in space group P$\overline{4}$ can be obtained through the induction from the maximal subgroup P2. The transformation matrix is also given by P=((1,0,0),(0,0,-1),(0,1,0)) and $\mathbf{p}=(0,0,0)$. The relations between the coordinates (and the labels) of the maximal $\mathbf{k}$-vectors in both settings are the exactly the same relations as in the previous section. At all the maximal $\mathbf{k}$-vectors, the number of irreps and the relations with the irreps at the corresponding $\mathbf{k}$-vectors in the subgroup are exactly the same. 

What it is different here with respect to the P2$\rightarrow$P4 induction are the little groups of the intermediate paths between maximal $\mathbf{k}$-vectors and, then, the sets of compatibility relations. In the $\Gamma\leftrightarrow\Lambda\leftrightarrow$Z path, for example, the line $\Lambda$ has as little group in P4 the whole space group without TR. Without TR, there is a 1 to 1 relation between the irreps at $\Lambda$ and the irreps at the two special points in this line, $\Gamma$ and Z. Therefore there is a 1 to 1 relation between the irreps at $\Gamma$ and Z without TR. When TR is added, two pairs of mutually conjugated irreps are merged into a single irrep at $\Gamma$ and at Z, but the 1 to 1 relation between the irreps at both points is kept.

In space group P$\overline{4}$, the little group of the intermediate line $\Lambda$ is P2 with TR, because the $\overline{4}(S_4)$ symmetry operation acts as the inversion in this line, whereas the 4-fold axis in P4 keeps every point of the line invariant. The compatibility relations now are less restrictive than in P4 and more different connections between the irreps at $\Gamma$ and Z are allowed. The same applies to the path A$\leftrightarrow$V$\leftrightarrow$M. The only restrictions imposed by the compatibility relations along these paths is the total multiplicity, that obviously must be the same everywhere to reproduce a connected band. At the other two maximal $\mathbf{k}$-vectors, X and R there is only one irrep induced from P2 and then the compatibility relations do not add extra restrictions between the irreps at these $\mathbf{k}$-vectors and all the other maximal $\mathbf{k}$-vectors. As every irrep of P2 at $\Gamma$, Z, A and M $\mathbf{k}$-vectors, induces two possible irreps in P$\overline{4}$, the single basic band in P2 induce $2^4=16$ possible combinations at these points and all correspond to basic bands. They are all the basic bands in P$\overline{4}$. Given the irreps at the maximal $\mathbf{k}$-vectors in this order,
\begin{equation}
\label{eq:irreps81}
\left(\irr{A}{5}\irr{A}{7},\irr{A}{6}\irr{A}{8},\irrg{5}\irrg{7},\irrg{6}\irrg{8},\irr{M}{5}\irr{M}{7},\irr{M}{6}\irr{M}{8},\irr{Z}{5}\irr{Z}{7},\irr{Z}{6}\irr{Z}{8},\irr{R}{3}\irr{R}{4},\irr{X}{3}\irr{X}{4}\right)
\end{equation}
The basic bands are,
\begin{equation}
\label{eq:multi81}
\left(n(\irr{A}{5}\irr{A}{7}),1-n(\irr{A}{5}\irr{A}{7}),n(\irrg{5}\irrg{7}),1-n(\irrg{5}\irrg{7}),n(\irr{M}{5}\irr{M}{7}),1-n(\irr{M}{5}\irr{M}{7}),n(\irr{Z}{5}\irr{Z}{7}),1-n(\irr{Z}{5}\irr{Z}{7}),1,1\right)^T
\end{equation}
with $0\le n(\irr{A}{5}\irr{A}{7}),n(\irrg{5}\irrg{7}),n(\irr{M}{5}\irr{M}{7}),n(\irr{Z}{5}\irr{Z}{7})\le1$. Out of these 16 basic bands of dimension 2, those bands that satisfy the relation $n(\irr{A}{5}\irr{A}{7})+n(\irrg{5}\irrg{7})+n(\irr{M}{5}\irr{M}{7})+n(\irr{Z}{5}\irr{Z}{7})=$even are EBRs, and the other 8 basic bands can not be expressed as linear integer (positive or negative) combinations of EBRs. These 8 basic bands are thus strong topological bands. In fact, there is only one strong topological index in this space group that can be defined as,
\begin{equation}
\label{eq:multi81last}
c_2=n(\irr{A}{5}\irr{A}{7})+n(\irrg{5}\irrg{7})+n(\irr{M}{5}\irr{M}{7})+n(\irr{Z}{5}\irr{Z}{7})\,\,\textrm{mod }2
\end{equation}
The 8 basic strong topological bands are equivalent, i.e., the difference of two of these strong bands is a linear combination of EBRs (with positive and negative integers).

The fragile root bands (if any) must be combinations of these 8 strong topological bands. It is immediate to see that there are 8 fragile bands of dimension 4 which are the result of combinations of two basic strong bands. The other bands of dimension 4 that are the result of combinations of two strong bands are also combinations of two EBRs so they are trivial. The 8 fragile bands of dimension 4 are,
\begin{equation}
\label{eq:fragile81}
(2a,2(1-a),2b,2(1-b),2c,2(1-c),2d,2(1-d),2,2)^T
\end{equation}
with $0\le a,b,c,d\le1$ and $a+b+c+d=1$ mod 2. These are all the fragile roots in this space group. Any other fragile band is a linear integer combination of these fragile roots and EBRs.

\subsection{Basic bands in the space group F$ddd$ (N. 70)}
\label{sub:SG70}
In this section we derive the basic bands and the fragile roots of the space group F$ddd$ (N. 70) following the group-subgroup chain,
\begin{equation}
\label{eq:transf15to70}
P\overline{1}\xrightarrow{(\mathbf{a}/2+\mathbf{b}/2,-\mathbf{a}/2\mathbf{b}/2,\mathbf{c};0,0,0)}C2/c\xrightarrow{(-\mathbf{c},-\mathbf{a},\mathbf{b}/2+\mathbf{c}/2;0,0,0)}Fddd
\end{equation}
Above each arrow it is indicated the transformation matrix of the group-subgroup pair.

The starting point are the 256 basic bands obtained previously for the space group P$\overline{1}$ (eqs. \ref{eq:labelsSG2} and \ref{eq:multSG2}), which can be used to induce the basic bands in space group C$2/c$ (N. 15). With the help of DCORREL we obtain the correlations between the irreps of the little groups of the maximal $\mathbf{k}$-vectors in P$\overline{1}$ and C$2/c$. In our case, the relevant relations are (on the left of the arrow the irreps of C$2/c$ and on the right the irreps of P$\overline{1}$):
\begin{equation}
\label{eq:subducton70}
\begin{array}{lcl}
\irrg{3}\irrg{4}\rightarrow\irrg{3}\irrg{3}& \hspace{2cm} &\irrg{5}\irrg{6}\rightarrow\irrg{2}\irrg{2}\\
\irr{A}{2}\irr{A}{2}\rightarrow\irr{Z}{2}\irr{Z}{2}\oplus\irr{Z}{3}\irr{Z}{3}&&\\
\irr{M}{2}\irr{M}{2}\rightarrow\irr{R}{2}\irr{R}{2}\oplus\irr{R}{3}\irr{R}{3}&&\\
\irr{Y}{3}\irr{Y}{4}\rightarrow\irr{V}{3}\irr{V}{3}&&\irr{Y}{5}\irr{Y}{6}\rightarrow\irr{V}{2}\irr{V}{2}\\
\irr{L}{2}\irr{L}{2}\rightarrow\irr{T}{2}\irr{T}{2}&&\irr{L}{3}\irr{L}{3}\rightarrow\irr{T}{3}\irr{T}{3}\\
\irr{L}{2}\irr{L}{2}\rightarrow\irr{U}{3}\irr{U}{3}&&\irr{L}{3}\irr{L}{3}\rightarrow\irr{U}{2}\irr{U}{2}\\
\irr{V}{2}\irr{V}{2}\rightarrow\irr{X}{3}\irr{X}{3}&&\irr{V}{3}\irr{V}{3}\rightarrow\irr{X}{2}\irr{X}{2}\\
\irr{V}{2}\irr{V}{2}\rightarrow\irr{Y}{3}\irr{Y}{3}&&\irr{V}{3}\irr{V}{3}\rightarrow\irr{Y}{2}\irr{Y}{2}
\end{array}
\end{equation}

At the $\Gamma$ point, the little group is the whole group both in C$2/c$ and in P$\overline{1}$, then the little group in C$2/c$ is a supergroup of the little group in P$\overline{1}$. Without TR, the irreps $\irrg{2}$ and $\irrg{3}$ of P$\overline{1}$ are self-conjugated in C$2/c$ and every irrep can induce two possible irreps in C$2/c$, $\irrg{2}\rightarrow\irrg{5}|\irrg{6}$ and $\irrg{3}\rightarrow\irrg{3}|\irrg{4}$. If TR is considered, the irreps double in P$\overline{1}$, and in C$2/c$ there are two pairs of complex conjugated irreps, $(\irrg{3},\irrg{4})$ and $(\irrg{5},\irrg{6})$. Then, there is 1 to 1 relation between the irreps at the $\Gamma$ point in both space groups, as shown in Eq. (\ref{eq:subducton70}). The same arguments can be applied to the Y point (V point in the subgroup).

The little group of points L and V in the space groups P$\overline{1}$ and C$2/c$ is the same so there is a 1 to 1 relationship between the irreps in both groups at these points.

The little groups of point A (Z point in the setting of the subgroup) are different in the group and in the subgroup: in each case it is the whole space group. Without TR, the two irreps in the subgroup P$\overline{1}$ belong to the same orbit with respect to C$2/c$ and, then, the pair of irreps $\irr{Z}{2}$ and $\irr{Z}{3}$ merge to get the irrep $\irr{A}{2}$ in C$2/c$. With TR all the irreps double. The merging of two irreps in the subgroup to give a single irrep in the supergroup means that not all the basic bands in the subgroup can induce a band in the supergroup. If the multiplicities of the irreps $\irr{Z}{2}$ and $\irr{Z}{3}$ are different in a basic band, this basic band \emph{alone} does not induce a band in the supergroup, because the multiplicity of 
$\irr{A}{2}$ is not well defined. It is necessary to combine, at least, two basic bands whose total multiplicity satisfies $n(\irr{Z}{2})=n(\irr{Z}{3})$ to make possible the induction. The same arguments can be applied to the point M (R point in the subgroup).

None of the basic bands in P$\overline{1}$ satisfies the condition $n(\irr{Z}{2})=n(\irr{Z}{3})$ or the condition $n(\irr{R}{2})=n(\irr{R}{3})$. Therefore, the basic bands in the supergroup must be induced from pairs of basic bands whose total multiplicities satisfy these two conditions. 

Taking into consideration the above restrictions, if we order the irreps in C$2/c$ in this way,

\begin{equation}
\label{eq:labelsSG15}
\left(\irrg{3}\irrg{4},\irrg{5}\irrg{6},\irr{A}{2}\irr{A}{2},\irr{M}{2}\irr{M}{2},\irr{Y}{3}\irr{Y}{4},\irr{Y}{5}\irr{Y}{6},\irr{L}{2}\irr{L}{2},\irr{L}{3}\irr{L}{3},\irr{V}{2}\irr{V}{2},\irr{V}{3}\irr{V}{3}\right)
\end{equation}
the induced basic bands are given by the multiplicities,
\begin{equation}
\label{eq:multSG15}
\left(n(\irrg{3}\irrg{4}),n(\irrg{5}\irrg{6}),1,1,n(\irr{Y}{3}\irr{Y}{4}),n(\irr{Y}{5}\irr{Y}{6}),n(\irr{L}{2}\irr{L}{2}),n(\irr{L}{3}\irr{L}{3}),n(\irr{V}{2}\irr{V}{2}),n(\irr{V}{3}\irr{V}{3})\right)^T
\end{equation}
with,
\begin{equation}
\label{eq:condSG15}
n(\irrg{3}\irrg{4})+n(\irrg{5}\irrg{6})=n(\irr{Y}{3}\irr{Y}{4})+n(\irr{Y}{5}\irr{Y}{6})=n(\irr{L}{2}\irr{L}{2})+n(\irr{L}{3}\irr{L}{3})=n(\irr{V}{2}\irr{V}{2})+n(\irr{V}{3}\irr{V}{3})=2.
\end{equation}

Then, there are 81 basic bands in C$2/c$, all of dimension 4, that correspond to all the possible choices of the multiplicities in (\ref{eq:multSG15}) under the restrictions (\ref{eq:condSG15}). 9 basic bands are EBRs and the other 72 can not be expressed as integer linear combination of the EBRs, then they are strong topological bands. This space group has no fragile bands of dimension 4. All the fragile root bands in this space group must be combinations of strong topological bands. If we compute all the possible combinations of two strong basic bands with complementary indices and check which combinations are not the combination of two EBRs, we find that there are 44 such combinations that correspond to fragile roots. Next we take combinations of three basic bands and check which ones can be expressed as integer linear combinations of EBRs (using positive and negative integers) but are not integer linear combinations with non-negative indices of EBRs and fragile roots of lower dimension. There are 16 bands that fulfill these conditions and they are fragile roots of dimension 12. None of the combinations of four or higher number of basic bands fulfill the condition, so the number of fragile roots in this space group is 44+16=60 \cite{song2019}.

The basic bands in space group F$ddd$ (N. 70) can be obtained from the 81 basic bands in space group C$2/c$. Introducing the numbers of both space groups and the transformation matrix shown in Eq. (\ref{eq:transf15to70}) in the program DCORREL, we get the correlations between the irreps in both groups. If we select the relevant ones for our purposes, we get the following relations (F$ddd$ on the left and C$2/c$ on the right):

\begin{equation}
\begin{array}{lcl}
\irrg{5}\rightarrow\irrg{3}\irrg{4}& \hspace{2cm} &\irrg{6}\rightarrow\irrg{5}\irrg{6}\\
\irr{T}{3}\irr{T}{4}\rightarrow\irr{Y}{3}\irr{Y}{4}\oplus\irr{Y}{5}\irr{Y}{6}&&\\
\irr{Y}{3}\irr{Y}{4}\rightarrow\irr{M}{2}\irr{M}{2}&&\\
\irr{Z}{3}\irr{Z}{4}\rightarrow\irr{A}{2}\irr{A}{2}&&\\
\irr{L}{2}\irr{L}{2}\rightarrow\irr{L}{3}\irr{L}{3}&&\irr{L}{3}\irr{L}{3}\rightarrow\irr{L}{2}\irr{L}{2}
\end{array}
\end{equation}
At the $\Gamma$, Y, Z, and L points, there is a 1 to 1 relation between the TR-invariant irreps in the subgroup and in the supergroup. However, at the T point (Y point in the subgroup) two irreps in the subgroup merge to give a single irrep in the supergroup. Again, not all the basic bands in the subgroup induce a band in the supergroup. Only those which fulfill the condition  $n(\irr{Y}{3}\irr{Y}{4})=n(\irr{Y}{5}\irr{Y}{6})$ do. The rest of basic bands in the subgroup must be combined in such a way that the total multiplicities satisfy the condition $n(\irr{Y}{3}\irr{Y}{4})=n(\irr{Y}{5}\irr{Y}{6})$. Considering the basic bands of space group C$2/c$ given by eqs. (\ref{eq:labelsSG15}) and (\ref{eq:multSG15}) and the mentioned restriction, if the irreps of the space group F$ddd$ at maximal $\mathbf{k}$-vectors are ordered in this way,
\begin{equation}
\label{eq:labelsSG70}
\left(\irrg{5},\irrg{6},\irr{T}{3}\irr{T}{4},\irr{Y}{3}\irr{Y}{4},\irr{Z}{3}\irr{Z}{4},\irr{L}{2}\irr{L}{2},\irr{L}{3}\irr{L}{3}\right)
\end{equation}
the induced basic bands are given by the multiplicities,
\begin{equation}
\label{eq:multSG70}
\left(n(\irrg{5}),n(\irrg{6}),1,1,1,n(\irr{L}{2}\irr{L}{2}),n(\irr{L}{3}\irr{L}{3})\right)^T
\end{equation}
with,
\begin{equation}
\label{eq:condSG70}
n(\irrg{5})+n(\irrg{6})=n(\irr{L}{2}\irr{L}{2})+n(\irr{L}{3}\irr{L}{3})=2.
\end{equation}
There are 9 basic bands in this space group and just one of them is an EBR. The other 8 are strong topological bands. Following the same steps described above in the P$\overline{1}\rightarrow$C$2/c$ induction, we find 2 different fragile bands of dimension 8 (then they are fragile roots) as combination of 2 strong basic bands that are not combinations of EBRs, 4 fragile roots of dimension 12 as combinations of 3 basic strong bands and 4 fragile roots of dimension 16 as combinations of 4 basic strong bands. This space group has 10 fragile roots given by the multiplicities,
\begin{equation}
\begin{array}{rrrrrrrrl}
2&2&2&2&2&0&4&\hspace{1cm}&d=8\\
2&2&2&2&2&4&0&\hspace{1cm}&d=8\\
1&5&3&3&3&0&6&\hspace{1cm}&d=12\\
1&5&3&3&3&6&0&\hspace{1cm}&d=12\\
5&1&3&3&3&0&6&\hspace{1cm}&d=12\\
5&1&3&3&3&6&0&\hspace{1cm}&d=12\\
0&8&4&4&4&0&8&\hspace{1cm}&d=16\\
0&8&4&4&4&8&0&\hspace{1cm}&d=16\\
8&0&4&4&4&0&8&\hspace{1cm}&d=16\\
8&0&4&4&4&8&0&\hspace{1cm}&d=16
\end{array}
\end{equation}
where $d$ is the dimension of the band.

\subsection{Bands in the space group I$2_13$ (N. 199)}
\label{bands199fragile}
The basic bands of the space group I$2_1$ can be obtained following this group-subgroup chain,
\begin{equation}
\label{eq:transf1to199}
P1\xrightarrow{(\mathbf{a}/2+\mathbf{b}/2,-\mathbf{a}/2+\mathbf{b}/2,\mathbf{c};0,0,0)}C2\xrightarrow{(\mathbf{a}+\mathbf{c},\mathbf{b},\mathbf{c};1/4,0,0)}I2_12_12_1\xrightarrow{(\mathbf{a},\mathbf{b},\mathbf{c};0,0,0)}I2_13
\end{equation}
We will get just the basic bands in each intermediate subgroup and, finally, the fragile roots in the space group I$2_13$.

From the unique basic band in P1, with the help of DCORREL, we get a single basic band in C2. The maximal $\mathbf{k}$-vectors V:(1/2,1/2,0) and L:(1/2,1/2,1/2) have the same little group (P1) both in P1 and C2, so there is a 1 to 1 relation between the irreps at these $\mathbf{k}$-vectors in the group-subgroup pair. The other four maximal $\mathbf{k}$-vectors $\Gamma$:(0,0,0), A:(0,0,1/2), M:(0,1,1/2) and Y:(0,1,0) have different little groups in P1 and C2 and the unique irrep in the subgroup at each point is then self-conjugated. Without TR each irrep can induce two different complex irreps into C2 which are mutually conjugated. Therefore, when TR is added, the two irreps at each point in the supergroup merge into one irrep and there is finally a 1 to 1 correspondence between the irreps in the subgroup and the irreps in the supergroup at these points. Ordering the irreps in C2 at the maximal $\mathbf{k}$-vectors as,
\begin{equation}
\label{eq:labelsSG5}
\left(\irrg{3}\irrg{4},\irr{A}{3}\irr{A}{4},\irr{M}{3}\irr{M}{4},\irr{Y}{3}\irr{Y}{4},\irr{L}{2}\irr{L}{2},\irr{V}{2}\irr{V}{2}\right),
\end{equation}
the multiplicities of the single basic band are,
\begin{equation}
\label{eq:multSG5}
\left(1,1,1,1,1,1\right)^T.
\end{equation}
Now from this band we induce the basic bands in space group I$2_12_12_1$. This is the example used in section (\ref{sup:inducedbands}) to explain in detail the general procedure. The irreps at maximal $\mathbf{k}$-vectors and the calculated single basic band (which is also an EBR) are given in eqs. (\ref{eq:labelsSG24}) and (\ref{eq:multSG24}), respectively.

Finally, we consider the last group-subgroup pair in Eq. (\ref{eq:transf1to199}) to get the basic bands in the space group I$2_13$. The maximal $\mathbf{k}$-vectors are $\Gamma$:(0,0,0), H:(1,1,1), N=(1/2,1/2,0), P:(1/2,1/2,1/2) and PA:(-1/2,-1/2,-1/2). The little group of N is the same in both space groups and there is a 1 to 1 correlation between the unique double-valued irrep in I$2_12_12_1$ and in I$2_13$. The little groups of $\Gamma$ and X are different in the group and in the subgroup, being the group-subgroup index 3. As the unique double-valued irreps in $\Gamma$ and X are $\irrg{5}$ and $\irr{X}{5}$, respectively, and as both are self-conjugated, they can induce three different irreps without TR into the supergroup: $\irrg{5}|\irrg{6}|\irrg{7}$ the first one and $\irr{H}{5}|\irr{H}{6}|\irr{H}{7}$ the second one. When TR is added, the irreps in the supergroup are $\irrg{5}\irrg{6}$, $\irrg{7}$ at $\Gamma$ and $\irr{H}{5}\irr{H}{6}$, $\irr{H}{7}$ at H. As the multiplicity of $\irrg{5}$ and $\irr{X}{5}$ in the unique basic band of the space group  I$2_12_12_1$ is 2, the two irreps can induce the following irreps into I$2_13$,
\begin{equation}
	\begin{array}{lcl}
		2\,\irrg{5}\rightarrow2\,\irrg{5}&\textrm{or}&2\,\irrg{5}\rightarrow\irrg{6}\irrg{7}\\ 
		2\,\irr{X}{5}\rightarrow2\,\irr{H}{5}&\textrm{or}&2\,\irr{X}{5}\rightarrow\irr{H}{6}\irr{H}{7}
	\end{array}
\end{equation}

The little groups of the two last maximal $\mathbf{k}$-vectors, P and PA, are also different in the group and in the subgroup, but the little groups do not contain TR (or any other symmetry operation combined with TR). There are 4 double-valued irreps at P in the subgroup (in the subgroup the label of this point is W): $\irr{W}{2}$,$\irr{W}{3}$,$\irr{W}{4}$ and $\irr{W}{5}$. The first three irreps belong to the same orbit and, thus, they induce together a single irrep into the supergroup, $\irr{W}{2}\oplus\irr{W}{3}\oplus\irr{W}{4}\rightarrow\irr{P}{7}$. The last one is self-conjugated and it can induce three different irreps into the supergroup, $\irr{W}{5}\rightarrow\irr{P}{4}|\irr{P}{5}|\irr{P}{6}$.

At the point PA we get exactly the same correlations if we change the labels of the irreps as W$\rightarrow$WA and P$\rightarrow$PA. 

The compatibility relations (\cite{dcomprel}) along the path P$\leftrightarrow\Lambda\leftrightarrow$PA give the following restrictions:
\begin{eqnarray}
	\label{eq:comprel199PPA}
	n(\irr{P}{4})+n(\irr{P}{7})&=&n(\irr{PA}{4})+n(\irr{PA}{7})\nonumber\\
	n(\irr{P}{5})+n(\irr{P}{7})&=&n(\irr{PA}{6})+n(\irr{PA}{7})\\
	n(\irr{P}{6})+n(\irr{P}{7})&=&n(\irr{PA}{5})+n(\irr{PA}{7})\nonumber
\end{eqnarray}
and the TR symmetry forces the relation $n(\irr{P}{i})=n(\irr{PA}{i})$ for $i=4,5,6,7$. Given the irreps at maximal $\mathbf{k}$-vectors in the space group I$2_13$ in this order,
\begin{equation}
\label{eq:labelsSG199}
\left(\irrg{5},\irrg{6}\irrg{7},\irr{H}{5},\irr{H}{6}\irr{H}{7},\irr{P}{4},\irr{P}{5},\irr{P}{6},\irr{P}{7},\irr{PA}{4},\irr{PA}{5},\irr{PA}{6},\irr{PA}{7},\irr{N}{3}\irr{N}{4}\right),
\end{equation}
and considering the above restrictions, before we take into account the constrains imposed by additional compatibility relations, the induced sets of irreps are,
\begin{eqnarray}
	\label{eq:solutions191prev}
\left(n(\irrg{5}),n(\irrg{6}\irrg{7}),n(\irr{H}{5}),n(\irr{H}{6}\irr{H}{7}),n(\irr{P}{4}),n(\irr{P}{5}),n(\irr{P}{6}),n(\irr{P}{7}),n(\irr{P}{4}),n(\irr{P}{5}),n(\irr{P}{6}),n(\irr{P}{7}),2N\right)^T
\end{eqnarray}
with
\begin{eqnarray}
	\label{eq:solutions191sum}
2n(\irrg{5})+4n(\irrg{6}\irrg{7})=2n(\irr{H}{5})+4n(\irr{H}{6}\irr{H}{7})=n(\irr{P}{4})+n(\irr{P}{5})+n(\irr{P}{6})+3n(\irr{P}{7})=4N,
\end{eqnarray}
where $N>0$ is an integer and $4N$ is the dimension of the band. 

Next we check the compatibility relations along the path  PA$\leftrightarrow\Lambda\leftrightarrow\Gamma\leftrightarrow\Lambda\leftrightarrow$P$\leftrightarrow\Lambda\leftrightarrow$H. There are three irreps at the line $\Lambda$: $\irrl{4}$,$\irrl{5}$ and $\irrl{6}$. As the multiplicity of these irreps along the path must not change, we can establish these additional restrictions (one restriction for each irrep at $\Lambda$):
\begin{eqnarray}
	\label{eq:solutions191comp}
\irrl{4}&\rightarrow&n(\irr{P}{4})+n(\irr{P}{7})=2n(\irrg{6}\irrg{7})=n(\irr{P}{4})+n(\irr{P}{7})=2n(\irr{H}{6}\irr{H}{7})\nonumber\\
\irrl{5}&\rightarrow&n(\irr{P}{6})+n(\irr{P}{7})=n(\irrg{5})+n(\irrg{6}\irrg{7})=n(\irr{P}{5})+n(\irr{P}{7})=n(\irr{H}{5})+n(\irr{H}{6}\irr{H}{7})\\
\irrl{6}&\rightarrow&n(\irr{P}{5})+n(\irr{P}{7})=n(\irrg{5})+n(\irrg{6}\irrg{7})=n(\irr{P}{6})+n(\irr{P}{7})=n(\irr{H}{5})+n(\irr{H}{6}\irr{H}{7})\nonumber
\end{eqnarray}
and $n(\irr{P}{7})=n(\irr{P}{4})+n(\irr{P}{5})+n(\irr{P}{6})$.

These restrictions can be simplified to,
\begin{equation}
\begin{array}{lcl}
n(\irrg{5})+n(\irrg{6}\irrg{7})&=&n(\irr{H}{5})+n(\irr{H}{6}\irr{H}{7})=n(\irr{P}{5})+n(\irr{P}{7})\nonumber\\	
2n(\irrg{6}\irrg{7})&=&2n(\irr{H}{6}\irr{H}{7})=n(\irr{P}{4})+n(\irr{P}{7})\nonumber\\
n(\irr{P}{5})&=&n(\irr{P}{6})\nonumber\\
n(\irr{P}{4})+2n(\irr{P}{5})&=&N
\end{array}
\end{equation}
Then, all the possible pairs of multiplicities $\left(n(\irr{P}{4}),n(\irr{P}{5})\right)$ give all the possible bands in this space group. There is only one solution for $N=1$, $\left(n(\irr{P}{4}),n(\irr{P}{5})\right)=(1,0)$ and the resulting basic band is,
\begin{equation}
\label{eq:band191dim4}
(1,0,1,0,1,0,0,1,1,0,0,1,2)^T
\end{equation}
This band can be expressed as linear combination of EBRs where, at least, one integer must be negative, and then it is a fragile root band.

There are only two solutions for $N=2$, $\left(n(\irr{P}{4}),n(\irr{P}{5})\right)=(2,0)$ and $\left(n(\irr{P}{4}),n(\irr{P}{5})\right)=(0,1)$. The first one is 2 times the band given in Eq. \ref{eq:band191dim4} and corresponds to one EBR. The second one is a basic band and also an EBR,
\begin{equation}
\label{eq:band191dim8}
(2,1,2,1,0,1,1,2,0,1,1,2,4)^T
\end{equation}
Any other band in the group is a linear combination of the two basic bands \ref{eq:band191dim4} and \ref{eq:band191dim8}, and then there is only one fragile root \ref{eq:band191dim4} and no strong bands.

\subsection{Bands in the space group P$4/m$ (No. 83).}
\label{bands:fragile83}
Finally, we will calculate the basic bands in the space group P$4/m$. We take as maximal subgroup the space group P4 (N. 75), whose four basic bands have been calculated in section \ref{subsec:P4}. The order of the irreps in P4 are given in Eq. \ref{eq:irr75} and the multiplicities of the basic bands in Eq. \ref{eq:multi75}. The transformation matrix of the group-subgroup pair is the identity and the set of maximal $\mathbf{k}$-vectors and their labels are exactly the same. Introducing these data in DCORREL, we get the following results:
\begin{equation}
	\begin{array}{l}
	\irrr{5}\irrr{7}\rightarrow\irrr{5}\irrr{7}\hspace{0.5cm}\textrm{or}\hspace{0.5cm}\irrr{{11}}\irrr{9}\\
	\irrr{6}\irrr{8}\rightarrow\irrr{6}\irrr{8}\hspace{0.5cm}\textrm{or}\hspace{0.5cm}\irrr{{10}}\irrr{{12}}
	\end{array}
\end{equation}
for $\rho=\Gamma$, A, M, Z, and
\begin{equation}
	\irrr{3}\irrr{4}\rightarrow\irrr{3}\irrr{4}\hspace{0.5cm}\textrm{or}\hspace{0.5cm}\irrr{5}\irrr{6}
\end{equation}
for $\rho$=R,X.

Every double-valued irrep at the maximal $\mathbf{k}$-vectors $\Gamma$, A, M, Z, R, X in the space group P4 is self-conjugated and induces two possible irreps in P$4/m$. Therefore, every basic band in space group P4 can induce $2^6=64$ basic bands in space group P$4/m$. Then, the 4 basic bands in P4 induce 256 basic bands in P$4/m$, and they are all the basic bands in this space group. 16 of these basic bands are EBRs. The other 240 are strong topological bands. Any other strong topological band is a linear combination of these 240 bands.

To find the fragile roots in this space group we must consider combinations of strong basic bands increasing successively the number of strong bands to combine. At first sight, it seems an intractable system if a high number of combinations of basic strong bands should be considered. However, in the intermediate steps a large number of combinations can be removed. For instance, out of $240^2=57600$ possible combinations of 2 strong basic bands, there are only 6808 different sets of combinations of irreps that cannot be expressed as a linear combination with non-negative integers of an EBR and another strong band. 80 combinations give rise to fragile (and then root) bands and the remaining 6728 sets of irreps are the only sets of irreps that must be considered in the next step, i.e., the combinations of 3 strong basic bands, that can be constructed combining these 6728 subsets with the 240 basic strong bands. Again, if we remove from the resulting subset of $240\times6728=1614720$ combinations the ones that can be expressed as the linear combination of a non-strong band (an EBR or a fragile root identified in the previous steps) and strong bands (including the case of 0 strong bands), the number of remaining distinct bands is 55808, which are divided into 800 fragile roots and 55008 sets of irreps to consider in the next step, etc...

In following steps (addition of an increasing number of basic strong bands) the number of subsets that remain to be considered in subsequent steps increases up to a maximum value of relevant 326496 bands as the result of the combinations of 7 basic strong bands. For larger numbers of combinations of strong bands, the number of remaining sets of irreps decreases. Finally, any combination of 13 bands can be put as a linear combination of a no strong (EBR or fragile) band a strong bands. Therefore, all the 58840 fragile roots found as combinations of 2,3,\ldots,12 strong bands are all the fragile roots in the space group.

\section{Smith Form of the Compatibility Matrix}
\label{sup:compasmith}
In section (\ref{compatibilityfirst}) we describe the usefulness of the compatibility relations in the analysis of the band connections along high-symmetry lines or planes between maximal $\mathbf{k}$-vectors. In this section we apply these conditions systematically to obtain the compatibility matrix, i. e., a matrix that contains all the information about the compatibility relations in a space group. For a symmetry data vector $B$ describing a band in Eq. (\ref{Bvector}) (note that we have changed slightly the notation of the multiplicities to make explicit the different $\mathbf{k}$-vectors in the list, which is necessary for the discussion in this section),
\begin{equation}
\label{Bvectorkvecs}
B = (m_1^{\bf{k}_1},\ldots, m_{N_{k_1}}^{\bf{k}_1},m_1^{\bf{k}_2},\ldots, m_{N_{k_2}}^{\bf{k}_2}, \ldots,\ldots,\ldots, m_1^{\bf{k}_N},\ldots, m_{N_{k_N}}^{\bf{k}_N})^T
\end{equation}
 to satisfy the compatibility relations, we must have that

\begin{equation}
\sum_{i=1}^{N_{\bf{k}_{\alpha}}} m_i^{\bf{k}_{\alpha}} \rho_{\mathcal{G}^{\bf{k}_{\alpha}}}^i\downarrow \mathcal{G}^{\bf{k}_l}=\sum_{i=1}^{N_{\bf{k}_{\beta}}} m_i^{\bf{k}_{\beta}} \rho_{\mathcal{G}^{\bf{k}_{\beta}}}^i\downarrow \mathcal{G}^{\bf{k}_l}
\label{representationreduction2points}
\end{equation} 
where $\bf{k}_{\alpha}$, $\bf{k}_{\beta}$ are two high-symmetry points in the space group $\mathcal{G}$, $\bf{k}_l$ is a symmetry line connecting the two maximal $k$-vectors and $m_i^{\bf{k}_{\alpha}}$ is the multiplicity of the  $i$ representation $\rho_{\mathcal{G}^{\bf{k}_{\alpha}}}^i$ at the point $\bf{k}_{\alpha}$. Using Eq. (\ref{representationreduction}), this becomes a linear equation. We write the Eq. (\ref{representationreduction}) in a way that clearly differentiates the $m_{ij}^{\bf{k}_{\alpha},\bf{k}_l}$ integer multiplicities in the decomposition of $\rho_i$: 

\begin{equation}
\rho_{\mathcal{G}^{\bf{k}_{\alpha}}}^i\downarrow\mathcal{G}^{\bf{k}_l}=\bigoplus_{i=1}^sm_{ij}^{\bf{k}_{\alpha},\bf{k}_l}\rho_{\mathcal{G}^{\bf{k}_l}}^j
\end{equation}
 where $m_{ij}^{\bf{k}_{\alpha},\bf{k}_l}$ are known from the compatibility relations on the BCS. Hence the Eq. (\ref{representationreduction2points}) becomes:

\begin{equation}
\sum_{i=1}^{N_{\bf{k}_{\alpha}}} m_i^{\bf{k}_{\alpha}}m_{ij}^{\bf{k}_{\alpha},\bf{k}_l}=\sum_{i=1}^{N_{\bf{k}_{\beta}}} m_i^{\bf{k}_{\beta}}m_{ij}^{\bf{k}_{\beta},\bf{k}_l}
\label{representationreduction2pointssimple1}
\end{equation}
 Note that the index $j$ on the left and right hand side is the same. This equation is valid between any two maximal $\mathbf{k}$-vectors. We can now form the integer-valued compatibility row vector $Comp_{\bf{k}_{\alpha},\bf{k}_{\beta},\bf{k}_l}$ for every triplet $\bf{k}_{\alpha}$, $\bf{k}_{\beta}$, $\bf{k}_l$ such that at the position of $m_i^{\bf{k}_{\alpha}}$ in Eq. (\ref{Bvectorkvecs}) the vector-component is $m_{ij}^{\bf{k}_{\alpha},\bf{k}_l}$, at the position of $m_i^{\bf{k}_{\beta}}$  in Eq. (\ref{Bvectorkvecs}) the vector-component is $-m_{ij}^{\bf{k}_{\beta},\bf{k}_l}$ and 0 otherwise.
 
 The compatibility matrix $Comp$ has as many rows as irreps in all the possible different paths that can be defined between all the pairs of maximal $\mathbf{k}$-vectors.
 The matrix equation that checks if the band vector $B$ in eqs. (\ref{Bvector}) and (\ref{Bvectorkvecs}) satisfies the compatibility relations reads (Eq. \ref{comprel} in the main text),
\begin{equation}
	\label{comprelsupmat}
Comp\cdot B=0
\end{equation}
In general, not all the equations are linearly independent, but one of the advantages of the Smith Decomposition procedure is that its simplicity to identify the redundancies in the set of equations.
We now have obtained an integer equation which would give us the general solution for the band vector $B$ in Eq. (\ref{Bvector}).

The determination of the possible bands given in general form in Eq. (\ref{Bvector}) that satisfy the compatibility relations is equivalent to the procedure followed in the main text for the set of equations, $Comp\cdot B=0$. In fact, as it is stressed
in the main text, as rank($Comp$)=rank($\widetilde{L}$)=rank($Comp\cup \widetilde{L}$), the sets of equations (\ref{comprelsupmat}) and (\ref{Lmatcomp}) in the main text are equivalent.

\section{Example: Strong Indices From Topological Classes: Inversion Symmetry in 2 and 3 D}
\label{sup:strongtopo}
In two dimensions the high-symmetry points in this space group are $\Gamma$:(0,0), X:(1/2,0), Y:(0,1/2) and V:(1/2,1/2) and the irreps at these high symmetry points are $\irrg{2}$, $\irrg{3}$, $\irr{X}{2}$, $\irr{X}{3}$, $\irr{Y}{2}$, $\irr{Y}{3}$ and $\irr{V}{2}$, $\irr{V}{3}$ without TR and the same symbols but doubled if TR is also considered (Kramers degeneracy). In this context, it is typical the use of an alternative notation, using the components of the $\mathbf{k}$-vector (with the $2\pi$ factor) to denote the $\mathbf{k}$-vector and the "+" and "-" signs for the irreps with subscript 2 and 3, respectively (with or without TR). For example, the symbol $(+,(\pi,0))$ represents the irrep $\irr{X}{2}$ without TR or $\irr{X}{2}\irr{X}{2}$ with TR and $(-,(\pi,\pi))$ represents the irrep $\irr{V}{3}$ or $\irr{V}{3}\irr{V}{3}$ without or with TR, respectively.

In three dimensions the high symmetry points are  $\Gamma$:(0,0,0), X:(1/2,0,0), Y:(0,1/2,0), V:(1/2,1/2,0), Z:(0,0,1/2), U:(1/2,0,1/2), T:(0,1/2,1/2) and R:(1/2,1/2,1/2) and the same conventions can be applied.

We can then express any set of bands as a $2*2^D$-component vector in $D$ dimensions. Adding spinfull TR only doubles the representations to make them Kramers pairs, so the current classification is valid with and without TR.

\subsection{Example: Inversion in 2D}

We ask: can a set of bands given by
\begin{eqnarray}
B= ( n_{+,(0,0)},n_{-,(0,0)}, n_{+,(\pi,0)},n_{-,(\pi,0)}, n_{+,(\pi,\pi)},n_{-,(\pi,\pi)}, n_{+,(0,\pi)},n_{-,(0,\pi)})
\end{eqnarray}
be written as a sum of EBRs or not and then it is topological (or a semimetal in SOC-free groups?)

We first enumerate all the $EBR$'s in 2D: 
\begin{eqnarray}
&s_{1a}\uparrow SG=  \text{ebr}_1=(1,0,1,0,1,0,1,0)\noindent \\  & p_{1a}\uparrow SG=  \text{ebr}_2=(0,1,0,1,0,1,0,1) \noindent \\  &s_{1b}\uparrow SG=  \text{ebr}_3=(1,0,0,1,0,1,1,0) \noindent \\  & p_{1b}\uparrow SG=  \text{ebr}_4=(0,1,1,0,1,0,0,1)\noindent \\  & s_{1c}\uparrow SG=  \text{ebr}_5=(1,0,1,0,0,1,0,1)\noindent \\  & p_{1c}\uparrow SG= \text{ebr}_6=(0,1,0,1,1,0,1,0) \noindent \\  &s_{1d}\uparrow SG= \text{ebr}_7=(1,0,0,1,1,0,0,1) \noindent \\  &p_{1d}\uparrow SG= \text{ebr}_8=(0,1,1,0,0,1,1,0)
\end{eqnarray}
We note that the order (basis) of $K$ points in the $B$ vector and in each EBR has to be the same: here it is $\Gamma$, X, V, Y.

We then join the EBRs in a matrix $EBR_{2D}= (ebr_1^T, ebr_2^T,ebr_3^T,ebr_4^T,ebr_5^T,ebr_6^T,ebr_7^T,ebr_8^T)$:
\begin{equation}
EBR_{\text{Inversion 2D}} =\left(
\begin{array}{cccccccc}
 1 & 0 & 1 & 0 & 1 & 0 & 1 & 0 \\
 0 & 1 & 0 & 1 & 0 & 1 & 0 & 1 \\
 1 & 0 & 0 & 1 & 1 & 0 & 0 & 1 \\
 0 & 1 & 1 & 0 & 0 & 1 & 1 & 0 \\
 1 & 0 & 0 & 1 & 0 & 1 & 1 & 0 \\
 0 & 1 & 1 & 0 & 1 & 0 & 0 & 1 \\
 1 & 0 & 1 & 0 & 0 & 1 & 0 & 1 \\
 0 & 1 & 0 & 1 & 1 & 0 & 1 & 0 \\
\end{array}
\right)
\end{equation}
We perform the Smith decomposition to get the $\Delta$ in Eq(\ref{schmidtequation}):

\begin{equation}
\Delta_{\text{Inversion 2D}}=\left(
\begin{array}{cccccccc}
 1 & 0 & 0 & 0 & 0 & 0 & 0 & 0 \\
 0 & 1 & 0 & 0 & 0 & 0 & 0 & 0 \\
 0 & 0 & 1 & 0 & 0 & 0 & 0 & 0 \\
 0 & 0 & 0 & 1 & 0 & 0 & 0 & 0 \\
 0 & 0 & 0 & 0 & 2 & 0 & 0 & 0 \\
 0 & 0 & 0 & 0 & 0 & 0 & 0 & 0 \\
 0 & 0 & 0 & 0 & 0 & 0 & 0 & 0 \\
 0 & 0 & 0 & 0 & 0 & 0 & 0 & 0 \\
\end{array}
\right)
\end{equation} 
Hence the classification of strong topological phases with inversion in $2$ D is $Z_2$  (Chern insulators or Quantum Spin Hall, without or with TR, respectively). This is because, if a set of bands represented by an $8$-component vector $B$ in $2$ dimensions can be written as a sum of EBRs (it is not topological), then $C= L\cdot B$ where $L$ is obtained from Eq. (\ref{schmidtequation}) of the main text by the Smith decomposition. The component $c_5$ of the $C$ vector must be even, and  $c_6, c_7, c_8$ must be zero. Otherwise it \emph{cannot} be expressed as a sum of EBRs and it is topological. If  $c_6, c_7, c_8$ are nonzero then the set of bands $B$ does not satisfy the compatibility relations (in this case and this space group, it means that the irreps at different maximal $\mathbf{k}$-vectors are not all inter-connected). If $c_5$ is not even then the set of bands $B$ cannot be expressed in terms of an $EBR$ and hence it would need to be topological. The topological classification is $Z_2$ and the index can be computed (once a particular choice of the matrix $L$ in Eq(\ref{schmidtequation}) has been obtained through the Smith decomposition) by:
\begin{eqnarray}
&Z_2= c_5 \; \text{mod} \; 2= {n_{+,(0,0)} - n_{+,(0,\pi)} + n_{+,(\pi,0)} - n_{+,(\pi,\pi)}}\; \text{mod} \;2 =  {n_{+,(0,0)} + n_{+,(0,\pi)} + n_{+,(\pi,0)} + n_{+,(\pi,\pi)}}\; \text{mod} \; 2= \nonumber \\ & =  {n_{-,(0,0)} + n_{-,(0,\pi)} + n_{-,(\pi,0)} + n_{-,(\pi,\pi)}}\; \text{mod} \; 2
 \end{eqnarray}
If $c_5$ is odd, we have a TI (a Chern insulator without TR and a QSH with TR, where with TR we only count the Kramers pair eigenvalue once). It is clear that this can be generalized further.

 \subsection{Example: Inversion in 3D}
 
 We choose the momentum inversion symmetric vectors in the same order as in Section (\ref{subsec:sp2}) ($\Gamma$, X, Y, V, Z, U, T and R), whose components are:
 \begin{equation}
 \left(
\begin{array}{ccc}
 (0 & 0 & 0) \\
 (\pi & 0 & 0)  \\
 (0 & \pi  & 0) \\
 (\pi & \pi  & 0)  \\
 (0  & 0 & \pi) \\
 (\pi  & 0 & \pi)  \\
 (0  & \pi  & \pi) \\
 (\pi  & \pi  & \pi)  \\
\end{array}
\right)
\end{equation}

We ask: can a set of bands given by
\begin{eqnarray}
& B= ( n_{+,(000)},n_{-,(000)}, n_{+,(\pi00)},n_{-,(\pi00)}, n_{+,(0\pi0)},n_{-,(0\pi0)}, n_{+,(\pi\pi0)},n_{-,(\pi\pi0)}, \nonumber \\ &n_{+,(00\pi)},n_{-,(00\pi)}, n_{+,(\pi0\pi)},n_{-,(\pi0\pi)}, n_{+,(0\pi\pi)},n_{-,(0\pi\pi)}, n_{+,(\pi\pi\pi)},n_{-,(\pi\pi\pi)} )
\end{eqnarray}
be written as a sum of EBRs or not (and then is it topological or a semimetal?).

The EBR matrix is given in Eq. \ref{eq:EBRsSG2}, where each EBR correspond to a column of the matrix. 
 
We perform the Smith decomposition of the matrix \ref{eq:EBRsSG2} to get the Smith Normal Form $\Delta$ in Eq(\ref{schmidtequation}) of the main text:

\begin{equation}
\Delta_{\text{Inversion 3D}}=\left(
\begin{array}{cccccccccccccccc}
 1 & 0 & 0 & 0 & 0 & 0 & 0 & 0 & 0 & 0 & 0 & 0 & 0 & 0 & 0 & 0 \\
 0 & 1 & 0 & 0 & 0 & 0 & 0 & 0 & 0 & 0 & 0 & 0 & 0 & 0 & 0 & 0 \\
 0 & 0 & 1 & 0 & 0 & 0 & 0 & 0 & 0 & 0 & 0 & 0 & 0 & 0 & 0 & 0 \\
 0 & 0 & 0 & 1 & 0 & 0 & 0 & 0 & 0 & 0 & 0 & 0 & 0 & 0 & 0 & 0 \\
 0 & 0 & 0 & 0 & 1 & 0 & 0 & 0 & 0 & 0 & 0 & 0 & 0 & 0 & 0 & 0 \\
 0 & 0 & 0 & 0 & 0 & 2 & 0 & 0 & 0 & 0 & 0 & 0 & 0 & 0 & 0 & 0 \\
 0 & 0 & 0 & 0 & 0 & 0 & 2 & 0 & 0 & 0 & 0 & 0 & 0 & 0 & 0 & 0 \\
 0 & 0 & 0 & 0 & 0 & 0 & 0 & 2 & 0 & 0 & 0 & 0 & 0 & 0 & 0 & 0 \\
 0 & 0 & 0 & 0 & 0 & 0 & 0 & 0 & 4 & 0 & 0 & 0 & 0 & 0 & 0 & 0 \\
 0 & 0 & 0 & 0 & 0 & 0 & 0 & 0 & 0 & 0 & 0 & 0 & 0 & 0 & 0 & 0 \\
 0 & 0 & 0 & 0 & 0 & 0 & 0 & 0 & 0 & 0 & 0 & 0 & 0 & 0 & 0 & 0 \\
 0 & 0 & 0 & 0 & 0 & 0 & 0 & 0 & 0 & 0 & 0 & 0 & 0 & 0 & 0 & 0 \\
 0 & 0 & 0 & 0 & 0 & 0 & 0 & 0 & 0 & 0 & 0 & 0 & 0 & 0 & 0 & 0 \\
 0 & 0 & 0 & 0 & 0 & 0 & 0 & 0 & 0 & 0 & 0 & 0 & 0 & 0 & 0 & 0 \\
 0 & 0 & 0 & 0 & 0 & 0 & 0 & 0 & 0 & 0 & 0 & 0 & 0 & 0 & 0 & 0 \\
 0 & 0 & 0 & 0 & 0 & 0 & 0 & 0 & 0 & 0 & 0 & 0 & 0 & 0 & 0 & 0 \\
\end{array}
\right)
 \end{equation}
 
\noindent to obtain the classification $Z_2 \times Z_2 \times Z_2 \times Z_4$. This is the correct classification. Furthermore the topological indices can be computed  as  $c_6 \; \text{mod}  \;2$,  $c_7 \; \text{mod}  \;2$, $c_8 \; \text{mod}  \;2$, $c_9 \; \text{mod}  \;4$ from  $C= L\cdot B$   (because the matrix $L$ in Eq(\ref{schmidtequation}) of the main text can be chosen as, 

\begin{eqnarray}
&c_6 \; \text{mod}  \;2 =  n_{+,(0,\pi ,0)}-n_{+,(0,\pi ,\pi )}-n_{+,(\pi ,0,0)}+n_{+,\{\pi ,0,\pi)} \; \text{mod}  \;2 = \nonumber \\ &= n_{+,(0,\pi ,0)}+n_{+,(0,\pi ,\pi )}+n_{+,(\pi ,0,0)}+n_{+,\{\pi ,0,\pi)} \; \text{mod}  \;2   = \nonumber \\ &= n_{-,(0,\pi ,0)}+n_{-,(0,\pi ,\pi )}+n_{-,(\pi ,0,0)}+n_{-,\{\pi ,0,\pi)} \; \text{mod}  \;2   \nonumber \\ 
&c_7 \; \text{mod}  \;2   = n_{+,(0,0,\pi )}-n_{+,(0,\pi ,\pi )}-n_{+,(\pi ,0,0)}+n_{+,(\pi ,\pi ,0)}  \; \text{mod}  \;2 = \nonumber \\   & = n_{+,(0,0,\pi )}+n_{+,(0,\pi ,\pi )}+n_{+,(\pi ,0,0)}+n_{+,(\pi ,\pi ,0)}  \; \text{mod}  \;2 = \nonumber \\ &=   n_{-,(0,0,\pi )}+n_{-,(0,\pi ,\pi )}+n_{-,(\pi ,0,0)}+n_{-,(\pi ,\pi ,0)}  \; \text{mod}  \;2  \nonumber \\ 
&c_8 \; \text{mod}  \;2  = -n_{+,(0,0,0)}+n_{+,(0,0,\pi)}+n_{+,(0,\pi ,0)}-n_{+,(0,\pi ,\pi)}+2 n_{2,(0,0,0)}  \; \text{mod}  \;2 = \nonumber \\ &= n_{+,(0,0,0)}+n_{+,(0,0,\pi)}+n_{+,(0,\pi ,0)}+n_{+,(0,\pi ,\pi)}  \; \text{mod}  \;2 = \nonumber \\ &=   n_{-,(0,0,0)}+n_{-,(0,0,\pi)}+n_{-,(0,\pi ,0)}+n_{-,(0,\pi ,\pi)}  \; \text{mod}  \;2 \nonumber \\ 
& c_9 \; \text{mod}  \;4 = -n_{+,(0,0,0)}+n_{+,(0,0,\pi )}+n_{+,(0,\pi ,0)}-n_{+,(0,\pi ,\pi )}+\nonumber \\ & + n_{+,(\pi ,0,0)}-n_{+,(\pi ,0,\pi )}-n_{+,(\pi ,\pi ,0)}+n_{+,(\pi ,\pi ,\pi )}+4 n_{-,(0,0,0)}   \; \text{mod}  \;4  \nonumber \\ & = n_{-,(0,0,0)}+n_{+,(0,0,\pi )}+n_{+,(0,\pi ,0)}+n_{-,(0,\pi ,\pi )}+\nonumber \\ & + n_{+,(\pi ,0,0)}+n_{-,(\pi ,0,\pi )}+n_{-,(\pi ,\pi ,0)}+n_{+,(\pi ,\pi ,\pi )}   \; \text{mod}  \;4 
\end{eqnarray}

\section{Examples of determination of the fragile root bands through the Smith Decomposition}
\label{fragileroots:Smith}
In this section we apply the Smith decomposition in the determination of the fragile root bands following the general procedure explained in section \ref{fragileroots:main} of the main text. We  start from rank=1 up to rank 3 space groups with TR. Only double-valued bands will be considered; the analysis of the single-valued bands is equivalent.

\subsection{Rank 1 EBR matrix space groups}
There are 70 space groups whose $EBR$ matrix with SOC has rank 1 (see table \ref{tb:ranksdoubles} in the main text). Since they are all rank 1, neither of them supports strong topological phases. As stressed in the main text, this does not mean that no Fragile phases exist. It just means that, if they exist, they cannot be identified by symmetry indices. The groups that allow Berry Fragile Phases and whose EBR matrix has Rank 1 are listed below, as well as their EBR matrices:

\[\textbf{SG }  21:  EBR_{21}= \left(
\begin{array}{ccccc}
1 & 1 & 1 & 1 & 2 \\
1 & 1 & 1 & 1 & 2 \\
1 & 1 & 1 & 1 & 2 \\
1 & 1 & 1 & 1 & 2 \\
1 & 1 & 1 & 1 & 2 \\
1 & 1 & 1 & 1 & 2 \\
\end{array}
\right),\;\;\; \textbf{SG }  35:  EBR_{35}=  \left(
\begin{array}{ccc}
1 & 1 & 2 \\
1 & 1 & 2 \\
1 & 1 & 2 \\
1 & 1 & 2 \\
1 & 1 & 2 \\
1 & 1 & 2 \\
\end{array}
\right), \;\;\; \textbf{SG }  42:  EBR=_{42} 
\left(
\begin{array}{cc}
1 & 2 \\
1 & 2 \\
1 & 2 \\
1 & 2 \\
1 & 2 \\
\end{array}
\right)\]

\[ \textbf{SG }  94:  EBR_{94}=  \left(
\begin{array}{ccc}
1 & 1 & 2 \\
1 & 1 & 2 \\
1 & 1 & 2 \\
1 & 1 & 2 \\
1 & 1 & 2 \\
1 & 1 & 2 \\
1 & 1 & 2 \\
1 & 1 & 2 \\
1 & 1 & 2 \\
1 & 1 & 2 \\
\end{array}
\right),\;\;\; \textbf{SG }    98 :  EBR_{98}= \left(
\begin{array}{ccc}
1 & 1 & 2 \\
1 & 1 & 2 \\
1 & 1 & 2 \\
1 & 1 & 2 \\
1 & 1 & 2 \\
1 & 1 & 2 \\
1 & 1 & 2 \\
2 & 2 & 4 \\
2 & 2 & 4 \\
\end{array}
\right), \;\;\;  \textbf{SG }  101 :  EBR_{101}= \left(
\begin{array}{ccc}
1 & 1 & 2 \\
1 & 1 & 2 \\
1 & 1 & 2 \\
1 & 1 & 2 \\
1 & 1 & 2 \\
1 & 1 & 2 \\
1 & 1 & 2 \\
2 & 2 & 4 \\
\end{array}
\right)\]

\[\textbf{SG }   102:  EBR_{102}=   \left(
\begin{array}{cc}
1 & 2 \\
1 & 2 \\
1 & 2 \\
1 & 2 \\
1 & 2 \\
1 & 2 \\
1 & 2 \\
1 & 2 \\
1 & 2 \\
1 & 2 \\
\end{array}
\right)\]
The EBR matrices have all rank 1, and are composed from more than one EBR. This means that at least one of the EBRs is decomposable in sums of the others. Hence there exist Fragile Phases (decomposable EBRs), but their origin has to come from Berry Phases; we hence call them Berry Fragile Phases and analyze them in a further publication.

\subsection{Rank 2 EBR matrix space groups}

There are 21 space groups whose EBR matrix is of rank 2, as shown in Table \ref{tb:ranksdoubles} of the main text. We have checked that none of the groups of rank $2$ support strong topological phases.  

We find that in some of the groups in the table, where  the EBR matrix has rank $2$, there can be eigenvalue-identified fragile phases. Per group, we identify (a) the EBR matrices, (b) the $B$ vector satisfying the compatibility relations, $B = R_{comp} \cdot Y$ (Eq. (\ref{eq:finalBvec}) of the main text, where here and in the following we omit the $p$ subindex, assuming that the $B$-vectors considered always fulfill the compatibility relations) and (c) the EBR vector $V_{EBR}=R\cdot \Delta^{-1}\cdot L\cdot R_{\text{comp}}\cdot Y$ which expresses any vector $B$ in terms of $EBR$s. The conditions for the existence of a fragile state are,

\begin{equation}
\mathbf{Condition: } B \ge 0, \; \text{At least one component of } V_{EBR} \text{ has to be} <0, \text{i.e., there is no way to write } B \text{ as } V_{EBR} \ge 0.
\end{equation}

We now proceed group by group to identify the number of fragile states. 

\[ \textbf{SG }79 : \text{EBR}_{79}= \left(
\begin{array}{ccc}
0 & 1 & 1 \\
1 & 0 & 1 \\
0 & 1 & 1 \\
1 & 0 & 1 \\
1 & 1 & 2 \\
1 & 1 & 2 \\
1 & 1 & 2 \\
\end{array}
\right),\text{B=} \left(
\begin{array}{c}
y_2-y_1 \\
y_1 \\
y_2-y_1 \\
y_1 \\
y_2 \\
y_2 \\
y_2 \\
\end{array}
\right),V_{\text{EBR}_{79}}= \left(
\begin{array}{c}
y_1 \\
y_2-y_1 \\
0 \\
\end{array}
\right)\]
We see that $B\ge 0$ implies $V_{\text{EBR}_{79}} \ge 0$ hence no fragile phases.

\[ \textbf{SG }90 : \text{EBR}_{90}= \left(
\begin{array}{cccc}
1 & 1 & 0 & 2 \\
1 & 1 & 2 & 0 \\
1 & 1 & 1 & 1 \\
1 & 1 & 1 & 1 \\
1 & 1 & 0 & 2 \\
1 & 1 & 2 & 0 \\
1 & 1 & 1 & 1 \\
1 & 1 & 1 & 1 \\
1 & 1 & 1 & 1 \\
1 & 1 & 1 & 1 \\
\end{array}
\right),\text{B=} \left(
\begin{array}{c}
2 y_2-y_1 \\
y_1 \\
y_2 \\
y_2 \\
2 y_2-y_1 \\
y_1 \\
y_2 \\
y_2 \\
y_2 \\
y_2 \\
\end{array}
\right),V_{\text{EBR}_{90}}= \left(
\begin{array}{c}
2 y_2-y_1 \\
0 \\
y_1-y_2 \\
0 \\
\end{array}
\right)\]
We prove that, given $B\ge 0$, we can always write any set of bands as combinations of EBRs. $B\ge 0$ implies $2 y_2\ge y_1 \ge 0$, and hence the first component of the $V_{\text{EBR}_{90}}$ is always non-negative. Therefore, the only possibility to have a fragile phase is for the third component of  $V_{\text{EBR}_{90}}$ to be negative ,  $y_1 - y_2 =- k, \; \; k>0$. The vector $B$ can be expressed as:
\[ B = (2y_2-y_1) ebr_1 + (y_1-y_2) ebr_3 = (k+ y_2) \cdot ebr_1- k\cdot ebr_3\]

Since  $B\ge 0$ implies $2 y_2\ge y_1 \ge 0$ this further implies $y_2 \ge k$. This in turn means $k+y_2\ge 2k$. We let $k+y_2 = 2k + \delta$, $\delta\ge 0$. We hence can express the vector $B$:
\[ B = \delta \cdot ebr_1+ k\cdot (2 \cdot ebr_1-ebr_3). \] 

From the EBR$_{90}$ matrix, we see that $ebr_1= ebr_2$ and $ebr_3+ ebr_4= 2 ebr_1$. Hence we can re-write $B$ as a linear combination of EBRs with positive coefficients $B = \delta \cdot ebr_1+ k\cdot ebr_4 $. In this space group an Eigenvalue Fragile Phase cannot exist.

\[ \textbf{SG }97 : \text{EBR}_{97}= \left(
\begin{array}{cccccc}
1 & 0 & 1 & 0 & 1 & 1 \\
0 & 1 & 0 & 1 & 1 & 1 \\
1 & 0 & 1 & 0 & 1 & 1 \\
0 & 1 & 0 & 1 & 1 & 1 \\
1 & 1 & 1 & 1 & 2 & 2 \\
1 & 1 & 1 & 1 & 2 & 2 \\
1 & 1 & 1 & 1 & 2 & 2 \\
\end{array}
\right),\text{B=} \left(
\begin{array}{c}
y_2-y_1 \\
y_1 \\
y_2-y_1 \\
y_1 \\
y_2 \\
y_2 \\
y_2 \\
\end{array}
\right),V_{\text{EBR}_{97}}= \left(
\begin{array}{c}
y_2-y_1 \\
y_1 \\
0 \\
0 \\
0 \\
0 \\
\end{array}
\right)\]
We see that $B\ge 0$ implies $V_{\text{EBR}_{97}} \ge 0$ hence there are no fragile phases.

\[ \textbf{SG }100 : \text{EBR}_{100}= \left(
\begin{array}{ccc}
0 & 2 & 1 \\
2 & 0 & 1 \\
1 & 1 & 1 \\
1 & 1 & 1 \\
0 & 2 & 1 \\
2 & 0 & 1 \\
1 & 1 & 1 \\
1 & 1 & 1 \\
1 & 1 & 1 \\
1 & 1 & 1 \\
\end{array}
\right),\text{B=} \left(
\begin{array}{c}
2 y_2-y_1 \\
y_1 \\
y_2 \\
y_2 \\
2 y_2-y_1 \\
y_1 \\
y_2 \\
y_2 \\
y_2 \\
y_2 \\
\end{array}
\right),V_{\text{EBR}_{100}}= \left(
\begin{array}{c}
y_1-y_2 \\
0 \\
2 y_2-y_1 \\
\end{array}
\right)\]

The set of different EBRs is exactly the same as in SG 90 and then the same result is obtained: no Eigenvalue Fragile Phase exists.

\[ \textbf{SG }104 : \text{EBR}_{104}= \left(
\begin{array}{ccc}
0 & 2 & 2 \\
2 & 0 & 2 \\
1 & 1 & 2 \\
1 & 1 & 2 \\
0 & 1 & 1 \\
1 & 0 & 1 \\
1 & 1 & 2 \\
1 & 1 & 2 \\
1 & 1 & 2 \\
1 & 1 & 2 \\
\end{array}
\right),\text{B=} \left(
\begin{array}{c}
2 y_2-2 y_1 \\
2 y_1 \\
y_2 \\
y_2 \\
y_2-y_1 \\
y_1 \\
y_2 \\
y_2 \\
y_2 \\
y_2 \\
\end{array}
\right),V_{\text{EBR}_{104}}= \left(
\begin{array}{c}
y_1 \\
y_2-y_1 \\
0 \\
\end{array}
\right)\]
We see that $B\ge 0$ implies $V_{\text{EBR}_{104}} \ge 0$ hence no fragile phases exist.

\[ \textbf{SG }107 : \text{EBR}_{107}= \left(
\begin{array}{ccc}
1 & 0 & 1 \\
0 & 1 & 1 \\
1 & 0 & 1 \\
0 & 1 & 1 \\
1 & 1 & 2 \\
1 & 1 & 2 \\
1 & 1 & 2 \\
\end{array}
\right),\text{B=} \left(
\begin{array}{c}
y_2-y_1 \\
y_1 \\
y_2-y_1 \\
y_1 \\
y_2 \\
y_2 \\
y_2 \\
\end{array}
\right),V_{\text{EBR}_{107}}= \left(
\begin{array}{c}
y_2-y_1 \\
y_1 \\
0 \\
\end{array}
\right)\]
We see that $B\ge 0$ implies $V_{\text{EBR}_{107}} \ge 0$ hence no fragile phases.

\[ \textbf{SG }108 : \text{EBR}_{108}= \left(
\begin{array}{ccc}
0 & 2 & 1 \\
2 & 0 & 1 \\
0 & 2 & 1 \\
2 & 0 & 1 \\
1 & 1 & 1 \\
2 & 2 & 2 \\
1 & 1 & 1 \\
1 & 1 & 1 \\
\end{array}
\right),\text{B=} \left(
\begin{array}{c}
2 y_2-y_1 \\
y_1 \\
2 y_2-y_1 \\
y_1 \\
y_2 \\
2 y_2 \\
y_2 \\
y_2 \\
\end{array}
\right),V_{\text{EBR}_{108}}= \left(
\begin{array}{c}
y_1-y_2 \\
0 \\
2 y_2-y_1 \\
\end{array}
\right)\]
The proof that fragile phases do not exist in SG 108 is identical to the proof in SG 100.

\[ \textbf{SG }146 \textrm{ and } \textbf{SG }160: \text{EBR}_{146}=\text{EBR}_{160}= \left(
\begin{array}{cc}
1 & 0 \\
0 & 1 \\
1 & 0 \\
0 & 1 \\
1 & 1 \\
1 & 1 \\
\end{array}
\right),\text{B=} \left(
\begin{array}{c}
y_2-y_1 \\
y_1 \\
y_2-y_1 \\
y_1 \\
y_2 \\
y_2 \\
\end{array}
\right),V_{\text{EBR}_{146}}=V_{\text{EBR}_{160}}= \left(
\begin{array}{c}
y_2-y_1 \\
y_1 \\
\end{array}
\right)\]
We see that $B\ge 0$ implies $V_{\text{EBR}_{146}}=V_{\text{EBR}_{160}} \ge 0$ hence no fragile phases.

\[ \textbf{SG }155 : \text{EBR}_{155}= \left(
\begin{array}{cccc}
0 & 1 & 0 & 1 \\
1 & 0 & 1 & 0 \\
0 & 1 & 0 & 1 \\
1 & 0 & 1 & 0 \\
1 & 1 & 1 & 1 \\
1 & 1 & 1 & 1 \\
\end{array}
\right),\text{B=} \left(
\begin{array}{c}
y_2-y_1 \\
y_1 \\
y_2-y_1 \\
y_1 \\
y_2 \\
y_2 \\
\end{array}
\right),V_{\text{EBR}_{155}}= \left(
\begin{array}{c}
y_1 \\
y_2-y_1 \\
0 \\
0 \\
\end{array}
\right)\]
We see that $B\ge 0$ implies $V_{\text{EBR}_{155}} \ge 0$ hence no fragile phases.

\[ \textbf{SG }161 : \text{EBR}_{161}= \left(
\begin{array}{cc}
2 & 0 \\
0 & 2 \\
1 & 0 \\
1 & 0 \\
0 & 1 \\
2 & 2 \\
1 & 1 \\
1 & 1 \\
\end{array}
\right),\text{B=} \left(
\begin{array}{c}
2 y_2-2 y_1 \\
2 y_1 \\
y_2-y_1 \\
y_2-y_1 \\
y_1 \\
2 y_2 \\
y_2 \\
y_2 \\
\end{array}
\right),V_{\text{EBR}_{161}}= \left(
\begin{array}{c}
y_2-y_1 \\
y_1 \\
\end{array}
\right)\]
We see that $B\ge 0$ implies $V_{\text{EBR}_{161}} \ge 0$ hence no fragile phases.
\[ \textbf{SG }195 : \text{EBR}_{195}= \left(
\begin{array}{cccccc}
1 & 0 & 1 & 0 & 1 & 1 \\
0 & 1 & 0 & 1 & 1 & 1 \\
1 & 0 & 1 & 0 & 1 & 1 \\
0 & 1 & 0 & 1 & 1 & 1 \\
1 & 2 & 1 & 2 & 3 & 3 \\
1 & 2 & 1 & 2 & 3 & 3 \\
\end{array}
\right),\text{B=} \left(
\begin{array}{c}
y_2-2 y_1 \\
y_1 \\
y_2-2 y_1 \\
y_1 \\
y_2 \\
y_2 \\
\end{array}
\right),V_{\text{EBR}_{195}}= \left(
\begin{array}{c}
y_2-2 y_1 \\
y_1 \\
0 \\
0 \\
0 \\
0 \\
\end{array}
\right)\]
We see that $B\ge 0$ implies $V_{\text{EBR}_{195}} \ge 0$ hence no fragile phases.
\[ \textbf{SG }196 : \text{EBR}_{196}= \left(
\begin{array}{cccccccc}
1 & 0 & 1 & 0 & 1 & 0 & 1 & 0 \\
0 & 1 & 0 & 1 & 0 & 1 & 0 & 1 \\
1 & 2 & 1 & 2 & 1 & 2 & 1 & 2 \\
0 & 1 & 0 & 1 & 0 & 1 & 0 & 1 \\
1 & 1 & 1 & 1 & 1 & 1 & 1 & 1 \\
\end{array}
\right),\text{B=} \left(
\begin{array}{c}
y_2-y_1 \\
y_1 \\
y_1+y_2 \\
y_1 \\
y_2 \\
\end{array}
\right),V_{\text{EBR}_{196}}= \left(
\begin{array}{c}
y_2-y_1 \\
y_1 \\
0 \\
0 \\
0 \\
0 \\
0 \\
0 \\
\end{array}
\right)\]
We see that $B\ge 0$ implies $V_{\text{EBR}_{196}} \ge 0$ hence no fragile phases.
\[ \textbf{SG }197 : \text{EBR}_{197}= \left(
\begin{array}{ccc}
1 & 0 & 1 \\
0 & 1 & 1 \\
1 & 0 & 1 \\
0 & 1 & 1 \\
1 & 0 & 1 \\
0 & 1 & 1 \\
0 & 1 & 1 \\
1 & 0 & 1 \\
0 & 1 & 1 \\
0 & 1 & 1 \\
1 & 2 & 3 \\
\end{array}
\right),\text{B=} \left(
\begin{array}{c}
y_2-2 y_1 \\
y_1 \\
y_2-2 y_1 \\
y_1 \\
y_2-2 y_1 \\
y_1 \\
y_1 \\
y_2-2 y_1 \\
y_1 \\
y_1 \\
y_2 \\
\end{array}
\right),V_{\text{EBR}_{197}}= \left(
\begin{array}{c}
y_2-2 y_1 \\
y_1 \\
0 \\
\end{array}
\right)\]
We see that $B\ge 0$ implies $V_{\text{EBR}_{197}} \ge 0$ hence no fragile phases.
\[ \textbf{SG }198 : \text{EBR}_{198}= \left(
\begin{array}{cc}
0 & 2 \\
2 & 1 \\
1 & 0 \\
0 & 1 \\
1 & 1 \\
2 & 2 \\
2 & 2 \\
2 & 2 \\
\end{array}
\right),\text{B=} \left(
\begin{array}{c}
2 y_1 \\
2 y_2-y_1 \\
y_2-y_1 \\
y_1 \\
y_2 \\
2 y_2 \\
2 y_2 \\
2 y_2 \\
\end{array}
\right),V_{\text{EBR}_{198}}= \left(
\begin{array}{c}
y_2-y_1 \\
y_1 \\
\end{array}
\right)\]
We see that $B\ge 0$ implies $V_{\text{EBR}_{198}} \ge 0$ hence no fragile phases.
\[ \textbf{SG }199 : \text{EBR}_{199}= \left(
\begin{array}{ccc}
0 & 2 & 2 \\
2 & 1 & 2 \\
0 & 2 & 2 \\
2 & 1 & 2 \\
2 & 0 & 1 \\
0 & 1 & 1 \\
0 & 1 & 1 \\
2 & 2 & 3 \\
2 & 0 & 1 \\
0 & 1 & 1 \\
0 & 1 & 1 \\
2 & 2 & 3 \\
4 & 4 & 6 \\
\end{array}
\right),\text{B=} \left(
\begin{array}{c}
2 y_1 \\
y_2-y_1 \\
2 y_1 \\
y_2-y_1 \\
y_2-2 y_1 \\
y_1 \\
y_1 \\
y_2 \\
y_2-2 y_1 \\
y_1 \\
y_1 \\
y_2 \\
2 y_2 \\
\end{array}
\right),V_{\text{EBR}_{199}}= \left(
\begin{array}{c}
0 \\
3 y_1-y_2 \\
y_2-2 y_1 \\
\end{array}
\right)\]
This is an instructive example. Since $B\ge 0$, we have that $y_2 \ge 2 y_1 \ge 0$. Hence the coefficient of $ebr_3$ is always positive. In order to have a fragile phase, we must have 
\[
3 y_1- y_2 = -k, \;\; k\in \mathbb{Z}, \;\; k>0
\] With this, we can re-write $B$ as:
\[ B = y_1 \cdot ebr_3 + k\cdot (ebr_3- ebr_2)\] The EBR linear dependence is
\[
2 (ebr_3-ebr_2)= ebr_1 
\]  Hence if $k$ is even, we have $B= y_1 \cdot ebr_3+ k/2 \cdot ebr_1$. Hence, for a fragile phase, we must have $k$ odd, or $3 y_1- y_2 <0, \;\; (3 y_1- y_2 )\mod 2= 1$. This is just one of the conditions. We can further write $ B =( y_1+k)  \cdot ebr_3 - k\cdot ebr_2$. Hence if $k+y_1$ is even, then, using the linear dependence, we have:  $ B =( y_1+k)  \cdot ebr_2 - k\cdot ebr_2+ ( y_1+k)/2 \cdot ebr_1 = y_1  \cdot ebr_2 + ( y_1+k)/2 \cdot ebr_1$ and hence no fragile phase. When $k+y_1$ odd (or $y_1$ even, since we already found $k$ must be odd, we can re-write $B = (k+y_1-1)\cdot  ebr_2 - k \cdot ebr_2  + (y_1+k-1)/2\cdot ebr_1 + ebr_3= (y_1-1)\cdot  ebr_2  + (y_1+k-1)/2\cdot ebr_1 + ebr_3$. Hence for a fragile phase we must have $y_1=0$. Hence the topological indices of a fragile phase are:

\[\boxed{y_1=0,\;\; y_2 \mod 2=1}\]
The root of the fragile state is 
\[\boxed{V_{EBR_{199}}^{\text{fragile, root}} = (0,-1,1)}\]
We get the same result as in section \ref{bands199fragile} using the induction procedure.

All fragile states in the system can be written as
\[\boxed{k\cdot V_{EBR_{199}}^{\text{fragile, root}} =k\cdot  (0,-1,1)}, \;\;\;\;  k \;\; \textbf{odd} \]

\[ \textbf{SG }208 : \text{EBR}_{208}= \left(
\begin{array}{ccccccccc}
1 & 0 & 1 & 0 & 1 & 0 & 1 & 1 & 1 \\
1 & 0 & 1 & 0 & 1 & 0 & 1 & 1 & 1 \\
0 & 2 & 1 & 2 & 1 & 2 & 2 & 2 & 2 \\
1 & 0 & 1 & 0 & 1 & 0 & 1 & 1 & 1 \\
1 & 0 & 1 & 0 & 1 & 0 & 1 & 1 & 1 \\
0 & 2 & 1 & 2 & 1 & 2 & 2 & 2 & 2 \\
1 & 2 & 2 & 2 & 2 & 2 & 3 & 3 & 3 \\
1 & 2 & 2 & 2 & 2 & 2 & 3 & 3 & 3 \\
1 & 2 & 2 & 2 & 2 & 2 & 3 & 3 & 3 \\
1 & 2 & 2 & 2 & 2 & 2 & 3 & 3 & 3 \\
\end{array}
\right),\text{B=} \left(
\begin{array}{c}
y_2-y_1 \\
y_2-y_1 \\
y_1 \\
y_2-y_1 \\
y_2-y_1 \\
y_1 \\
y_2 \\
y_2 \\
y_2 \\
y_2 \\
\end{array}
\right),V_{\text{EBR}_{208}}= \left(
\begin{array}{c}
y_2-2 y_1 \\
0 \\
0 \\
0 \\
y_1 \\
0 \\
0 \\
0 \\
0 \\
\end{array}
\right)\]
We will now prove the existence of a fragile state in SG $208$, find its root, as well as its symmetry index. $B\ge 0$ implies $y_2\ge y_1 \ge 0$. For the fragile state to exist while $B\ge 0$, we need $y_2- 2y_1= -k, \;\; k> 0, \; k\in \mathbb{Z}$. From this we have $2 y_1 -k\ge y_1$ and hence $y_1 \ge k$. We then re-write $B$:
\[
B= - k\cdot ebr_1 + \frac{y_2+k}{2} \cdot ebr_5
\] From the EBR$_{208}$ matrix we have the following set of restrictions:
\[ebr_2 = ebr_4= ebr_6; \;\; ebr_3= ebr_5; \;\; ebr_7= ebr_8= ebr_9;\;\;  ebr_1+ ebr_2= ebr_7;\;\; ebr_2 + 2 ebr_3 = 2 ebr_7
\] Since $y_2\ge y_1 \ge k$ (necessary condition for a fragile phase to exist), we write $y_2= k+ \delta$, $\delta \ge 0$, $\delta \in \mathbb{Z}$. The $B$ vector becomes:
\[
B= - k\cdot ebr_1 + (k+ \delta/2)\cdot ebr_3 = k\cdot ( ebr_3 - ebr_1) + \frac{\delta}{2}\cdot ebr_3
\] In order to find the possible Fragile phases, we have to find if we can express $ebr_3- ebr_1$ in terms of a sum of EBRs. From the EBR linear dependency relations,  we see that 
\[ -ebr_1+ ebr_3= ebr_7- ebr_3\] And we can hence re-write the vector $B$ as:
\[B= k \cdot ebr_7 + (\frac{\delta}{2}- k)\cdot ebr_3 = k\cdot (ebr_7- ebr_3) + \frac{\delta}{2}ebr_3
\] In order for the phase to be fragile, we see that we need $\delta/2 <k$ (otherwise we can write it as sums of positive EBRs). This equation leads to $ 4 y_1+ k> 3 y_2 \ge 3 y_1 \ge 3k$. We also see that $2 (ebr_7- ebr_3) = ebr_2$ and hence only $k$ odd can have fragile phases. We now call $\delta/2- k= - r,\;\; r>0$ for a fragile phase to exist and re-write $B$ as:
\[B = (\frac{\delta}{2} +r)\cdot ebr_7- r \cdot  ebr_3\] Since $2\cdot ebr_7- 2 \cdot ebr_3 = ebr_2$ and $2\cdot ebr_7-   ebr_3 = ebr_2+ ebr_3$, we see that for a fragile phase we must have $\delta=0$; any other $\delta$, irrespective of $r$, would allow us to re-express $B$ in terms of sums of EBRs. Hence for a fragile phase to exist the index is

\[\boxed{y_2=y_1=k = 1 \mod2}\]

The fragile root vector is:

\[\boxed{V_{EBR_{208}}^{\text{fragile, root}} = (-1,0,0,0,1,0,0,0,0)}\]

All fragile states in the system can be written as
\[\boxed{k\cdot V_{EBR_{208}}^{\text{fragile, root}} =k\cdot  (-1,0,0,0,1,0,0,0,0)}, \;\;\;\;  k \;\; \textbf{odd} \]

\[ \textbf{SG }210 : \text{EBR}_{210}= \left(
\begin{array}{cccccccc}
1 & 0 & 1 & 0 & 1 & 0 & 1 & 0 \\
1 & 0 & 1 & 0 & 1 & 0 & 1 & 0 \\
0 & 2 & 0 & 2 & 1 & 2 & 1 & 2 \\
1 & 2 & 1 & 2 & 2 & 2 & 2 & 2 \\
1 & 2 & 1 & 2 & 2 & 2 & 2 & 2 \\
0 & 2 & 0 & 2 & 1 & 2 & 1 & 2 \\
2 & 2 & 2 & 2 & 3 & 2 & 3 & 2 \\
1 & 2 & 1 & 2 & 2 & 2 & 2 & 2 \\
1 & 2 & 1 & 2 & 2 & 2 & 2 & 2 \\
1 & 2 & 1 & 2 & 2 & 2 & 2 & 2 \\
\end{array}
\right),\text{B=} \left(
\begin{array}{c}
y_1-y_2 \\
y_1-y_2 \\
2 y_2-y_1 \\
y_2 \\
y_2 \\
2 y_2-y_1 \\
y_1 \\
y_2 \\
y_2 \\
y_2 \\
\end{array}
\right),V_{\text{EBR}_{210}}= \left(
\begin{array}{c}
2 y_1-3 y_2 \\
0 \\
0 \\
0 \\
0 \\
0 \\
2 y_2-y_1 \\
0 \\
\end{array}
\right)\]

$B\ge 0$ implies $2 y_2 \ge y_1 \ge y_2 \ge 0$. The only way that the phase can be fragile is for the first component of the $V_{\text{EBR}_{210}}$ to be negative. We substitute $2 y_1-3 y_2 = -k, \;\; k >0$  since $y_1\ge y_2$, we have that the two inequalities give $y_2 \ge k$. We can hence re-write $B$ as:
\[ 
B= - k \cdot ebr_1 + \frac{y_2+ k}{2} \cdot ebr_7
\] since $y_2\ge k$, we let $y_2 = k+ \delta$ where $\delta \ge 0$, $\delta \in \mathbb{Z}$. From the EBR$_{210}$ matrix we also have the following linear dependencies (we write all of them even though we only use some of them):
\[ ebr_1= ebr_3,\;\; ebr_2= ebr_4= ebr_6= ebr_8,\;\; ebr_5= ebr_7,\;\; 2\cdot ebr_1 + ebr_2 = 2 \cdot ebr_5\]
Using these relations, we re-write $B$ as:
\[ B= k\cdot (ebr_5- ebr_1) + \frac{\delta}{2} \cdot ebr_5. \] If $k$ is even then we can substitute $ ebr_2 = 2 \cdot (ebr_5- ebr_1)$ and write $B$ as sums of EBRs. So $k$ must be odd for fragile phases. Also, if $\delta>0$ then we can use $(k+1)\cdot ebr_5 - k \cdot ebr_1=  ebr_1 + (k+1)/2 \cdot ebr_2$. Hence the conditions for the fragile phases are $y_2=y_1=k$ and odd to give the index:

\[\boxed{y_2= y_1=1 \mod 2}\] 

The fragile root vector is:
\[\boxed{V_{EBR_{210}}^{\text{fragile, root}} =(-1,0,0,0,0,0,1,0)}\]

All fragile states in the system can be written as
\[\boxed{k\cdot V_{EBR_{210}}^{\text{fragile, root}} =k\cdot  (-1,0,0,0,0,0,1,0)}, \;\;\;\;  k \;\; \textbf{odd} \]

\[ \textbf{SG }212 : \text{EBR}_{212}= \left(
\begin{array}{cccc}
1 & 0 & 1 & 0 \\
1 & 0 & 1 & 0 \\
1 & 2 & 1 & 2 \\
0 & 1 & 0 & 1 \\
1 & 0 & 1 & 0 \\
1 & 1 & 1 & 1 \\
2 & 2 & 2 & 2 \\
1 & 1 & 1 & 1 \\
1 & 1 & 1 & 1 \\
2 & 2 & 2 & 2 \\
\end{array}
\right),\text{B=} \left(
\begin{array}{c}
y_1 \\
y_1 \\
2 y_2-y_1 \\
y_2-y_1 \\
y_1 \\
y_2 \\
2 y_2 \\
y_2 \\
y_2 \\
2 y_2 \\
\end{array}
\right),V_{\text{EBR}_{212}}= \left(
\begin{array}{c}
y_1 \\
y_2-y_1 \\
0 \\
0 \\
\end{array}
\right)\]
\noindent We see that $B\ge 0$ implies $V_{\text{EBR}_{212}} \ge 0$ hence no fragile phases.

\[ \textbf{SG }213 : \text{EBR}_{213}= \left(
\begin{array}{cccc}
1 & 0 & 1 & 0 \\
1 & 0 & 1 & 0 \\
1 & 2 & 1 & 2 \\
0 & 1 & 0 & 1 \\
1 & 0 & 1 & 0 \\
1 & 1 & 1 & 1 \\
2 & 2 & 2 & 2 \\
1 & 1 & 1 & 1 \\
1 & 1 & 1 & 1 \\
2 & 2 & 2 & 2 \\
\end{array}
\right),\text{B=} \left(
\begin{array}{c}
y_1 \\
y_1 \\
2 y_2-y_1 \\
y_2-y_1 \\
y_1 \\
y_2 \\
2 y_2 \\
y_2 \\
y_2 \\
2 y_2 \\
\end{array}
\right),V_{\text{EBR}_{213}}= \left(
\begin{array}{c}
y_1 \\
y_2-y_1 \\
0 \\
0 \\
\end{array}
\right)\]
We see that $B\ge 0$ implies $V_{\text{EBR}_{213}} \ge 0$ hence no fragile phases.

\[ \textbf{SG }214 : \text{EBR}_{214}= \left(
\begin{array}{cccccc}
1 & 0 & 1 & 0 & 1 & 1 \\
1 & 0 & 1 & 0 & 1 & 1 \\
1 & 2 & 1 & 2 & 2 & 2 \\
1 & 0 & 1 & 0 & 1 & 1 \\
1 & 0 & 1 & 0 & 1 & 1 \\
1 & 2 & 1 & 2 & 2 & 2 \\
0 & 2 & 0 & 2 & 1 & 1 \\
1 & 0 & 1 & 0 & 1 & 1 \\
1 & 0 & 1 & 0 & 1 & 1 \\
2 & 2 & 2 & 2 & 3 & 3 \\
4 & 4 & 4 & 4 & 6 & 6 \\
\end{array}
\right),\text{B=} \left(
\begin{array}{c}
y_1 \\
y_1 \\
y_2-y_1 \\
y_1 \\
y_1 \\
y_2-y_1 \\
y_2-2 y_1 \\
y_1 \\
y_1 \\
y_2 \\
2 y_2 \\
\end{array}
\right),V_{\text{EBR}_{214}}= \left(
\begin{array}{c}
3 y_1-y_2 \\
0 \\
0 \\
0 \\
0 \\
y_2-2 y_1 \\
\end{array}
\right)\]
The condition $B\ge 0$ gives  $y_2 \ge 2 y_1 \ge y_1\ge 0$. For a  fragile state, the first component of $V_{\text{EBR}_{214}}$ must be negative (the other component is always positive). We let $ 3y_1- y_2 = - k, \; k>0, \; k\in \mathbb{Z}$. We re-write $B$ as:
\[ B= (3y_1- y_2)\cdot ebr_1 + (y_2-2 y_1) \cdot ebr_6 = y_1 \cdot ebr_6 + k\cdot(ebr_6 - ebr_1).\] The linear dependencies in the EBR$_{214}$ matrix are: 
\[ ebr_1= ebr_3,\;\; ebr_2 = ebr_4,\;\; ebr_5 = ebr_6,\;\; 2\cdot(ebr_6-ebr_1) = ebr_2. \] Hence if $k$ is even $B$ can always be written in terms of a sum of ebrs  ($B= y_1\cdot ebr_6 + k \cdot ebr_2$). For a fragile phase this means that $k$ must be odd. Also, if $y_1>0$, it means that the state is not fragile, as $2 \cdot ebr_6 - ebr_1 = ebr_2+ ebr_1$. Hence the index is:
\[ \boxed{ y_2 = 1 \mod 2,\;\;\;  y_1=0}\]

\noindent The fragile root vector is:
\[\boxed{V_{EBR_{214}}^{\text{fragile, root}} =(-1,0,0,0,0,1)}\]

\noindent All fragile states in SG 214 can be written as
\[\boxed{k\cdot V_{EBR_{214}}^{\text{fragile, root}} =k\cdot  (-1,0,0,0,0,0,1)}, \;\;\;\;  k \;\; \textbf{odd} \]
where $k= y_2$.

\subsection{Rank 3 EBR matrix space groups}
There are 36 space groups whose EBR matrix has rank 3, as shown in Table \ref{tb:ranksdoubles} of the main text. Here we find also the first groups which support strong topological phases. As such, we present their Smith Normal Form $\Delta$ which also tells us what the topological classification of the state is.

\[ \textbf{SG }48 : \text{EBR}_{48}= \left(
\begin{array}{cccccccc}
1 & 1 & 1 & 1 & 0 & 4 & 0 & 4 \\
1 & 1 & 1 & 1 & 4 & 0 & 4 & 0 \\
1 & 1 & 1 & 1 & 4 & 0 & 0 & 4 \\
1 & 1 & 1 & 1 & 0 & 4 & 4 & 0 \\
1 & 1 & 1 & 1 & 2 & 2 & 2 & 2 \\
1 & 1 & 1 & 1 & 2 & 2 & 2 & 2 \\
1 & 1 & 1 & 1 & 2 & 2 & 2 & 2 \\
1 & 1 & 1 & 1 & 2 & 2 & 2 & 2 \\
1 & 1 & 1 & 1 & 2 & 2 & 2 & 2 \\
1 & 1 & 1 & 1 & 2 & 2 & 2 & 2 \\
\end{array}
\right),\Delta= \left(
\begin{array}{cccccccc}
1 & 0 & 0 & 0 & 0 & 0 & 0 & 0 \\
0 & 2 & 0 & 0 & 0 & 0 & 0 & 0 \\
0 & 0 & 4 & 0 & 0 & 0 & 0 & 0 \\
0 & 0 & 0 & 0 & 0 & 0 & 0 & 0 \\
0 & 0 & 0 & 0 & 0 & 0 & 0 & 0 \\
0 & 0 & 0 & 0 & 0 & 0 & 0 & 0 \\
0 & 0 & 0 & 0 & 0 & 0 & 0 & 0 \\
0 & 0 & 0 & 0 & 0 & 0 & 0 & 0 \\
0 & 0 & 0 & 0 & 0 & 0 & 0 & 0 \\
0 & 0 & 0 & 0 & 0 & 0 & 0 & 0 \\
\end{array}
\right),\text{B=} \left(
\begin{array}{c}
2 y_3-y_2 \\
y_2 \\
2 y_3-y_1 \\
y_1 \\
y_3 \\
y_3 \\
y_3 \\
y_3 \\
y_3 \\
y_3 \\
\end{array}
\right),V_{\text{EBR}_{48}}= \left(
\begin{array}{c}
2 y_3-y_2 \\
0 \\
0 \\
0 \\
\frac{1}{4} \left(y_2-y_1\right) \\
0 \\
\frac{1}{4} \left(y_1+y_2-2 y_3\right) \\
0 \\
\end{array}
\right)\]
The EBR linear dependence relations, which we will use are: 
\[ebr_1= ebr_2 =ebr_3= ebr_4,\;\;\; ebr_5+ ebr_6 = ebr_7+ ebr_8= 4\cdot ebr_1
\]

\noindent $B\ge 0$ implies $2 y_3 \ge  y_2 \ge 0$ and $2 y_3 \ge  y_1 \ge 0$. In order not to have a strong phase, we must have   \[y_2- y_1 = - 4 k_1,  \text{and} \;\;\;  y_1+ y_2 - 2 y_3 = - 4 k_2 \]

\noindent If the coefficient on the left hand side in both equations is not multiple of $4$ the band is strong topological since we would have fractional numbers in $V_{\text{EBR}_{48}}$. This in turn implies that the only fragile phases we can possibly have are:

\[  k_1>0, \;\; k_1 \in \mathbb{Z}\;\;\;  \text{and/or}\;\;  k_2>0, \;\; k_2 \in \mathbb{Z}\]

\noindent We see we can now have 3 possibilities for fragile phases: \textbf{A:} $ k_1>0, \;\; k_2>0$; \textbf{B:} $k_1>0, \;\; k_2<0$ \textbf{C:} $k_1<0, \;\; k_2>0$. We now investigate each of these cases separately.

\textbf{A:} $k_1>0, \;\; k_2>0$. We re-express the vector $B$:
\begin{eqnarray}
&B= (4k_2+ y_1) \cdot ebr_1 - k_1 \cdot ebr_5 - k_2 \cdot ebr_7 = k_2 \cdot ebr_8 + y_1 \cdot ebr_1 - k_1 \cdot ebr_5 =  \nonumber \\ &= k_2 \cdot ebr_8 + y_2 \cdot ebr_1 + k_1 \cdot (4 \cdot ebr_1 -ebr_5) =  k_2 \cdot ebr_8 + y_2 \cdot ebr_1 + k_1 \cdot ebr_6
\end{eqnarray}
Hence it is impossible to have fragile phases in case \textbf{A}.

\textbf{B:} $  k_1\ge 0, \;\; k_2<0$. We now use $y_1= - 4 k_2 + 2 y_3 -y_2$ to give:
\begin{eqnarray}
& B= k_2 \cdot ebr_8 + ( - 4 k_2 + 2 y_3 -y_1)  \cdot ebr_1 + k_1 \cdot ebr_6  = \nonumber \\ &= - k_2 ( 4 \cdot ebr_1 -  ebr_8) + (2 y_3 -y_1)  \cdot ebr_1 + k_1 \cdot ebr_6  = \nonumber \\ &= - k_2 \cdot ebr_7+  (2 y_3 -y_1)  \cdot ebr_1 + k_1 \cdot ebr_6   \end{eqnarray}
we have that $B$ is expressed only in terms of positive EBRs (since $-k_2>0$) and hence it is impossible to have fragile phases in case \textbf{B}.

\textbf{C:} $  k_1<0, \;\; k_2\ge0$. From the expression of $B$ as (see above) 
\[B = k_2 \cdot ebr_8 + y_1 \cdot ebr_1 - k_1 \cdot ebr_5  \]
we have that $B$ is expressed only in terms of positive EBRs (since $-k_1>0$) and hence it is impossible to have fragile phases in case \textbf{C}.

We have hence proved that there are not fragile phases in SG 48.

\[ \textbf{SG }50 \textrm{ and } \textbf{SG }59:\text{EBR}_{50}=\text{EBR}_{59}= \left(
\begin{array}{cccccccc}
1 & 1 & 1 & 1 & 0 & 4 & 0 & 4 \\
1 & 1 & 1 & 1 & 4 & 0 & 4 & 0 \\
1 & 1 & 1 & 1 & 2 & 2 & 2 & 2 \\
1 & 1 & 1 & 1 & 2 & 2 & 2 & 2 \\
1 & 1 & 1 & 1 & 2 & 2 & 2 & 2 \\
1 & 1 & 1 & 1 & 2 & 2 & 2 & 2 \\
1 & 1 & 1 & 1 & 2 & 2 & 2 & 2 \\
1 & 1 & 1 & 1 & 2 & 2 & 2 & 2 \\
1 & 1 & 1 & 1 & 0 & 4 & 4 & 0 \\
1 & 1 & 1 & 1 & 4 & 0 & 0 & 4 \\
\end{array}
\right),\Delta= \left(
\begin{array}{cccccccc}
1 & 0 & 0 & 0 & 0 & 0 & 0 & 0 \\
0 & 2 & 0 & 0 & 0 & 0 & 0 & 0 \\
0 & 0 & 4 & 0 & 0 & 0 & 0 & 0 \\
0 & 0 & 0 & 0 & 0 & 0 & 0 & 0 \\
0 & 0 & 0 & 0 & 0 & 0 & 0 & 0 \\
0 & 0 & 0 & 0 & 0 & 0 & 0 & 0 \\
0 & 0 & 0 & 0 & 0 & 0 & 0 & 0 \\
0 & 0 & 0 & 0 & 0 & 0 & 0 & 0 \\
0 & 0 & 0 & 0 & 0 & 0 & 0 & 0 \\
0 & 0 & 0 & 0 & 0 & 0 & 0 & 0 \\
\end{array}
\right),
\]
\[
\text{B=} \left(
\begin{array}{c}
-y_1+2 y_2+2 y_3 \\
y_1 \\
y_2+y_3 \\
y_2+y_3 \\
y_2+y_3 \\
y_2+y_3 \\
y_2+y_3 \\
y_2+y_3 \\
y_2 \\
y_2+2 y_3 \\
\end{array}
\right),V_{\text{EBR}_{50}}=V_{\text{EBR}_{59}}= \left(
\begin{array}{c}
2 \left(y_2+y_3\right)-y_1 \\
0 \\
0 \\
0 \\
\frac{1}{4} \left(y_1-y_2\right) \\
0 \\
\frac{1}{4} \left(y_1-y_2-2 y_3\right) \\
0 \\
\end{array}
\right)\]
The EBR linear independence conditions are: 
\[ ebr_1= ebr_2= ebr_3=ebr_4,\;\;\; ebr_5+ ebr_6 = ebr_7+ ebr_8= 4 \cdot ebr_1\]

\noindent The condition $B\ge 0$ implies
\[ 2(y_2+ y_3)\ge y_1 \ge 0,\;\; y_2 \ge 0 \;\; y_2+ y_3 \ge 0, \;\; y_2+ 2 y_3 \ge 0. \] In order to not have a strong topological state, we must have:
\[ y_1- y_2 = - 4 k_1,\;\;\; y_1 - y_2 - 2 y_3 = - 4 k_2,\;\;\; k_1, k_2 \in \mathbb{Z}.\]

\noindent This in turn implies that the only fragile phases we can possibly have are:

\[  k_1>0, \;\; k_1 \in \mathbb{Z}\;\;\;  \text{and/or}\;\;  k_2>0, \;\; k_2 \in \mathbb{Z}.\]

\noindent We see we can now have 3 possibilities for fragile phases: \textbf{A:} $ k_1>0, \;\; k_2>0$; \textbf{B:} $k_1>0, \;\; k_2<0$ \textbf{C:} $k_1<0, \;\; k_2>0$. We now investigate each of these cases separately.

\textbf{A:} $k_1>0, \;\; k_2>0$. We re-express the vector $B$:
\begin{eqnarray} & B= - k_2 \cdot ebr_7 - k_1 \cdot ebr_5 + (4k_2+ y_2) \cdot ebr_1  = \nonumber \\ &= k_2 \cdot ebr_8 + y_2 \cdot ebr_1 - k_1 \cdot ebr_5 = \nonumber \\ &=  k_2 \cdot ebr_8 + (y_1+ 4 k_1) \cdot ebr_1 - k_1 \cdot ebr_5 = \nonumber \\ &=   k_2 \cdot ebr_8  +y_1 \cdot ebr_1 +  k_1 \cdot ebr_6  \end{eqnarray}
Hence it is impossible to have fragile phases in case \textbf{A}, since $y_1\ge0$.

\textbf{B:} $  k_1\ge 0, \;\; k_2<0$. We now express B as:
\begin{eqnarray} & B= - k_2 \cdot ebr_7 - k_1 \cdot ebr_5 + (-y_1 + 2 y_2+ 2 y_3) \cdot ebr_1  = \nonumber \\ &=  - k_2 \cdot ebr_7 +  k_1 \cdot( 4 \cdot ebr_1- ebr_5 )+ (y_1 + 2 y_3+ 4 k_1) \cdot ebr_1 = \nonumber \\ &=  - k_2 \cdot ebr_7 +  k_1 ebr_6+ (y_1 + 2 y_3+ 4 k_1) \cdot ebr_1= \nonumber \\ &=  - k_2 \cdot ebr_7 +  k_1 ebr_6+ (y_2 + 2 y_3) \cdot ebr_1 \end{eqnarray}
we have that $B$ is expressed only in terms of positive EBRs (since $-k_2>0$) and hence it is impossible to have fragile phases in case \textbf{B}.

\textbf{C:} $  k_1<0, \;\; k_2\ge0$. From the expression of $B$ as (see above) 
\begin{eqnarray} & B= - k_2 \cdot ebr_7 - k_1 \cdot ebr_5 + (4k_2+ y_2) \cdot ebr_1  = \nonumber \\ &= k_2 \cdot ebr_8 - k_1 \cdot ebr_5 +  y_2 \cdot ebr_1   \end{eqnarray} 
we have that $B$ is expressed only in terms of positive EBRs (since $-k_1>0$) and hence it is impossible to have fragile phases in case \textbf{C}. We have hence proved that there are not fragile phases in SG 50.

\[ \textbf{SG }52 \textrm{ and } \textbf{SG }56 : \text{EBR}_{52}=\text{EBR}_{56}= \left(
\begin{array}{cccccc}
0 & 4 & 0 & 4 & 2 & 2 \\
4 & 0 & 4 & 0 & 2 & 2 \\
2 & 2 & 2 & 2 & 2 & 2 \\
1 & 1 & 1 & 1 & 1 & 1 \\
1 & 1 & 1 & 1 & 1 & 1 \\
0 & 2 & 2 & 0 & 1 & 1 \\
0 & 2 & 2 & 0 & 1 & 1 \\
2 & 0 & 0 & 2 & 1 & 1 \\
2 & 0 & 0 & 2 & 1 & 1 \\
2 & 2 & 2 & 2 & 2 & 2 \\
2 & 2 & 2 & 2 & 2 & 2 \\
2 & 2 & 2 & 2 & 2 & 2 \\
2 & 2 & 2 & 2 & 2 & 2 \\
\end{array}
\right),\Delta= \left(
\begin{array}{cccccc}
1 & 0 & 0 & 0 & 0 & 0 \\
0 & 1 & 0 & 0 & 0 & 0 \\
0 & 0 & 4 & 0 & 0 & 0 \\
0 & 0 & 0 & 0 & 0 & 0 \\
0 & 0 & 0 & 0 & 0 & 0 \\
0 & 0 & 0 & 0 & 0 & 0 \\
0 & 0 & 0 & 0 & 0 & 0 \\
0 & 0 & 0 & 0 & 0 & 0 \\
0 & 0 & 0 & 0 & 0 & 0 \\
0 & 0 & 0 & 0 & 0 & 0 \\
0 & 0 & 0 & 0 & 0 & 0 \\
0 & 0 & 0 & 0 & 0 & 0 \\
0 & 0 & 0 & 0 & 0 & 0 \\
\end{array}
\right),\text{B=} \left(
\begin{array}{c}
4 y_3-y_2 \\
y_2 \\
2 y_3 \\
y_3 \\
y_3 \\
2 y_3-y_1 \\
2 y_3-y_1 \\
y_1 \\
y_1 \\
2 y_3 \\
2 y_3 \\
2 y_3 \\
2 y_3 \\
\end{array}
\right)
\],\\
\[V_{\text{EBR}_{52}}=V_{\text{EBR}_{56}}= \left(
\begin{array}{c}
y_1-y_3 \\
\frac{1}{4} \left(2 y_1-y_2\right) \\
\frac{1}{4} \left(y_2-2 y_1\right) \\
0 \\
0 \\
2 y_3-y_1 \\
\end{array}
\right)\]
These groups lack of fragile topological phases. The $B\ge 0$ equation reads:
\[
2y_3\ge y_1 \ge 0, \;\;\; 4 y_3\ge y_2\ge 0\]
The EBR linear dependencies read:
\[ ebr_5= ebr_6,\;\;\; ebr_1+ ebr_2 = ebr_3+ ebr_4 = 2\cdot ebr_5 \] In order to not have a strong topological state, we must have:
\[ \;\;\; 2y_1 - y_2  = - 4 k_2,\;\;\; k_2 \in \mathbb{Z}\]
We also denote:
\[y_1- y_3 = -  k_1,\;\;\; k_1 \in \mathbb{Z}\]

\noindent This in turn implies that the only fragile phases we can possibly have are:

\[  k_1>0, \;\; k_1 \in \mathbb{Z}\;\;\;  \text{and/or}\;\;  k_2>0, \;\; k_2 \in \mathbb{Z}\]

\noindent We now have 3 possibilities for fragile phases: \textbf{A:} $ k_1>0, \;\; k_2>0$; \textbf{B:} $k_1>0, \;\; k_2<0$; \textbf{C:} $k_1<0, \;\; k_2>0$. We investigate each of these cases separately.

\textbf{A:} $k_1>0, \;\; k_2>0$. We re-express the vector $B$:
\[B=- k_1\cdot ebr_1 - k_2 \cdot ebr_2 + k_2 \cdot ebr_3  + (y_3+ k_1) \cdot ebr_5 \]
In order to use $ebr_3+ ebr_4 = 2\cdot ebr_5$ we must find some inequalities. 
\begin{eqnarray}
& y_1= y_3-k_1\ge 0 \Rightarrow y_3 \ge k_1  \\ & 2y_1= y_2 - 4 k_2 \ge 0 \Rightarrow y_2 \ge 4 k_2 \Rightarrow 4 y_3 \ge y_2 \ge 4 k_2 \Rightarrow  y_3 \ge k_2 \nonumber \\ &  y_1= y_3-k_1, \;  2y_1 - y_2  = - 4 k_2, \; \Rightarrow y_2 = 2 y_3- 2 k_1+ 4 k_2 \ge 0, \Rightarrow y_3 \ge - 2 k_2 +k_1   \nonumber \\ & y_1= y_3-k_1, \;  2y_1 - y_2  = - 4 k_2, \; \Rightarrow 2 y_3 = 2 k_1- 4 k_2+ y_2, \; 4y_3=4 k_1- 8 k_2+ 2 y_2  \ge y_2, \Rightarrow y_2 \ge 8 k_2- 4 k_1, \; y_3\ge 2 k_2-k_1 \nonumber \label{inequalitiesSG52}
\end{eqnarray}

\noindent These 3 inequalities will help us to obtain the fact that no fragile phases exist in this group. 
If $k_1\ge k_2$ we use $y_3= k_1+ \delta, \;\; \delta \ge 0$ to rewrite
\begin{eqnarray}
&B= k_1\cdot( 2 ebr_5- ebr_1) - k_2\cdot ebr_2+ k_2\cdot ebr_3 + \delta\cdot ebr_5 = (k_1- k_2) \cdot ebr_2 + k_2 \cdot ebr_3+ \delta\cdot ebr_5
\end{eqnarray} Hence there are no fragile phases. 
For the case $k_2 \ge k_1$ we use the last of the inequalities in Eq. (\ref{inequalitiesSG52}) to re-write $y_3+k_1= 2 k_2+ \delta, \; \delta \in \mathbb{Z}$. This then gives us the decomposition of $B$
\begin{eqnarray}
&B= -k_1\cdot ebr_1 + k_2\cdot( 2ebr_5- ebr_2) + k_2\cdot ebr_3 + \delta\cdot ebr_5 = (k_2- k_1) \cdot ebr_1 + k_2 \cdot ebr_3+ \delta\cdot ebr_5
\end{eqnarray} 
Hence there are no fragile phases in case \textbf{A}. We now move to the next cases:

\textbf{B:} $  k_1\ge 0, \;\; k_2<0$.  We now express B as
\begin{eqnarray}
&B=- k_1\cdot ebr_1 - k_2 \cdot (ebr_2-  ebr_3)  + (y_3+ k_1) \cdot ebr_5 = \nonumber \\ & = - k_1\cdot ebr_1 - k_2 \cdot (ebr_4-  ebr_1)  + (y_3+k_1) \cdot ebr_5 = \nonumber \\ & = (k_2-k_1)\cdot ebr_1 - k_2\cdot ebr_4 +(y_3+ k_1) \cdot ebr_5  \end{eqnarray}
From the third equation in Eq. (\ref{inequalitiesSG52}) we can write $y_3+ k_1=  - 2 k_2 + 2 k_1 +  \delta, \;\; \delta \ge 0 $, which gives: 
\begin{eqnarray}
&B= (k_2-k_1)\cdot ebr_1 - k_2\cdot ebr_4 +2( k_1-k_2 + \delta/2 ) \cdot ebr_5  = \nonumber \\ &= (k_1- k_2) \cdot(2 ebr_5- ebr_1) - k_2 \cdot ebr_4 + \delta \cdot ebr_5  = \nonumber \\ & = (k_1- k_2) \cdot ebr_2 - k_2 \cdot ebr_4 + \delta \cdot ebr_5 \end{eqnarray}
for which all the coefficients are positive and hence we find no eigenvalue fragile phases in case \textbf{B}.

\textbf{C:} $  k_1< 0, \;\; k_2\ge 0$.

\noindent From the last equation in Eq. (\ref{inequalitiesSG52}) we can write $y_3+ k_1= 2 k_2+ \delta, \;\; \delta \ge 0 $. 

\noindent We now express B as:
\begin{eqnarray}
&B=- k_1\cdot ebr_1 - k_2 \cdot ebr_2+ k_2 \cdot   ebr_3  + (2 k_2+ \delta ) \cdot ebr_5 = \nonumber \\ & =- k_1\cdot ebr_1 + k_2\cdot (2  ebr_5 -  ebr_2) + k_2 \cdot   ebr_3  +  \delta  \cdot ebr_5 = - k_1\cdot ebr_1 + k_2 \cdot ebr_1 + k_2 \cdot   ebr_3  +  \delta  \cdot ebr_5 
\end{eqnarray}
All the coefficients are positive and hence we find no eigenvalue fragile phases in case \textbf{C}. Therefore, there are no fragile phases in SGs 52 and 56.

\[ \textbf{SG }54 : \text{EBR}_{54}= \left(
\begin{array}{ccccccc}
0 & 4 & 0 & 4 & 2 & 2 & 2 \\
4 & 0 & 4 & 0 & 2 & 2 & 2 \\
1 & 1 & 1 & 1 & 1 & 1 & 1 \\
1 & 1 & 1 & 1 & 1 & 1 & 1 \\
2 & 2 & 2 & 2 & 2 & 2 & 2 \\
2 & 2 & 2 & 2 & 2 & 2 & 2 \\
1 & 1 & 1 & 1 & 1 & 1 & 1 \\
1 & 1 & 1 & 1 & 1 & 1 & 1 \\
2 & 2 & 2 & 2 & 2 & 2 & 2 \\
0 & 4 & 4 & 0 & 2 & 2 & 2 \\
4 & 0 & 0 & 4 & 2 & 2 & 2 \\
2 & 2 & 2 & 2 & 2 & 2 & 2 \\
\end{array}
\right),\Delta= \left(
\begin{array}{ccccccc}
1 & 0 & 0 & 0 & 0 & 0 & 0 \\
0 & 2 & 0 & 0 & 0 & 0 & 0 \\
0 & 0 & 4 & 0 & 0 & 0 & 0 \\
0 & 0 & 0 & 0 & 0 & 0 & 0 \\
0 & 0 & 0 & 0 & 0 & 0 & 0 \\
0 & 0 & 0 & 0 & 0 & 0 & 0 \\
0 & 0 & 0 & 0 & 0 & 0 & 0 \\
0 & 0 & 0 & 0 & 0 & 0 & 0 \\
0 & 0 & 0 & 0 & 0 & 0 & 0 \\
0 & 0 & 0 & 0 & 0 & 0 & 0 \\
0 & 0 & 0 & 0 & 0 & 0 & 0 \\
0 & 0 & 0 & 0 & 0 & 0 & 0 \\
\end{array}
\right),\text{B=} \left(
\begin{array}{c}
4 y_3-y_1 \\
y_1 \\
y_3 \\
y_3 \\
2 y_3 \\
2 y_3 \\
y_3 \\
y_3 \\
2 y_3 \\
4 y_3-y_2 \\
y_2 \\
2 y_3 \\
\end{array}
\right),V_{\text{EBR}_{54}}= \left(
\begin{array}{c}
\frac{1}{4} \left(y_1+y_2-4 y_3\right) \\
0 \\
\frac{1}{4} \left(y_1-y_2\right) \\
0 \\
0 \\
0 \\
2 y_3-\frac{y_1}{2} \\
\end{array}
\right)\]
If we re-label the EBRs in SG 48 in this way,
\[ ebr_5\rightarrow ebr_3, \;\; ebr_7 \rightarrow ebr_1, \;\;  ebr_1 \rightarrow ebr_5/2, \;\;  y_3 \rightarrow 2 y_3, \;\;  y_1 \rightarrow y_2, \;\; y_2 \rightarrow y_1, \]we obtain the EBRs of SG 54. Hence we find no eigenvalue Fragile phases in SG 54.

\[ \textbf{SG }57 : \text{EBR}_{57}= \left(
\begin{array}{cccccc}
0 & 4 & 0 & 4 & 2 & 2 \\
4 & 0 & 4 & 0 & 2 & 2 \\
1 & 1 & 1 & 1 & 1 & 1 \\
1 & 1 & 1 & 1 & 1 & 1 \\
2 & 2 & 2 & 2 & 2 & 2 \\
1 & 1 & 1 & 1 & 1 & 1 \\
1 & 1 & 1 & 1 & 1 & 1 \\
2 & 2 & 2 & 2 & 2 & 2 \\
0 & 4 & 4 & 0 & 2 & 2 \\
4 & 0 & 0 & 4 & 2 & 2 \\
2 & 2 & 2 & 2 & 2 & 2 \\
2 & 2 & 2 & 2 & 2 & 2 \\
\end{array}
\right),\Delta= \left(
\begin{array}{cccccc}
1 & 0 & 0 & 0 & 0 & 0 \\
0 & 2 & 0 & 0 & 0 & 0 \\
0 & 0 & 4 & 0 & 0 & 0 \\
0 & 0 & 0 & 0 & 0 & 0 \\
0 & 0 & 0 & 0 & 0 & 0 \\
0 & 0 & 0 & 0 & 0 & 0 \\
0 & 0 & 0 & 0 & 0 & 0 \\
0 & 0 & 0 & 0 & 0 & 0 \\
0 & 0 & 0 & 0 & 0 & 0 \\
0 & 0 & 0 & 0 & 0 & 0 \\
0 & 0 & 0 & 0 & 0 & 0 \\
0 & 0 & 0 & 0 & 0 & 0 \\
\end{array}
\right),\text{B=} \left(
\begin{array}{c}
4 y_3-y_2 \\
y_2 \\
y_3 \\
y_3 \\
2 y_3 \\
y_3 \\
y_3 \\
2 y_3 \\
4 y_3-y_1 \\
y_1 \\
2 y_3 \\
2 y_3 \\
\end{array}
\right),V_{\text{EBR}_{57}}= \left(
\begin{array}{c}
\frac{1}{4} \left(y_1+y_2-4 y_3\right) \\
0 \\
\frac{1}{4} \left(y_2-y_1\right) \\
0 \\
0 \\
2 y_3-\frac{y_2}{2} \\
\end{array}
\right)\]
With the substitution $y_1 \leftrightarrow y_2$ the proof of the absence of fragile phases becomes identical to that on SG 54.

\[ \textbf{SG }60 : \text{EBR}_{60}= \left(
\begin{array}{ccccc}
0 & 4 & 0 & 4 & 2 \\
4 & 0 & 4 & 0 & 2 \\
1 & 1 & 1 & 1 & 1 \\
1 & 1 & 1 & 1 & 1 \\
0 & 2 & 2 & 0 & 1 \\
0 & 2 & 2 & 0 & 1 \\
2 & 0 & 0 & 2 & 1 \\
2 & 0 & 0 & 2 & 1 \\
1 & 1 & 1 & 1 & 1 \\
1 & 1 & 1 & 1 & 1 \\
1 & 1 & 1 & 1 & 1 \\
1 & 1 & 1 & 1 & 1 \\
2 & 2 & 2 & 2 & 2 \\
2 & 2 & 2 & 2 & 2 \\
2 & 2 & 2 & 2 & 2 \\
\end{array}
\right),\Delta= \left(
\begin{array}{ccccc}
1 & 0 & 0 & 0 & 0 \\
0 & 1 & 0 & 0 & 0 \\
0 & 0 & 4 & 0 & 0 \\
0 & 0 & 0 & 0 & 0 \\
0 & 0 & 0 & 0 & 0 \\
0 & 0 & 0 & 0 & 0 \\
0 & 0 & 0 & 0 & 0 \\
0 & 0 & 0 & 0 & 0 \\
0 & 0 & 0 & 0 & 0 \\
0 & 0 & 0 & 0 & 0 \\
0 & 0 & 0 & 0 & 0 \\
0 & 0 & 0 & 0 & 0 \\
0 & 0 & 0 & 0 & 0 \\
0 & 0 & 0 & 0 & 0 \\
0 & 0 & 0 & 0 & 0 \\
\end{array}
\right),\text{B=} \left(
\begin{array}{c}
4 y_3-y_1 \\
y_1 \\
y_3 \\
y_3 \\
2 y_3-y_2 \\
2 y_3-y_2 \\
y_2 \\
y_2 \\
y_3 \\
y_3 \\
y_3 \\
y_3 \\
2 y_3 \\
2 y_3 \\
2 y_3 \\
\end{array}
\right),V_{\text{EBR}_{60}}= \left(
\begin{array}{c}
y_2-y_3 \\
\frac{1}{4} \left(2 y_2-y_1\right) \\
\frac{1}{4} \left(y_1-2 y_2\right) \\
0 \\
2 y_3-y_2 \\
\end{array}
\right)\]
With the substitution $y_1 \leftrightarrow y_2$, the proof to show the absence of fragile phases in SG 60 becomes identical to that in SG 52.

\[ \textbf{SG }61 : \text{EBR}_{61}= \left(
\begin{array}{cccc}
0 & 4 & 0 & 4 \\
4 & 0 & 4 & 0 \\
0 & 1 & 1 & 0 \\
0 & 1 & 1 & 0 \\
0 & 1 & 1 & 0 \\
0 & 1 & 1 & 0 \\
1 & 0 & 0 & 1 \\
1 & 0 & 0 & 1 \\
1 & 0 & 0 & 1 \\
1 & 0 & 0 & 1 \\
1 & 1 & 1 & 1 \\
1 & 1 & 1 & 1 \\
1 & 1 & 1 & 1 \\
1 & 1 & 1 & 1 \\
1 & 1 & 1 & 1 \\
1 & 1 & 1 & 1 \\
2 & 2 & 2 & 2 \\
2 & 2 & 2 & 2 \\
2 & 2 & 2 & 2 \\
\end{array}
\right),\Delta= \left(
\begin{array}{cccc}
1 & 0 & 0 & 0 \\
0 & 1 & 0 & 0 \\
0 & 0 & 4 & 0 \\
0 & 0 & 0 & 0 \\
0 & 0 & 0 & 0 \\
0 & 0 & 0 & 0 \\
0 & 0 & 0 & 0 \\
0 & 0 & 0 & 0 \\
0 & 0 & 0 & 0 \\
0 & 0 & 0 & 0 \\
0 & 0 & 0 & 0 \\
0 & 0 & 0 & 0 \\
0 & 0 & 0 & 0 \\
0 & 0 & 0 & 0 \\
0 & 0 & 0 & 0 \\
0 & 0 & 0 & 0 \\
0 & 0 & 0 & 0 \\
0 & 0 & 0 & 0 \\
0 & 0 & 0 & 0 \\
\end{array}
\right),\text{B=} \left(
\begin{array}{c}
4 y_3-y_1 \\
y_1 \\
y_3-y_2 \\
y_3-y_2 \\
y_3-y_2 \\
y_3-y_2 \\
y_2 \\
y_2 \\
y_2 \\
y_2 \\
y_3 \\
y_3 \\
y_3 \\
y_3 \\
y_3 \\
y_3 \\
2 y_3 \\
2 y_3 \\
2 y_3 \\
\end{array}
\right),V_{\text{EBR}_{61}}= \left(
\begin{array}{c}
y_2 \\
y_3-\frac{y_1}{4} \\
\frac{1}{4} \left(y_1-4 y_2\right) \\
0 \\
\end{array}
\right)\]
We have the EBR linear dependencies:
\[ ebr_1+ ebr_2 = ebr_3+ ebr_4.\] From $B\ge 0$ we find:
\[ 4 y_3 \ge y_1 \ge 0, \;\; y_3 \ge y_2 \ge 0 \]  In order for $B$ to have a fragile phase (and not a strong topological state), we need:
\[ y_1 - 4y_2 = - 4 k, \;\; k>0, \; k\in \mathbb{Z}.\] The vector $B$ then becomes:
\[B= y_2\cdot ebr_1 + (y_3- \frac{y_1}{4})\cdot ebr_2 - k\cdot ebr_3\]
We have  $ y_2 = k+ y_1/4$, then $y_3 \ge y_2$ gives $ y_3 - y_1/4 \ge k$. We then write $y_3 - y_1/4 = k+\delta, \; \; \delta \ge 0, \; \delta \in \mathbb{Z}$. Therefore $B$ becomes: 
\[B= k \cdot ebr_1 + k \cdot ebr_2 - k\cdot ebr_3 + \frac{y_1}{4} \cdot ebr_1 + \delta \cdot ebr_2= k \cdot ebr_4+ \frac{y_1}{4} \cdot ebr_1 + \delta \cdot ebr_2\]
All the coefficients in the above decomposition are positive, showing that Eigenvalue Fragile Phases cannot exist in SG 61.

\[ \textbf{SG }62 : \text{EBR}_{62}= \left(
\begin{array}{ccccc}
0 & 4 & 0 & 4 & 2 \\
4 & 0 & 4 & 0 & 2 \\
1 & 1 & 1 & 1 & 1 \\
1 & 1 & 1 & 1 & 1 \\
1 & 1 & 1 & 1 & 1 \\
1 & 1 & 1 & 1 & 1 \\
2 & 2 & 2 & 2 & 2 \\
0 & 2 & 2 & 0 & 1 \\
2 & 0 & 0 & 2 & 1 \\
2 & 2 & 2 & 2 & 2 \\
2 & 2 & 2 & 2 & 2 \\
2 & 2 & 2 & 2 & 2 \\
\end{array}
\right),\Delta= \left(
\begin{array}{ccccc}
1 & 0 & 0 & 0 & 0 \\
0 & 1 & 0 & 0 & 0 \\
0 & 0 & 4 & 0 & 0 \\
0 & 0 & 0 & 0 & 0 \\
0 & 0 & 0 & 0 & 0 \\
0 & 0 & 0 & 0 & 0 \\
0 & 0 & 0 & 0 & 0 \\
0 & 0 & 0 & 0 & 0 \\
0 & 0 & 0 & 0 & 0 \\
0 & 0 & 0 & 0 & 0 \\
0 & 0 & 0 & 0 & 0 \\
0 & 0 & 0 & 0 & 0 \\
\end{array}
\right),\text{B=} \left(
\begin{array}{c}
4 y_3-y_1 \\
y_1 \\
y_3 \\
y_3 \\
y_3 \\
y_3 \\
2 y_3 \\
2 y_3-y_2 \\
y_2 \\
2 y_3 \\
2 y_3 \\
2 y_3 \\
\end{array}
\right),V_{\text{EBR}_{62}}= \left(
\begin{array}{c}
y_2-y_3 \\
\frac{1}{4} \left(2 y_2-y_1\right) \\
\frac{1}{4} \left(y_1-2 y_2\right) \\
0 \\
2 y_3-y_2 \\
\end{array}
\right)\]
With the substitution $y_1 \leftrightarrow y_2$, the proof to show the absence of fragile phases in SG 62 becomes identical to that in SG 52. 

\[ \textbf{SG }68 : \text{EBR}_{68}= \left(
\begin{array}{ccccccc}
1 & 1 & 0 & 4 & 0 & 4 & 2 \\
1 & 1 & 4 & 0 & 4 & 0 & 2 \\
1 & 1 & 2 & 2 & 2 & 2 & 2 \\
1 & 1 & 4 & 0 & 0 & 4 & 2 \\
1 & 1 & 0 & 4 & 4 & 0 & 2 \\
1 & 1 & 2 & 2 & 2 & 2 & 2 \\
1 & 1 & 2 & 2 & 2 & 2 & 2 \\
1 & 1 & 2 & 2 & 2 & 2 & 2 \\
\end{array}
\right),\Delta= \left(
\begin{array}{ccccccc}
1 & 0 & 0 & 0 & 0 & 0 & 0 \\
0 & 2 & 0 & 0 & 0 & 0 & 0 \\
0 & 0 & 4 & 0 & 0 & 0 & 0 \\
0 & 0 & 0 & 0 & 0 & 0 & 0 \\
0 & 0 & 0 & 0 & 0 & 0 & 0 \\
0 & 0 & 0 & 0 & 0 & 0 & 0 \\
0 & 0 & 0 & 0 & 0 & 0 & 0 \\
0 & 0 & 0 & 0 & 0 & 0 & 0 \\
\end{array}
\right),\text{B=} \left(
\begin{array}{c}
2 y_3-y_1 \\
y_1 \\
y_3 \\
2 y_3-y_2 \\
y_2 \\
y_3 \\
y_3 \\
y_3 \\
\end{array}
\right),V_{\text{EBR}_{68}}= \left(
\begin{array}{c}
2 y_3-y_1 \\
0 \\
\frac{1}{4} \left(y_1-y_2\right) \\
0 \\
\frac{1}{4} \left(y_1+y_2-2 y_3\right) \\
0 \\
0 \\
\end{array}
\right)\]
With the substitution $y_1 \leftrightarrow y_2$ the proof of absence of fragile phases in SG 68 is identical to that of SG 48. 

\[ \textbf{SG }70 : \text{EBR}_{70}= \left(
\begin{array}{cccccc}
1 & 1 & 0 & 4 & 0 & 4 \\
1 & 1 & 4 & 0 & 4 & 0 \\
1 & 1 & 2 & 2 & 2 & 2 \\
1 & 1 & 2 & 2 & 2 & 2 \\
1 & 1 & 2 & 2 & 2 & 2 \\
1 & 1 & 3 & 1 & 1 & 3 \\
1 & 1 & 1 & 3 & 3 & 1 \\
\end{array}
\right),\Delta= \left(
\begin{array}{cccccc}
1 & 0 & 0 & 0 & 0 & 0 \\
0 & 1 & 0 & 0 & 0 & 0 \\
0 & 0 & 4 & 0 & 0 & 0 \\
0 & 0 & 0 & 0 & 0 & 0 \\
0 & 0 & 0 & 0 & 0 & 0 \\
0 & 0 & 0 & 0 & 0 & 0 \\
0 & 0 & 0 & 0 & 0 & 0 \\
\end{array}
\right),\text{B=} \left(
\begin{array}{c}
-y_1+2 y_2+2 y_3 \\
y_1 \\
y_2+y_3 \\
y_2+y_3 \\
y_2+y_3 \\
y_2 \\
y_2+2 y_3 \\
\end{array}
\right),V_{\text{EBR}_{70}}= \left(
\begin{array}{c}
2 \left(y_2+y_3\right)-y_1 \\
0 \\
\frac{1}{4} \left(y_1-y_2-3 y_3\right) \\
0 \\
\frac{1}{4} \left(y_1-y_2+y_3\right) \\
0 \\
\end{array}
\right)\]
This is a rather cumbersome case. We show that this group has 10 roots. The EBR linear dependence relations read:
\[ ebr_1= ebr_2, \;\;\; ebr_3+ ebr_4 = ebr_5+ ebr_6 = 4 ebr_1\]
$B\ge 0$ gives the conditions:
\[ 2 y_2+ 2 y_3 \ge y_1\ge 0, \;\; y_2+ 2 y_3 \ge 0, \;\; y_2 \ge 0. \] The vector $B$ can be written as:
\[ B= ( 2(y_2+ y_3)- y_1)\cdot ebr_1 + \frac{1}{4} \left(y_1-y_2-3 y_3\right)\cdot ebr_3+  \frac{1}{4} \left(y_1-y_2+y_3\right)\cdot ebr_5\]

\noindent The fragile (but not strong) topological phases are:
\[ \mathbf{A}: y_1 -y_2 - 3 y_3 = - 4 k_1;\;\; y_1 - y_2+ y_3 = - 4 k_2, \;\; k_1, k_2> 0 \]
\[ \mathbf{B}: y_1 -y_2 - 3 y_3 =  4 k_1;\;\; y_1 - y_2+ y_3 = - 4 k_2, \;\; k_1\ge 0,\;\;  k_2> 0 \]
\[ \mathbf{C}: y_1 -y_2 - 3 y_3 =-  4 k_1;\;\; y_1 - y_2+ y_3 =  4 k_2, \;\; k_1 >0,\;\;  k_2 \ge 0 \]

\noindent We analyze these cases: 

Case \textbf{A}: ($ y_1 -y_2 - 3 y_3 = - 4 k_1;\;\; y_1 - y_2+ y_3 = - 4 k_2, \;\; k_1, k_2> 0$):  $y_3= k_1 - k_2, \;\; \; y_1 - y_2 = -3 k_2- k_1$.
\[ B= ( 2(y_2+ y_3)- y_1) \cdot ebr_1 - k_1 \cdot ebr_3- k_2 \cdot ebr_5= (y_2+ k_2+ 3 k_1)\cdot  ebr_1 - k_1 \cdot ebr_3- k_2 \cdot ebr_5  \]
We have:  
\[ y_1 (\ge 0)= y_2 - 3 k_2- k_1 \;\;   \Rightarrow   y_2 \ge 3 k_2 + k_1 \Rightarrow y_2 = 3 k_2 + k_1 + \delta, \;\; \delta \ge 0, \; \delta \in \mathcal{Z} \] Hence we have:
\[ B=(4k_2+ 4 k_1)\cdot  ebr_1 - k_1 \cdot ebr_3 - k_2 \cdot  ebr_5 + \delta \cdot  ebr_1= k_1\cdot  ebr_4 + k_2 \cdot ebr_6 + \delta \cdot ebr_1 \]
One can write $B$ in terms of positive sums of EBRs and hence there are no fragile phases in case  \textbf{A}.    

Case \textbf{B}: ($ y_1 -y_2 - 3 y_3 =  4 k_1;\;\; y_1 - y_2+ y_3 = - 4 k_2, \;\; k_1\ge 0,\;\;  k_2> 0$):  $y_3=- k_1 - k_2, \;\; \; y_1 - y_2 = -3 k_2+  k_1$.
\[ B= ( 2(y_2+ y_3)- y_1)\cdot ebr_1 + k_1 \cdot ebr_3- k_2 \cdot ebr_5= (  y_2+ k_2-  3 k_1 )\cdot  ebr_1 + k_1 \cdot ebr_3- k_2 \cdot ebr_5  \] We now find all the conditions  imposed by $B>0$

\begin{eqnarray} 
& y_1\ge 0 \Rightarrow  y_2 \ge 3 k_2 - k_1 \nonumber \\ & y_2 + 2 y_3 \ge 0 \Rightarrow  y_2 \ge 2k_2+ 2 k_1 \nonumber \\ & y_2+ y_3 \ge 0 \Rightarrow y_2 \ge  k_1+ k_2 \;\;\; \text{It is redundant, because the equation $y_2 \ge  k_1+ k_2$ gives a tighter limit}\nonumber \\ &2y_2+ 2y_3\ge y_1 \Rightarrow y_2 \ge y_1-y_2- 2 y_3 = 3 k_1 - k_2
\end{eqnarray} We hence find that 
\[ y_2 \ge \max(3 k_2 - k_1, 3k_1- k_2, 2k_2 + 2 k_1)\]

Case \textbf{C}: ($ y_1 -y_2 - 3 y_3 = - 4 k_1;\;\; y_1 - y_2+ y_3 =  4 k_2, \;\; k_1> 0,\;\;  k_2\ge 0$) :  $y_3=k_1 + k_2, \;\; \; y_1 - y_2 = 3 k_2 -  k_1$.
\[ B= ( 2(y_2+ y_3)- y_1) ebr_1 - k_1 \cdot ebr_3+ k_2 \cdot ebr_5= (  y_2-  k_2+  3 k_1 )\cdot  ebr_1 - k_1 \cdot ebr_3+ k_2 \cdot ebr_5  \] We now find all the conditions  imposed by $B>0$ by sending $k_{1,2}\rightarrow - k_{1,2}$ in case \textbf{B}
\[ y_2 \ge \max(-3 k_2 + k_1, -3k_1 + k_2, -2k_2 - 2 k_1) =\max(-3 k_2 + k_1, -3k_1 + k_2, 0)\]

We now delve deeper into cases $\mathbf{B}$, $\mathbf{C}$, and make a relabeling of the bands.  We can neglect the $k_1, k_2$ notation that we have developed above. That notation was primarily used to show case $\mathbf{A}$ has no fragile phase. 

In order to find the roots, we will consider the number of bands $N$
\[ dim(B) = 20(y_2+ y_3)= 20 N;\;\;\; y_2+ y_3=N>0\] 
(note that $y_2 + y_3=0$ is \emph{not} an acceptable value for $N$ as it would mean zero bands). We will take consecutively $N=1, 2,3...$ and find the roots. 
Now we first find the inequalities that arise from $B>0$:
\begin{eqnarray}
&y_2 \ge 0 \Rightarrow y_3\le N\nonumber \\ &
y_2+2 y_3  \ge 0 \Rightarrow y_3\ge - N
\end{eqnarray} hence:

\[ \boxed{ - N\le y_3 \le N,\;\;\; 0 \le y_1 \le 2 N} \]

\noindent We re-write:

\[B= (2N- y_1) ebr_1+\frac{1}{4}(y_1 - N- 2 y_3) ebr_3 + \frac{1}{4} (y_1- N+ 2 y_3) ebr_5\]
Either the coefficient of $ebr_3$ or that of $ebr_5$ but not both have to be negative in order to obtain a fragile state.

The existence of fragile phases requires:

Case \textbf{B}:  $y_1- N- 2y_3 \ge 0, y_1- N+ 2 y_3 <0$. In order not to be able to write the $B$ in terms of $ebr_4$ with positive coefficients using $4ebr_1-ebr_3= ebr_4$, the fragile state also implies $2N-y_1< -4\frac{1}{4}(y_1 - N+ 2 y_3)$ or $ 2y_3 <-N$

Case \textbf{C}:  $y_1- N- 2y_3 <0, y_1- N+ 2 y_3 \ge 0$. In order not to be able to write the $B$ in terms of $ebr_4$ with positive coefficients using $4ebr_1-ebr_3= ebr_4$, the fragile state also implies $2N-y_1< -4\frac{1}{4}(y_1 - N- 2 y_3)$ or $ 2y_3 >N$  

We now have the conditions for the existence of fragile states:

\begin{eqnarray}
	\label{eq:SG70}
& y_2+ y_3= N>0;\;\;\;
- N\le y_3 \le N;\;\;\;\; 0 \le y_1 \le 2 N\nonumber \\ &
y_1 -N- 2y_3 = 0 \;\; \mod 4;\;\;\; y_1- N+ 2 y_3=0 \;\; \mod 4 \nonumber \\ & 
2 y_3 <-N \;\;\; \text{or} \;\;\; 2 y_3>N
\end{eqnarray}
The first line implements the condition $B\ge0$ and the number of bands is nonzero. The second line implements the absence of strong topological phases. The third line implements the fragile conditions. Some of the conditions are redundant. We can re-write the \textbf{necessary and sufficient fragile conditions}:
\begin{eqnarray}
&\boxed{ y_2+ y_3= N>0;\;\;\; 0 \le y_1 \le 2 N;\;\;\;\;   y_1 -N- 2y_3 = 0 \;\; \mod 4} \nonumber \\ & \boxed{- 2N\le 2y_3 <- N;\;\;\; \text{or} \;\;\; N <  2 y_3 \le 2N }
\end{eqnarray} The last row are the fragile conditions. Since $N=y_2+y_3$, we rewrite the fragile conditions:
\begin{eqnarray}
&\boxed{ y_2+ y_3>0;\;\ 0 \le y_1 \le 2 (y_2+ y_3);\;\; y_2\ge 0; \;\; y_2 + 2 y_3 \ge 0; \;\;\;\; y_1 -y_2- 3y_3 = 0 \; \mod 4} \nonumber \\ & \boxed{3y_3+ y_2 <0;\;\;\; \text{or} \;\;\; y_2  <   y_3 } \label{allfragilesg70}
\end{eqnarray} Where the last line represents the fragile conditions (the first line represents the $B>0$ condition)

We now find the roots as well as the decomposition of any fragile state into Roots. There are roots for $N=1,2,3,4$; all fragile states for $N>4$ can be expressed in terms of the roots plus EBRs. We show that the equations have a $y_3 \leftrightarrow -y_3$ symmetry. It is hence enough to solve the fragile phases for $N< 2 y_3 \le 2 N$ (where there should be $5$ roots) and by symmetry we will obtain the fragile phases and roots for the $- 2N\le 2y_3 <- N$ fragile phases.

The roots are (one can easily obtain them analytically by solving the $N=1,2,3,4$ fragile phases):
\begin{eqnarray}
&F_1= 2\cdot ebr_1+ ebr_3- ebr_5\nonumber\\  & F_3= 5\cdot ebr_1+ ebr_3- 2 \cdot ebr_5\nonumber\\  & F_5=  ebr_1+ 2 \cdot ebr_3- ebr_5\nonumber\\  &F_7= 8\cdot ebr_1+ ebr_3- 3 \cdot  ebr_5\nonumber\\  &F_9= 3 \cdot ebr_3- ebr_5
\end{eqnarray} and
\begin{eqnarray}
&F_2= 2\cdot ebr_1+ ebr_5- ebr_3\nonumber\\  & F_4= 5\cdot ebr_1+ ebr_5- 2\cdot ebr_3\nonumber\\  & F_6=  ebr_1+ 2 \cdot ebr_5- ebr_3\nonumber\\  &F_8= 8\cdot ebr_1+ ebr_5- 3 \cdot  ebr_3\nonumber\\  &F_{10}= 3 \cdot ebr_5- ebr_3
\end{eqnarray} 

We now have all fragile phases (Eq. \ref{allfragilesg70}), we have all roots, the only remaining thing is to analytically show they are indeed the roots: i.e. to express every fragile phase as a linear combination of EBRs and roots with positive coefficients. We choose to express the fragile phases $y_2<y_3$ (or alternatively $N< 2 y_3 \le 2 N$) in terms of $F_2, F_4, F_6, F_8, F_{10}$ (Case \textbf{B}).  By simple substitution, we then have the remaining fragile phases (Case \textbf{C}). 

Since $y_1\le 2N$ (see Eq. \ref{eq:SG70}) we write
\[ y_1 = 2N- \alpha, \;\;\; 0 \le \alpha \le 2N \]

\noindent The condition $N< 2 y_3 \le 2N $ implies
\[ 2 y_3 = N+1 + \beta,\;\;\; 0 \le \beta \le N-1;\;\;\;\;\; \beta \mod 2= N \mod 2 +1\] to give

\[B= \alpha \cdot ebr_1-\frac{1}{4}(\alpha+ \beta+ 1)\cdot ebr_3 + \frac{1}{4} (2N+1 + \beta- \alpha)\cdot ebr_5\]

We hence have two cases

Case \textbf{BB}: $N=2 k+1, \;\; \beta= 2 m,\;\; 0\le \alpha \le 4 k+2,\;\;\; 0\le m\le k$

\[B=\alpha\cdot ebr_1 -\frac{1}{4}(\alpha + 2m+ 1)\cdot ebr_3 + \frac{1}{4}(4k + 3- \alpha + 2 m)\cdot ebr_5\]

In order to implement the no strong phases condition, we have two further sub-cases
\begin{itemize}
	
	\item Case \textbf{BBB}: $m= 2r, \;\;\; \alpha = 4 p+3,\;\; 0\le 2 r\le k,\;\;\; 0\le p \le k-1 $ to give
	\[ B=(4 p+3)ebr_1- (p+r+1) ebr_3 +(k-p+r)ebr_5\]
	\begin{itemize}
		
		\item For $p$ even and $r \ge 1+ \frac{p}{2}$, we find the decomposition: 
		\[B= \frac{p}{2} F_8 + F_2+ F_6+ (r- 1- \frac{p}{2}) F_{10} + (k- 2 r)\cdot ebr_5\]
		
		\item For $p$ even and $r \le  \frac{p}{2}$, we find the decomposition: 
		
		\[B= (r-1) F_8 + 2 F_4+ebr_1+  (p- 2 r)\cdot ebr_4 +(k-p-1)\cdot ebr_5\]

		\item For $p$ odd and $r \ge \frac{p+1}{2}$ , we find the decomposition: 
		\[B= \frac{p-1}{2} F_8 + F_2+ F_4+ (r-1- \frac{p-1}{2}) F_{10} + (k- 2 r)\cdot ebr_5\]

		\item For $p$ odd and $r \le  \frac{p}{2}$, we find the decomposition: 
		
		\[B= (r-1) F_8 + 2 F_4+ebr_1+  (p- 2 r)\cdot ebr_4 +(k-p-1)\cdot ebr_5\]
		
		(it is easy to see that $r=0$, the only expression above with negative coefficients, actually has an expression in terms of only positive coefficient roots and EBRs.
		
	\end{itemize}

	\item  Case \textbf{BBC}: $m= 2r+1, \;\;\; \alpha = 4 p+1,\;\; 0\le 2 r+1\le k,\;\;\; 0\le p \le k $ to give
	\[ B=(4 p+1)ebr_1- (p+r+1) ebr_3 +(k+r-p+1)ebr_5\]
	\begin{itemize}
		
		\item For $r+1\ge p$ and  $ 2r+1 \ge p $, we find the decomposition: 
		\[B=F_8 + F_6 + 2 (p - 2) F_2 + (r - p + 1) F_{10} + (k - (2 r + 1)) \cdot ebr_5)\]
		
		(the negative $p=0,1$ are easily checked to have the decomposition:
		$p=0, \;\; B= F_6+ r F_{10}+ (k-(2r+1)) \cdot ebr_5$ and 
		$p=1, \;\; B= F_4+ r F_{10}+ (k-(2r+1)) \cdot ebr_5$)

		\item For  $r+1<p$ and $ 2r+1 \ge p $ , we find the decomposition: 
		
		\[B=(p - (r + 1)) F_8 + F_4  + 2 (2 r + 1 - p) F_2 + (k - (2 r + 1)) \cdot ebr_5 \]

		\item For  $ 2r+1 \le p $ , we find the decomposition:

		\[B= F_4 + r F_8 + (p - (2 r + 1)) \cdot ebr_4 + (k - p) \cdot ebr_5\]
		
	\end{itemize}
	
	This ends the completeness proof for $N$ odd; we are able to express any fragile band in terms of the roots and other EBRs.
	\end{itemize}
	
	Case \textbf{BC}: $N=2 k, \;\; \beta= 2 m+1,\;\; 0\le \alpha \le 4 k,\;\;\; 0\le m\le k-1$
	
	\[B = \alpha \cdot ebr_1 - \frac{1}{4}(\alpha +2m + 2) \cdot ebr_3 + \frac{1}{4} (4 k+ 2m +2 - \alpha) \cdot ebr_5\]
		\begin{itemize}
		
		\item Case \textbf{{BCB}: $m = 2r$ $\alpha = 4 p+2$, $0\le 2r \le k-1$, $ 0\le p \le k-1$  }
		
		\[B = (4p +2) \cdot ebr_1 -(p+r+1) \cdot ebr_3 +  ( k+ r-p) \cdot ebr_5\]
		
		\begin{itemize}
			
			\item For $2r \ge p$, (to which we will add $p$ even) we can immediately make the separation:
			
			\begin{eqnarray}
			&B = (4p +2) \cdot ebr_1 -(p+r+1) \cdot ebr_3 +  (3r+1 -p) \cdot ebr_5  + (k-1-2r) \cdot ebr_5
			\end{eqnarray} The last term is a positive (since $k-1\ge 2r$) sum of $ebr_5$.
			 If $p$ is even then $p = 2l$. Then $2r \ge p$ means $ r\ge l$. Hence we write $r= l+d$, $d\ge 0$
			
			\begin{eqnarray}
			&B = (8l +2) \cdot ebr_1 -(3l + d+1) \cdot ebr_3 +  ( l+ 3 d +1) \cdot ebr_5 + (k-1-2r) \cdot ebr_5 = \nonumber \\ &= l( 8 ebr_1 - 3 ebr_3+ ebr_5) + d(- ebr_3+ 3 ebr_5) + (2 ebr_1 - ebr_3+ ebr_5)  + (k-1-2r) \cdot ebr_5 = \nonumber \\ &= l F_8 + d F_{10} + F_2 +(k-1-2r) \cdot ebr_5 = \nonumber \\ & = \boxed{  \frac{p}{2} F_8 + (r - \frac{p}{2}) F_{10} + F_2 +(k-1-2r) \cdot ebr_5   } 
			\end{eqnarray} We point out that in the second line one can obtain the roots as the functions multiplying $l, d$. 
			
			\item For $2r \ge p$, (to which we will add $p$ odd), $p=2l+1$, the condition $2r \ge 2l+1$ implies $r \ge l+1$, or $r= l+1+ d$, $d>0$, we can immediately make the separation:

			\begin{eqnarray}
			&B = (8l +6) \cdot ebr_1 -(3l + d+3) \cdot ebr_3 +  ( l+ 3 d +3) \cdot ebr_5 + (k-1-2r) \cdot ebr_5 = \nonumber \\ &= l( 8 ebr_1 - 3 ebr_3+ ebr_5) + d(- ebr_3+ 3 ebr_5) + (6 ebr_1 - 3 ebr_3+ 3 ebr_5)  + (k-1-2r) \cdot ebr_5 = \nonumber \\ &= l F_8 + d F_{10} + F_4+ F_6 +(k-1-2r) \cdot ebr_5 = \nonumber \\ & = \boxed{  \frac{p-1}{2} F_8 + (r - \frac{p+1}{2}) F_{10} + F_4+ F_6 +(k-1-2r) \cdot ebr_5   } 
			\end{eqnarray} 
			
			\item For $p\ge 2 r$ we have: $p= 2r+ d$, $d \ge 0$; Since $k-1\ge p$ we rewrite:
			\begin{eqnarray}
			&B = (4p +2) \cdot ebr_1 -(p+r+1) \cdot ebr_3 +  (r+1) \cdot ebr_5  + (k-1-p) \cdot ebr_5
			\end{eqnarray} 
			We keep the last term, which is positive, but replace $p= 2r+ d $, $d \ge 0$ in the other terms
			
			\begin{eqnarray}
			&B = (8r +4 d+ 2) \cdot ebr_1 -(3r + d+3) \cdot ebr_3 +  (r+1) \cdot ebr_5 + (k-1-p) \cdot ebr_5 = \nonumber \\ &= r( 8 ebr_1 - 3 ebr_3+ ebr_5) + d(4 ebr_1 - ebr_3) + (2 ebr_1 -  ebr_3+  ebr_5)  + (k-1-p) \cdot ebr_5 = \nonumber \\ &= r F_8 + d\cdot ebr_4 + F_2 +(k-1-p) \cdot ebr_5 = \nonumber \\ & =   r F_8 + d\cdot ebr_4 + F_2 +(k-1-p) \cdot ebr_5   = \nonumber \\ & = \boxed{   r F_8 + (p-2r) \cdot ebr_4 + F_2 +(k-1-p) \cdot ebr_5   } 
			\end{eqnarray}

		\end{itemize}
		
		\item Case \textbf{BCC}: $m = 2r+1$ $\alpha = 4 p$, $0\le 2r \le k-2$, $ 0\le p \le k-2$ 
		
		\[B = 4p  \cdot ebr_1 -(p+r+1) \cdot ebr_3 +  ( k+ r+1 -p) \cdot ebr_5\]
		
		\begin{itemize}
			
			\item $ 2r \ge p$, (and $p$ even)

			\[B = 4p  \cdot ebr_1 -(p+r+1) \cdot ebr_3 +  (3r+3 -p) \cdot ebr_5+ ( k-2-2r) \cdot ebr_5\]
			As the last term is positive, do not change it in the next calculations,
			and as $p$ is even, $p=2l$ hence $r= l+ d$ with $d\ge 0$, 
			
			\begin{eqnarray}
			&B = 8l  \cdot ebr_1 -(3l + d+1) \cdot ebr_3 +  (l+ 3 d+ 3) \cdot ebr_5 + ( k-2-2r) \cdot ebr_5 = \nonumber \\ &= l( 8 ebr_1 - 3 ebr_3+ ebr_5) + (d+1) (3 ebr_5 - ebr_3) +  (k-2-2r) \cdot ebr_5 = \nonumber \\ &=\boxed{\frac{p}{2} F_8+ (r- \frac{p}{2}+1) F_{10} } 
			\end{eqnarray} 
			
			\item $ 2r \ge p$, (and $p$ odd)

			\[B = 4p  \cdot ebr_1 -(p+r+1) \cdot ebr_3 +  (3r+3 -p) \cdot ebr_5+ ( k-2-2r) \cdot ebr_5\]
			The last term is positive and we do not changr it in the next calculations
			and as $p$ is even, $p=2l+1$ hence $r= l+1+ d$ with $d\ge0$ 
			
			\begin{eqnarray}
			&B = (8l+4)   \cdot ebr_1 -(3l + d+3) \cdot ebr_3 +  (l+ 3 d+ 5) \cdot ebr_5 + ( k-2-2r) \cdot ebr_5 = \nonumber \\ &= l( 8 ebr_1 - 3 ebr_3+ ebr_5) + (d+1) (3 ebr_5 - ebr_3) + 2( 2 ebr_1- ebr_3+ ebr_5) +   (k-2-2r) \cdot ebr_5 = \nonumber \\ &=\boxed{\frac{p-1}{2} F_8+ (r- \frac{p-1}{2}) F_{10} + 2 F_2} 
			\end{eqnarray}

			\item $ 2r \le p$, we can write $p= 2 r+ d$ with $d\ge 0$ and we have the decomposition
			
			\begin{eqnarray}
			&B = (8r +4 d) \cdot ebr_1 -(3r + d+1) \cdot ebr_3 +  (r+3) \cdot ebr_5 + (k-2-p) \cdot ebr_5 = \nonumber \\ &= r( 8 ebr_1 - 3 ebr_3+ ebr_5) + d(4 ebr_1 - ebr_3) + (-  ebr_3+  3ebr_5)  + (k-2-p) \cdot ebr_5 = \nonumber \\ &= r F_8 + d\cdot ebr_4 + F_{10}+(k-2-p) \cdot ebr_5  = \nonumber \\ & = \boxed{   r F_8 + (p-2r) \cdot ebr_4 + F_{10} +(k-2-p) \cdot ebr_5   } 
			\end{eqnarray}

		\end{itemize}

\end{itemize}
This concludes the analysis of the indices, roots, and completeness of SG 70.

\[ \textbf{SG }73 : \text{EBR}_{73}= \left(
\begin{array}{ccccccc}
0 & 4 & 0 & 4 & 2 & 2 & 2 \\
4 & 0 & 4 & 0 & 2 & 2 & 2 \\
0 & 4 & 4 & 0 & 2 & 2 & 2 \\
4 & 0 & 0 & 4 & 2 & 2 & 2 \\
2 & 2 & 2 & 2 & 2 & 2 & 2 \\
2 & 2 & 2 & 2 & 2 & 2 & 2 \\
2 & 2 & 2 & 2 & 2 & 2 & 2 \\
1 & 1 & 1 & 1 & 1 & 1 & 1 \\
1 & 1 & 1 & 1 & 1 & 1 & 1 \\
1 & 1 & 1 & 1 & 1 & 1 & 1 \\
1 & 1 & 1 & 1 & 1 & 1 & 1 \\
\end{array}
\right),\Delta= \left(
\begin{array}{ccccccc}
1 & 0 & 0 & 0 & 0 & 0 & 0 \\
0 & 2 & 0 & 0 & 0 & 0 & 0 \\
0 & 0 & 4 & 0 & 0 & 0 & 0 \\
0 & 0 & 0 & 0 & 0 & 0 & 0 \\
0 & 0 & 0 & 0 & 0 & 0 & 0 \\
0 & 0 & 0 & 0 & 0 & 0 & 0 \\
0 & 0 & 0 & 0 & 0 & 0 & 0 \\
0 & 0 & 0 & 0 & 0 & 0 & 0 \\
0 & 0 & 0 & 0 & 0 & 0 & 0 \\
0 & 0 & 0 & 0 & 0 & 0 & 0 \\
0 & 0 & 0 & 0 & 0 & 0 & 0 \\
\end{array}
\right),\text{B=} \left(
\begin{array}{c}
4 y_3-y_1 \\
y_1 \\
4 y_3-y_2 \\
y_2 \\
2 y_3 \\
2 y_3 \\
2 y_3 \\
y_3 \\
y_3 \\
y_3 \\
y_3 \\
\end{array}
\right),V_{\text{EBR}_{73}}= \left(
\begin{array}{c}
\frac{1}{4} \left(y_1+y_2-4 y_3\right) \\
0 \\
\frac{1}{4} \left(y_1-y_2\right) \\
0 \\
0 \\
0 \\
2 y_3-\frac{y_1}{2} \\
\end{array}
\right)\]
The proof of absence of fragile phases in SG 73 is identical to that in SG 54.

\[ \textbf{SG }75 : \text{EBR}_{75}= \left(
\begin{array}{ccccc}
0 & 1 & 0 & 1 & 1 \\
1 & 0 & 1 & 0 & 1 \\
0 & 1 & 1 & 0 & 1 \\
1 & 0 & 0 & 1 & 1 \\
0 & 1 & 1 & 0 & 1 \\
1 & 0 & 0 & 1 & 1 \\
0 & 1 & 0 & 1 & 1 \\
1 & 0 & 1 & 0 & 1 \\
1 & 1 & 1 & 1 & 2 \\
1 & 1 & 1 & 1 & 2 \\
\end{array}
\right),\Delta= \left(
\begin{array}{ccccc}
1 & 0 & 0 & 0 & 0 \\
0 & 1 & 0 & 0 & 0 \\
0 & 0 & 1 & 0 & 0 \\
0 & 0 & 0 & 0 & 0 \\
0 & 0 & 0 & 0 & 0 \\
0 & 0 & 0 & 0 & 0 \\
0 & 0 & 0 & 0 & 0 \\
0 & 0 & 0 & 0 & 0 \\
0 & 0 & 0 & 0 & 0 \\
0 & 0 & 0 & 0 & 0 \\
\end{array}
\right),\text{B=} \left(
\begin{array}{c}
y_3-y_1 \\
y_1 \\
y_3-y_2 \\
y_2 \\
y_3-y_2 \\
y_2 \\
y_3-y_1 \\
y_1 \\
y_3 \\
y_3 \\
\end{array}
\right),V_{\text{EBR}_{75}}= \left(
\begin{array}{c}
y_2 \\
y_3-y_1 \\
y_1-y_2 \\
0 \\
0 \\
\end{array}
\right)\]
$B\ge 0$ gives the conditions 
\[ y_3 \ge y_1 \ge 0 , \;\;\; y_3\ge y_2 \ge 0\] The linear dependencies in the $EBR$ matrix read:
\[ ebr_1+ ebr_2= ebr_3 + ebr_4 = ebr_5\] The only possibility for a fragile phase is that the third component of the $V_{\text{EBR}}$  matrix be negative $y_1- y_2 = -k \;,\; k>0,\;\; k\in \mathbb{Z}$. This in turn gives $y_1= y_2 - k$ and since $y_1\ge 0$ it implies $y_2 \ge k$; since $y_3 \ge y_2$ it implies $y_3 - y_1\ge k$. Hence we will write $y_2= k+ \delta_2,\;\; \delta_2\ge 0$ and $y_3- y_1 = k+ \delta_3,\;\; \delta_3 \ge 0 $. Making these substitutions, we get:
\begin{eqnarray}
&B = k\cdot (ebr_1+ ebr_2- ebr_3) + \delta_1 \cdot ebr_1 + \delta_2 \cdot ebr_2= k \cdot ebr_4+ \delta_1 \cdot ebr_1 + \delta_2 \cdot ebr_2
\end{eqnarray}

All are positive coefficients, hence there are no Eigenvalue fragile phases in SG 75.

\[ \textbf{SG }89 : \text{EBR}_{89}= \left(
\begin{array}{cccccccccc}
1 & 0 & 1 & 0 & 1 & 0 & 1 & 0 & 1 & 1 \\
0 & 1 & 0 & 1 & 0 & 1 & 0 & 1 & 1 & 1 \\
1 & 0 & 1 & 0 & 0 & 1 & 0 & 1 & 1 & 1 \\
0 & 1 & 0 & 1 & 1 & 0 & 1 & 0 & 1 & 1 \\
1 & 0 & 1 & 0 & 0 & 1 & 0 & 1 & 1 & 1 \\
0 & 1 & 0 & 1 & 1 & 0 & 1 & 0 & 1 & 1 \\
1 & 0 & 1 & 0 & 1 & 0 & 1 & 0 & 1 & 1 \\
0 & 1 & 0 & 1 & 0 & 1 & 0 & 1 & 1 & 1 \\
1 & 1 & 1 & 1 & 1 & 1 & 1 & 1 & 2 & 2 \\
1 & 1 & 1 & 1 & 1 & 1 & 1 & 1 & 2 & 2 \\
\end{array}
\right),\Delta= \left(
\begin{array}{cccccccccc}
1 & 0 & 0 & 0 & 0 & 0 & 0 & 0 & 0 & 0 \\
0 & 1 & 0 & 0 & 0 & 0 & 0 & 0 & 0 & 0 \\
0 & 0 & 1 & 0 & 0 & 0 & 0 & 0 & 0 & 0 \\
0 & 0 & 0 & 0 & 0 & 0 & 0 & 0 & 0 & 0 \\
0 & 0 & 0 & 0 & 0 & 0 & 0 & 0 & 0 & 0 \\
0 & 0 & 0 & 0 & 0 & 0 & 0 & 0 & 0 & 0 \\
0 & 0 & 0 & 0 & 0 & 0 & 0 & 0 & 0 & 0 \\
0 & 0 & 0 & 0 & 0 & 0 & 0 & 0 & 0 & 0 \\
0 & 0 & 0 & 0 & 0 & 0 & 0 & 0 & 0 & 0 \\
0 & 0 & 0 & 0 & 0 & 0 & 0 & 0 & 0 & 0 \\
\end{array}
\right),\]
\[\text{B=} \left(
\begin{array}{c}
y_3-y_1 \\
y_1 \\
y_3-y_2 \\
y_2 \\
y_3-y_2 \\
y_2 \\
y_3-y_1 \\
y_1 \\
y_3 \\
y_3 \\
\end{array}
\right),V_{\text{EBR}_{89}}= \left(
\begin{array}{c}
y_3-y_2 \\
y_1 \\
0 \\
0 \\
y_2-y_1 \\
0 \\
0 \\
0 \\
0 \\
0 \\
\end{array}
\right)\]
With the substitution $y_2 \leftrightarrow y_1$ the proof that SG 89 has no fragile phases becomes identical to that of SG 75.

\[ \textbf{SG }99 : \text{EBR}_{99}= \left(
\begin{array}{ccccc}
1 & 0 & 1 & 0 & 1 \\
0 & 1 & 0 & 1 & 1 \\
1 & 0 & 0 & 1 & 1 \\
0 & 1 & 1 & 0 & 1 \\
1 & 0 & 0 & 1 & 1 \\
0 & 1 & 1 & 0 & 1 \\
1 & 0 & 1 & 0 & 1 \\
0 & 1 & 0 & 1 & 1 \\
1 & 1 & 1 & 1 & 2 \\
1 & 1 & 1 & 1 & 2 \\
\end{array}
\right),\Delta= \left(
\begin{array}{ccccc}
1 & 0 & 0 & 0 & 0 \\
0 & 1 & 0 & 0 & 0 \\
0 & 0 & 1 & 0 & 0 \\
0 & 0 & 0 & 0 & 0 \\
0 & 0 & 0 & 0 & 0 \\
0 & 0 & 0 & 0 & 0 \\
0 & 0 & 0 & 0 & 0 \\
0 & 0 & 0 & 0 & 0 \\
0 & 0 & 0 & 0 & 0 \\
0 & 0 & 0 & 0 & 0 \\
\end{array}
\right),\text{B=} \left(
\begin{array}{c}
y_3-y_1 \\
y_1 \\
y_3-y_2 \\
y_2 \\
y_3-y_2 \\
y_2 \\
y_3-y_1 \\
y_1 \\
y_3 \\
y_3 \\
\end{array}
\right),V_{\text{EBR}_{99}}= \left(
\begin{array}{c}
y_3-y_2 \\
y_1 \\
y_2-y_1 \\
0 \\
0 \\
\end{array}
\right)\]
With the substitution $y_2 \leftrightarrow y_1$ the proof that SG 99 has no fragile phases becomes identical to that of SG 75.

\[ \textbf{SG }103 : \text{EBR}_{103}= \left(
\begin{array}{ccccc}
0 & 2 & 0 & 2 & 2 \\
2 & 0 & 2 & 0 & 2 \\
0 & 1 & 1 & 0 & 1 \\
1 & 0 & 0 & 1 & 1 \\
0 & 2 & 2 & 0 & 2 \\
2 & 0 & 0 & 2 & 2 \\
0 & 1 & 0 & 1 & 1 \\
1 & 0 & 1 & 0 & 1 \\
1 & 1 & 1 & 1 & 2 \\
2 & 2 & 2 & 2 & 4 \\
\end{array}
\right),\Delta= \left(
\begin{array}{ccccc}
1 & 0 & 0 & 0 & 0 \\
0 & 1 & 0 & 0 & 0 \\
0 & 0 & 1 & 0 & 0 \\
0 & 0 & 0 & 0 & 0 \\
0 & 0 & 0 & 0 & 0 \\
0 & 0 & 0 & 0 & 0 \\
0 & 0 & 0 & 0 & 0 \\
0 & 0 & 0 & 0 & 0 \\
0 & 0 & 0 & 0 & 0 \\
0 & 0 & 0 & 0 & 0 \\
\end{array}
\right),\text{B=} \left(
\begin{array}{c}
2 y_3-2 y_1 \\
2 y_1 \\
y_3-y_2 \\
y_2 \\
2 y_3-2 y_2 \\
2 y_2 \\
y_3-y_1 \\
y_1 \\
y_3 \\
2 y_3 \\
\end{array}
\right),V_{\text{EBR}_{103}}= \left(
\begin{array}{c}
y_2 \\
y_3-y_1 \\
y_1-y_2 \\
0 \\
0 \\
\end{array}
\right)\]
With the substitution $y_2 \leftrightarrow y_1$ the proof that SG 103 has no fragile phases becomes identical to that of SG 75.

\[ \textbf{SG }112 : \text{EBR}_{112}= \left(
\begin{array}{cccccccc}
1 & 1 & 1 & 1 & 0 & 2 & 0 & 2 \\
1 & 1 & 1 & 1 & 2 & 0 & 2 & 0 \\
1 & 1 & 1 & 1 & 1 & 1 & 1 & 1 \\
1 & 1 & 1 & 1 & 0 & 2 & 2 & 0 \\
1 & 1 & 1 & 1 & 2 & 0 & 0 & 2 \\
1 & 1 & 1 & 1 & 1 & 1 & 1 & 1 \\
2 & 2 & 2 & 2 & 2 & 2 & 2 & 2 \\
2 & 2 & 2 & 2 & 2 & 2 & 2 & 2 \\
\end{array}
\right),\Delta= \left(
\begin{array}{cccccccc}
1 & 0 & 0 & 0 & 0 & 0 & 0 & 0 \\
0 & 1 & 0 & 0 & 0 & 0 & 0 & 0 \\
0 & 0 & 2 & 0 & 0 & 0 & 0 & 0 \\
0 & 0 & 0 & 0 & 0 & 0 & 0 & 0 \\
0 & 0 & 0 & 0 & 0 & 0 & 0 & 0 \\
0 & 0 & 0 & 0 & 0 & 0 & 0 & 0 \\
0 & 0 & 0 & 0 & 0 & 0 & 0 & 0 \\
0 & 0 & 0 & 0 & 0 & 0 & 0 & 0 \\
\end{array}
\right),\]
\[\text{B=} \left(
\begin{array}{c}
2 y_3-y_1 \\
y_1 \\
y_3 \\
2 y_3-y_2 \\
y_2 \\
y_3 \\
2 y_3 \\
2 y_3 \\
\end{array}
\right),V_{\text{EBR}_{112}}= \left(
\begin{array}{c}
2 y_3-y_1 \\
0 \\
0 \\
0 \\
\frac{1}{2} \left(y_1+y_2-2 y_3\right) \\
0 \\
\frac{1}{2} \left(y_1-y_2\right) \\
0 \\
\end{array}
\right)\],

\noindent $B>0$ gives the constraints $2 y_3 \ge y_1 \ge 0$ and  $2 y_3 \ge y_2 \ge 0$. The EBR linear dependence relations read:
\[ ebr_1 = ebr_2 = ebr_3 = ebr_4,\;\;\; ebr_5+ ebr_6= ebr_7+ ebr_8 = 2 ebr_1 \] We can also express $B$ as:
\[ B= ( 2 y_3-y_1) ebr_1 +  \frac{1}{2} \left(y_1+y_2-2 y_3\right)  ebr_5+  \frac{1}{2} \left(y_1-y_2\right) ebr_7\]

There are then 3 cases where $B$ can be a fragile state:
\begin{itemize} 
	\item Case \textbf{A}:  $y_1+ y_2 - 2 y_3 = - 2 k_1, \;\; k_1 >0$ and $ y_1- y_2 = - 2 k_2, \;\; k_2 >0$, which is equivalent to $y_1 = y_2 - 2 k_2 \ge 0$ hence $y_2 \ge 2 k_2$, $y_2 = 2 k_ 2+ \delta, \;\; \delta \ge 0$ and 
	\[
	 B= (y_2 +2 k_1) ebr_1 - k_1 ebr_5 - k_2 ebr_7 = \delta\cdot ebr_1 + k_1 ebr_6 + k_2 ebr_8
	\] This is a decomposition with positive coefficients (after using the linear dependence) of EBRs and there is no fragile case.
	
	\item Case \textbf{B}:   $y_1+ y_2 - 2 y_3 = - 2 k_1, \;\; k_1 >0$ and nd $ y_1- y_2 =  2 k_2, \;\; k_2 >0$ which gives $ 2 y_3 - y_1 = y_2 + 2 k_1$ and hence:
	\[ B= (y_2 +2 k_1) ebr_1 - k_1 ebr_5 + k_2 ebr_7 = \delta\cdot ebr_1 + k_1 ebr_6 + k_2 ebr_7\] This is a decomposition with positive coefficients (after using the linear dependence) of EBRs and there is no fragile case.

	\item Case \textbf{C}: $y_1+ y_2 - 2 y_3 =  2 k_1, \;\; k_1 >0$ and nd $ y_1- y_2 =  - 2 k_2, \;\; k_2 >0$ At this point we apply what we can denote as \emph{permutation symmetry}. Define $y_2' = 2 y_3 -y_2$. Upon the change:
	
	\[ y_2 \leftrightarrow y_2',\;\;  e_7 \leftrightarrow e_5,\;\;  e_6 \leftrightarrow e_8,\;\;  k_1 \leftrightarrow k_2 \]
	this becomes Case \textbf{B} and hence no fragile states exist.

\end{itemize}

\[ \textbf{SG }113 : \text{EBR}_{113}= \left(
\begin{array}{ccccc}
0 & 2 & 0 & 2 & 1 \\
2 & 0 & 2 & 0 & 1 \\
1 & 1 & 1 & 1 & 1 \\
1 & 1 & 1 & 1 & 1 \\
0 & 2 & 2 & 0 & 1 \\
2 & 0 & 0 & 2 & 1 \\
1 & 1 & 1 & 1 & 1 \\
1 & 1 & 1 & 1 & 1 \\
1 & 1 & 1 & 1 & 1 \\
1 & 1 & 1 & 1 & 1 \\
\end{array}
\right),\Delta= \left(
\begin{array}{ccccc}
1 & 0 & 0 & 0 & 0 \\
0 & 1 & 0 & 0 & 0 \\
0 & 0 & 2 & 0 & 0 \\
0 & 0 & 0 & 0 & 0 \\
0 & 0 & 0 & 0 & 0 \\
0 & 0 & 0 & 0 & 0 \\
0 & 0 & 0 & 0 & 0 \\
0 & 0 & 0 & 0 & 0 \\
0 & 0 & 0 & 0 & 0 \\
0 & 0 & 0 & 0 & 0 \\
\end{array}
\right),\text{B=} \left(
\begin{array}{c}
2 y_3-y_2 \\
y_2 \\
y_3 \\
y_3 \\
2 y_3-y_1 \\
y_1 \\
y_3 \\
y_3 \\
y_3 \\
y_3 \\
\end{array}
\right),V_{\text{EBR}_{113}}= \left(
\begin{array}{c}
\frac{1}{2} \left(y_1+y_2-2 y_3\right) \\
0 \\
\frac{1}{2} \left(y_2-y_1\right) \\
0 \\
2 y_3-y_2 \\
\end{array}
\right).\]
With the substitutions
\[ y_1 \leftrightarrow y_2,\;\; ebr_1 \leftrightarrow ebr_5,\; \; ebr_2 \leftrightarrow ebr_6,\;\; \leftrightarrow ebr_3 \leftrightarrow ebr_7,\;\; ebr_4 \leftrightarrow ebr_8 \] SG 113 becomes identical to SG 112 and has no fragile phases.

\[ \textbf{SG }114 : \text{EBR}_{114} \left(
\begin{array}{ccccc}
0 & 2 & 0 & 2 & 2 \\
2 & 0 & 2 & 0 & 2 \\
0 & 1 & 1 & 0 & 1 \\
1 & 0 & 0 & 1 & 1 \\
1 & 1 & 1 & 1 & 2 \\
1 & 1 & 1 & 1 & 2 \\
1 & 1 & 1 & 1 & 2 \\
1 & 1 & 1 & 1 & 2 \\
1 & 1 & 1 & 1 & 2 \\
1 & 1 & 1 & 1 & 2 \\
\end{array}
\right),\Delta= \left(
\begin{array}{ccccc}
1 & 0 & 0 & 0 & 0 \\
0 & 1 & 0 & 0 & 0 \\
0 & 0 & 2 & 0 & 0 \\
0 & 0 & 0 & 0 & 0 \\
0 & 0 & 0 & 0 & 0 \\
0 & 0 & 0 & 0 & 0 \\
0 & 0 & 0 & 0 & 0 \\
0 & 0 & 0 & 0 & 0 \\
0 & 0 & 0 & 0 & 0 \\
0 & 0 & 0 & 0 & 0 \\
\end{array}
\right),\text{B=} \left(
\begin{array}{c}
2 y_3-y_2 \\
y_2 \\
y_3-y_1 \\
y_1 \\
y_3 \\
y_3 \\
y_3 \\
y_3 \\
y_3 \\
y_3 \\
\end{array}
\right),V_{\text{EBR}_{114}}= \left(
\begin{array}{c}
y_1 \\
y_3-\frac{y_2}{2} \\
\frac{1}{2} \left(y_2-2 y_1\right) \\
0 \\
0 \\
\end{array}
\right).\]
$B\ge 0 $ give $ 2y_3 \ge y_2 \ge 0$ and $y_3 \ge y_1 \ge 0$. The EBR linear dependence reads
\[ ebr_1 + ebr_2 = ebr_3 + ebr_4 = ebr_5 \] and the $B$ decomposition into EBRs reads:
\[ B = y_1 ebr_1 + \frac{2 y_3 - y_2}{2} ebr_2 + \frac{y_2- 2 y_1}{2} ebr_3\]

\noindent In the absence of a strong phase, we have $2y_3- y_2 = 2 k_2 \ge 0$. For a fragile phase we have $y_2 - 2 y_1 = - 2 k_1,\;\; k_1 >0$. Hence $B$ becomes
\[ B= y_1 ebr_1+ k_2 ebr_2 - k_1 ebr_3 \]

\noindent From $2y_3-y_2=2k_2$ and $y_2-2y_1=-2k_1$, we have  $y_1= y_3 - k_2+ k_1$. Since $y_3\ge y_1$ we have $k_2 \ge k_1$. We re-write $B$ as
\[B= y_3 ebr_1 + k_2( ebr_2- ebr_1) + k_1(ebr_1- ebr_3) = (y_3 - k_2) ebr_1 + (k_2 - k_1) ebr_2 + k_1 ebr_4 \] But $y_3= k_2 +\frac{1}{2} y_2$, and we have

\[  B=  \frac{y_2}{2} ebr_1 + (k_2 - k_1) ebr_2 + k_1 ebr_4 \] 
where all the coefficients are positive and hence there are no fragile phases in SG 114.

\[ \textbf{SG }116 : \text{EBR}_{116}= \left(
\begin{array}{ccccccc}
1 & 1 & 0 & 2 & 0 & 2 & 2 \\
1 & 1 & 2 & 0 & 2 & 0 & 2 \\
1 & 1 & 1 & 1 & 1 & 1 & 2 \\
1 & 1 & 0 & 2 & 2 & 0 & 2 \\
1 & 1 & 2 & 0 & 0 & 2 & 2 \\
1 & 1 & 1 & 1 & 1 & 1 & 2 \\
1 & 1 & 1 & 1 & 1 & 1 & 2 \\
2 & 2 & 2 & 2 & 2 & 2 & 4 \\
\end{array}
\right),\Delta= \left(
\begin{array}{ccccccc}
1 & 0 & 0 & 0 & 0 & 0 & 0 \\
0 & 1 & 0 & 0 & 0 & 0 & 0 \\
0 & 0 & 2 & 0 & 0 & 0 & 0 \\
0 & 0 & 0 & 0 & 0 & 0 & 0 \\
0 & 0 & 0 & 0 & 0 & 0 & 0 \\
0 & 0 & 0 & 0 & 0 & 0 & 0 \\
0 & 0 & 0 & 0 & 0 & 0 & 0 \\
0 & 0 & 0 & 0 & 0 & 0 & 0 \\
\end{array}
\right),\]
\[\text{B=} \left(
\begin{array}{c}
2 y_3-y_1 \\
y_1 \\
y_3 \\
2 y_3-y_2 \\
y_2 \\
y_3 \\
y_3 \\
2 y_3 \\
\end{array}
\right),V_{\text{EBR}_{116}}= \left(
\begin{array}{c}
2 y_3-y_1 \\
0 \\
\frac{1}{2} \left(y_1+y_2-2 y_3\right) \\
0 \\
\frac{1}{2} \left(y_1-y_2\right) \\
0 \\
0 \\
\end{array}
\right)\]
By making the permutation $ebr_3\rightarrow ebr_5, ebr_4 \rightarrow ebr_6, ebr_5 \rightarrow ebr_7, ebr_6 \rightarrow ebr_8$, the proof of non-existence of fragile states in SG 116 becomes identical to that in SG 112. 

\[ \textbf{SG }117 : \text{EBR}_{117}= \left(
\begin{array}{cccccc}
0 & 2 & 0 & 2 & 1 & 1 \\
2 & 0 & 2 & 0 & 1 & 1 \\
1 & 1 & 1 & 1 & 1 & 1 \\
1 & 1 & 1 & 1 & 1 & 1 \\
0 & 2 & 2 & 0 & 1 & 1 \\
2 & 0 & 0 & 2 & 1 & 1 \\
1 & 1 & 1 & 1 & 1 & 1 \\
1 & 1 & 1 & 1 & 1 & 1 \\
1 & 1 & 1 & 1 & 1 & 1 \\
1 & 1 & 1 & 1 & 1 & 1 \\
\end{array}
\right),\Delta= \left(
\begin{array}{cccccc}
1 & 0 & 0 & 0 & 0 & 0 \\
0 & 1 & 0 & 0 & 0 & 0 \\
0 & 0 & 2 & 0 & 0 & 0 \\
0 & 0 & 0 & 0 & 0 & 0 \\
0 & 0 & 0 & 0 & 0 & 0 \\
0 & 0 & 0 & 0 & 0 & 0 \\
0 & 0 & 0 & 0 & 0 & 0 \\
0 & 0 & 0 & 0 & 0 & 0 \\
0 & 0 & 0 & 0 & 0 & 0 \\
0 & 0 & 0 & 0 & 0 & 0 \\
\end{array}
\right),\]
\[\text{B=} \left(
\begin{array}{c}
2 y_3-y_2 \\
y_2 \\
y_3 \\
y_3 \\
2 y_3-y_1 \\
y_1 \\
y_3 \\
y_3 \\
y_3 \\
y_3 \\
\end{array}
\right),V_{\text{EBR}_{117}}= \left(
\begin{array}{c}
\frac{1}{2} \left(y_1+y_2-2 y_3\right) \\
0 \\
\frac{1}{2} \left(y_2-y_1\right) \\
0 \\
0 \\
2 y_3-y_2 \\
\end{array}
\right)\]
By making the permutations $y_1 \leftrightarrow y_2$ and  permutation $ebr_6\rightarrow ebr_1, ebr_1 \rightarrow ebr_5, ebr_3 \rightarrow ebr_7, ebr_2 \rightarrow ebr_6, ebr_4 \rightarrow ebr_8$, the proof of non-existence of fragile states in SG 117 becomes identical to that in SG 112.

\[ \textbf{SG }118 : \text{EBR}_{118}= \left(
\begin{array}{cccccc}
0 & 2 & 0 & 2 & 1 & 1 \\
2 & 0 & 2 & 0 & 1 & 1 \\
0 & 2 & 2 & 0 & 1 & 1 \\
2 & 0 & 0 & 2 & 1 & 1 \\
1 & 1 & 1 & 1 & 1 & 1 \\
1 & 1 & 1 & 1 & 1 & 1 \\
1 & 1 & 1 & 1 & 1 & 1 \\
1 & 1 & 1 & 1 & 1 & 1 \\
1 & 1 & 1 & 1 & 1 & 1 \\
1 & 1 & 1 & 1 & 1 & 1 \\
\end{array}
\right),\Delta= \left(
\begin{array}{cccccc}
1 & 0 & 0 & 0 & 0 & 0 \\
0 & 1 & 0 & 0 & 0 & 0 \\
0 & 0 & 2 & 0 & 0 & 0 \\
0 & 0 & 0 & 0 & 0 & 0 \\
0 & 0 & 0 & 0 & 0 & 0 \\
0 & 0 & 0 & 0 & 0 & 0 \\
0 & 0 & 0 & 0 & 0 & 0 \\
0 & 0 & 0 & 0 & 0 & 0 \\
0 & 0 & 0 & 0 & 0 & 0 \\
0 & 0 & 0 & 0 & 0 & 0 \\
\end{array}
\right),\text{B=} \left(
\begin{array}{c}
2 y_3-y_2 \\
y_2 \\
2 y_3-y_1 \\
y_1 \\
y_3 \\
y_3 \\
y_3 \\
y_3 \\
y_3 \\
y_3 \\
\end{array}
\right),V_{\text{EBR}_{118}}= \left(
\begin{array}{c}
\frac{1}{2} \left(y_1+y_2-2 y_3\right) \\
0 \\
\frac{1}{2} \left(y_2-y_1\right) \\
0 \\
0 \\
2 y_3-y_2 \\
\end{array}
\right).\]
The proof of the absence of fragile states in SG 118 is identical to that in SG 117 (and by permutation, to SG 112). 

\[ \textbf{SG }120 : \text{EBR}_{120}= \left(
\begin{array}{cccccc}
1 & 0 & 2 & 0 & 2 & 1 \\
1 & 2 & 0 & 2 & 0 & 1 \\
1 & 0 & 2 & 2 & 0 & 1 \\
1 & 2 & 0 & 0 & 2 & 1 \\
1 & 1 & 1 & 1 & 1 & 1 \\
1 & 1 & 1 & 1 & 1 & 1 \\
2 & 2 & 2 & 2 & 2 & 2 \\
1 & 1 & 1 & 1 & 1 & 1 \\
1 & 1 & 1 & 1 & 1 & 1 \\
\end{array}
\right),\Delta= \left(
\begin{array}{cccccc}
1 & 0 & 0 & 0 & 0 & 0 \\
0 & 1 & 0 & 0 & 0 & 0 \\
0 & 0 & 2 & 0 & 0 & 0 \\
0 & 0 & 0 & 0 & 0 & 0 \\
0 & 0 & 0 & 0 & 0 & 0 \\
0 & 0 & 0 & 0 & 0 & 0 \\
0 & 0 & 0 & 0 & 0 & 0 \\
0 & 0 & 0 & 0 & 0 & 0 \\
0 & 0 & 0 & 0 & 0 & 0 \\
\end{array}
\right),\text{B=} \left(
\begin{array}{c}
2 y_3-y_1 \\
y_1 \\
2 y_3-y_2 \\
y_2 \\
y_3 \\
y_3 \\
2 y_3 \\
y_3 \\
y_3 \\
\end{array}
\right),V_{\text{EBR}_{120}}= \left(
\begin{array}{c}
2 y_3-y_1 \\
\frac{1}{2} \left(y_1+y_2-2 y_3\right) \\
0 \\
\frac{1}{2} \left(y_1-y_2\right) \\
0 \\
0 \\
\end{array}
\right)\]
By making the permutations  $ebr_2\rightarrow ebr_5, ebr_3 \rightarrow ebr_6, ebr_4 \rightarrow ebr_7, ebr_5 \rightarrow ebr_8$, the proof of non-existence of fragile states in SG 120 becomes identical to that in SG 112.

\[ \textbf{SG }122 : \text{EBR}_{122}= \left(
\begin{array}{ccccc}
0 & 2 & 0 & 2 & 2 \\
2 & 0 & 2 & 0 & 2 \\
1 & 1 & 1 & 1 & 2 \\
0 & 1 & 1 & 0 & 1 \\
1 & 0 & 0 & 1 & 1 \\
0 & 1 & 1 & 0 & 1 \\
1 & 0 & 0 & 1 & 1 \\
1 & 1 & 1 & 1 & 2 \\
0 & 1 & 1 & 0 & 1 \\
1 & 0 & 0 & 1 & 1 \\
0 & 1 & 1 & 0 & 1 \\
1 & 0 & 0 & 1 & 1 \\
1 & 1 & 1 & 1 & 2 \\
1 & 1 & 1 & 1 & 2 \\
1 & 1 & 1 & 1 & 2 \\
2 & 2 & 2 & 2 & 4 \\
\end{array}
\right),\Delta= \left(
\begin{array}{ccccc}
1 & 0 & 0 & 0 & 0 \\
0 & 1 & 0 & 0 & 0 \\
0 & 0 & 2 & 0 & 0 \\
0 & 0 & 0 & 0 & 0 \\
0 & 0 & 0 & 0 & 0 \\
0 & 0 & 0 & 0 & 0 \\
0 & 0 & 0 & 0 & 0 \\
0 & 0 & 0 & 0 & 0 \\
0 & 0 & 0 & 0 & 0 \\
0 & 0 & 0 & 0 & 0 \\
0 & 0 & 0 & 0 & 0 \\
0 & 0 & 0 & 0 & 0 \\
0 & 0 & 0 & 0 & 0 \\
0 & 0 & 0 & 0 & 0 \\
0 & 0 & 0 & 0 & 0 \\
0 & 0 & 0 & 0 & 0 \\
\end{array}
\right),\text{B=} \left(
\begin{array}{c}
2 y_3-y_2 \\
y_2 \\
y_3 \\
y_3-y_1 \\
y_1 \\
y_3-y_1 \\
y_1 \\
y_3 \\
y_3-y_1 \\
y_1 \\
y_3-y_1 \\
y_1 \\
y_3 \\
y_3 \\
y_3 \\
2 y_3 \\
\end{array}
\right),V_{\text{EBR}_{122}}= \left(
\begin{array}{c}
y_1 \\
y_3-\frac{y_2}{2} \\
\frac{1}{2} \left(y_2-2 y_1\right) \\
0 \\
0 \\
\end{array}
\right)\]
The proof of nonexistence of fragile phases in SG 122 is identical to that in SG 114.

\[ \textbf{SG }133 : \text{EBR}_{133}= \left(
\begin{array}{ccccccc}
1 & 1 & 1 & 0 & 2 & 0 & 4 \\
1 & 1 & 1 & 2 & 0 & 0 & 4 \\
1 & 1 & 1 & 2 & 0 & 4 & 0 \\
1 & 1 & 1 & 0 & 2 & 4 & 0 \\
1 & 1 & 1 & 1 & 1 & 2 & 2 \\
1 & 1 & 1 & 1 & 1 & 2 & 2 \\
2 & 2 & 2 & 2 & 2 & 4 & 4 \\
2 & 2 & 2 & 2 & 2 & 4 & 4 \\
2 & 2 & 2 & 2 & 2 & 4 & 4 \\
2 & 2 & 2 & 2 & 2 & 4 & 4 \\
\end{array}
\right),\Delta= \left(
\begin{array}{ccccccc}
1 & 0 & 0 & 0 & 0 & 0 & 0 \\
0 & 1 & 0 & 0 & 0 & 0 & 0 \\
0 & 0 & 4 & 0 & 0 & 0 & 0 \\
0 & 0 & 0 & 0 & 0 & 0 & 0 \\
0 & 0 & 0 & 0 & 0 & 0 & 0 \\
0 & 0 & 0 & 0 & 0 & 0 & 0 \\
0 & 0 & 0 & 0 & 0 & 0 & 0 \\
0 & 0 & 0 & 0 & 0 & 0 & 0 \\
0 & 0 & 0 & 0 & 0 & 0 & 0 \\
0 & 0 & 0 & 0 & 0 & 0 & 0 \\
\end{array}
\right),\text{B=} \left(
\begin{array}{c}
2 y_3-y_2 \\
2 y_3-y_1 \\
y_2 \\
y_1 \\
y_3 \\
y_3 \\
2 y_3 \\
2 y_3 \\
2 y_3 \\
2 y_3 \\
\end{array}
\right),V_{\text{EBR}_{133}}= \left(
\begin{array}{c}
2 y_3-y_2 \\
0 \\
0 \\
\frac{1}{2} \left(y_2-y_1\right) \\
0 \\
\frac{1}{4} \left(y_1+y_2-2 y_3\right) \\
0 \\
\end{array}
\right)\]
By making the permutations $y_1 \leftrightarrow y_2$ and  permutation $ebr_4\rightarrow ebr_7, ebr_5 \rightarrow ebr_8, ebr_6 \rightarrow 2\cdot ebr_5, ebr_7 \rightarrow 2\cdot ebr_8$, the proof of non-existence of fragile states in SG 133 becomes identical to that in SG 112.

\[ \textbf{SG }142 : \text{EBR}_{142}= \left(
\begin{array}{cccccc}
0 & 2 & 1 & 0 & 4 & 2 \\
2 & 0 & 1 & 0 & 4 & 2 \\
2 & 0 & 1 & 4 & 0 & 2 \\
0 & 2 & 1 & 4 & 0 & 2 \\
2 & 2 & 2 & 4 & 4 & 4 \\
1 & 1 & 1 & 2 & 2 & 2 \\
1 & 1 & 1 & 2 & 2 & 2 \\
1 & 1 & 1 & 2 & 2 & 2 \\
2 & 2 & 2 & 4 & 4 & 4 \\
2 & 2 & 2 & 4 & 4 & 4 \\
\end{array}
\right),\Delta= \left(
\begin{array}{cccccc}
1 & 0 & 0 & 0 & 0 & 0 \\
0 & 1 & 0 & 0 & 0 & 0 \\
0 & 0 & 4 & 0 & 0 & 0 \\
0 & 0 & 0 & 0 & 0 & 0 \\
0 & 0 & 0 & 0 & 0 & 0 \\
0 & 0 & 0 & 0 & 0 & 0 \\
0 & 0 & 0 & 0 & 0 & 0 \\
0 & 0 & 0 & 0 & 0 & 0 \\
0 & 0 & 0 & 0 & 0 & 0 \\
0 & 0 & 0 & 0 & 0 & 0 \\
\end{array}
\right),\text{B=} \left(
\begin{array}{c}
2 y_3-y_2 \\
2 y_3-y_1 \\
y_2 \\
y_1 \\
2 y_3 \\
y_3 \\
y_3 \\
y_3 \\
2 y_3 \\
2 y_3 \\
\end{array}
\right),V_{\text{EBR}_{142}}= \left(
\begin{array}{c}
\frac{1}{2} \left(y_2-y_1\right) \\
0 \\
2 y_3-y_2 \\
\frac{1}{4} \left(y_1+y_2-2 y_3\right) \\
0 \\
0 \\
\end{array}
\right)\] By making the permutations $y_1 \leftrightarrow y_2$ and  permutation $ebr_3 \rightarrow ebr_1, ebr_1 \rightarrow ebr_7, ebr_2 \rightarrow ebr_8, ebr_4 \rightarrow  2\cdot ebr_5, ebr_5 \rightarrow 2\cdot ebr_6$, the proof of non-existence of fragile states in SG 142 becomes identical to that in SG 112.

\[ \textbf{SG } 150 : \text{EBR}_{150}= \left(
\begin{array}{cccccc}
0 & 1 & 0 & 1 & 2 & 0 \\
1 & 0 & 1 & 0 & 0 & 2 \\
0 & 1 & 0 & 1 & 2 & 0 \\
1 & 0 & 1 & 0 & 0 & 2 \\
0 & 1 & 0 & 1 & 0 & 1 \\
0 & 1 & 0 & 1 & 0 & 1 \\
1 & 0 & 1 & 0 & 2 & 1 \\
0 & 1 & 0 & 1 & 0 & 1 \\
0 & 1 & 0 & 1 & 0 & 1 \\
1 & 0 & 1 & 0 & 2 & 1 \\
0 & 1 & 0 & 1 & 0 & 1 \\
0 & 1 & 0 & 1 & 0 & 1 \\
1 & 0 & 1 & 0 & 2 & 1 \\
0 & 1 & 0 & 1 & 0 & 1 \\
0 & 1 & 0 & 1 & 0 & 1 \\
1 & 0 & 1 & 0 & 2 & 1 \\
1 & 1 & 1 & 1 & 2 & 2 \\
1 & 1 & 1 & 1 & 2 & 2 \\
\end{array}
\right),\Delta= \left(
\begin{array}{cccccc}
1 & 0 & 0 & 0 & 0 & 0 \\
0 & 1 & 0 & 0 & 0 & 0 \\
0 & 0 & 1 & 0 & 0 & 0 \\
0 & 0 & 0 & 0 & 0 & 0 \\
0 & 0 & 0 & 0 & 0 & 0 \\
0 & 0 & 0 & 0 & 0 & 0 \\
0 & 0 & 0 & 0 & 0 & 0 \\
0 & 0 & 0 & 0 & 0 & 0 \\
0 & 0 & 0 & 0 & 0 & 0 \\
0 & 0 & 0 & 0 & 0 & 0 \\
0 & 0 & 0 & 0 & 0 & 0 \\
0 & 0 & 0 & 0 & 0 & 0 \\
0 & 0 & 0 & 0 & 0 & 0 \\
0 & 0 & 0 & 0 & 0 & 0 \\
0 & 0 & 0 & 0 & 0 & 0 \\
0 & 0 & 0 & 0 & 0 & 0 \\
0 & 0 & 0 & 0 & 0 & 0 \\
0 & 0 & 0 & 0 & 0 & 0 \\
\end{array}
\right),\text{B=} \left(
\begin{array}{c}
y_3-y_2 \\
y_2 \\
y_3-y_2 \\
y_2 \\
y_3-y_1 \\
y_3-y_1 \\
y_1 \\
y_3-y_1 \\
y_3-y_1 \\
y_1 \\
y_3-y_1 \\
y_3-y_1 \\
y_1 \\
y_3-y_1 \\
y_3-y_1 \\
y_1 \\
y_3 \\
y_3 \\
\end{array}
\right),V_{\text{EBR}_{150}}= \left(
\begin{array}{c}
3 y_2-2 y_1 \\
-2 y_1+y_2+y_3 \\
0 \\
0 \\
y_1-y_2 \\
y_1-y_2 \\
\end{array}
\right)\] With the replacements $y_1 \rightarrow y_3, \; y_2 \rightarrow y_1, \; y_3 \rightarrow y_2+ y_3$ and $ebr_5 \rightarrow ebr_ 3, \; ebr_6 \rightarrow ebr_4$ the proof and conditions for SG 150 becomes identical to the proof for SG 157 below. There are 2 fragile roots.

\[ \textbf{SG }157 : \text{EBR}_{157}= \left(
\begin{array}{cccc}
0 & 1 & 2 & 0 \\
1 & 0 & 0 & 2 \\
0 & 1 & 2 & 0 \\
1 & 0 & 0 & 2 \\
1 & 1 & 2 & 2 \\
1 & 1 & 2 & 2 \\
0 & 1 & 0 & 1 \\
0 & 1 & 0 & 1 \\
1 & 0 & 2 & 1 \\
0 & 1 & 0 & 1 \\
0 & 1 & 0 & 1 \\
1 & 0 & 2 & 1 \\
\end{array}
\right),\Delta= \left(
\begin{array}{cccc}
1 & 0 & 0 & 0 \\
0 & 1 & 0 & 0 \\
0 & 0 & 1 & 0 \\
0 & 0 & 0 & 0 \\
0 & 0 & 0 & 0 \\
0 & 0 & 0 & 0 \\
0 & 0 & 0 & 0 \\
0 & 0 & 0 & 0 \\
0 & 0 & 0 & 0 \\
0 & 0 & 0 & 0 \\
0 & 0 & 0 & 0 \\
0 & 0 & 0 & 0 \\
\end{array}
\right),\text{B=} \left(
\begin{array}{c}
-y_1+y_2+y_3 \\
y_1 \\
-y_1+y_2+y_3 \\
y_1 \\
y_2+y_3 \\
y_2+y_3 \\
y_2 \\
y_2 \\
y_3 \\
y_2 \\
y_2 \\
y_3 \\
\end{array}
\right),V_{\text{EBR}_{157}}= \left(
\begin{array}{c}
3 y_1-2 y_3 \\
y_1+y_2-y_3 \\
y_3-y_1 \\
y_3-y_1 \\
\end{array}
\right)\]
$B\ge 0 $ gives
\[ y_2+ y_3  \ge y_1 \ge 0, \; y_2 \ge 0, \; y_3 \ge 0\]
The EBR linear dependence reads:
\[ ebr_3 + 2\cdot  ebr_4 = 4 \cdot ebr_1 + 2 \cdot ebr_2. \] In order to simplify the calculation we first have to massage the decomposition of $B$ into EBRs:
\begin{eqnarray}
& B= (3y_1 - 2 y_3) ebr_1 + (y_1 + y_2 - y_3) ebr_2 + (y_3 -y_1)(ebr_3+ ebr_4)  =\nonumber \\ &= (3y_1 - 2 y_3) ebr_1 + (y_1 + y_2 - y_3) ebr_2 + (y_3 -y_1)(4 ebr_1+ 2 ebr_2) + (y_3 -y_1) ebr_4  =\nonumber \\ &=  (y_3+ y_2 - y_1) ebr_2 + (2 y_3- y_1) ebr_1 + (y_1 - y_3) ebr_4.\end{eqnarray} We now have $3$ cases for fragile phases:

\begin{itemize}
	
	\item Case \textbf{A}:  $2y_3 - y_1 = - k_1, \;\; y_1 - y_3 = - k_2$ $k_1, k_2 > 0$,  which means $y_3 = -k_1 - k_2, \;\; y_1= -k_1 - 2k_2$; this is impossible as $y_3 \ge 0$ from the $B\ge 0$ condition. Hence there are no fragile phases in case \textbf{A}.

	\item Case \textbf{B}:  $2y_3 - y_1 =  k_1, \;\; y_1 - y_3 = - k_2$  with $k_2 > 0$ $k_1 \ge 0$,  which means $y_3 = k_1 - k_2, \;\; y_1= k_1 - 2k_2$. The condition $y_1 \ge 0$ gives $k_1 \ge 2 k_2$. All other $B\ge 0$ conditions are weaker than this. Then $B$ becomes
	\[ B= (y_2 + k_2) ebr_2 + k_1 ebr_1 - k_2 ebr_4\] Since  $k_1 \ge 2 k_2$ we are going to write $k_1= 2 k_2 + \delta$ and we notice that $\delta = y_1$. We then have:
	\[ B= y_2 \cdot ebr_2 + y_1 \cdot ebr_1 + k_2( 2\cdot ebr_1 + ebr_2 - ebr_4) \]
	We see that the first fragile condition is $k_2 = 2n+1, \;\; n\ge 0$. (If $k_2$ is even, then the last term is  $\frac{k_2}{2} ( 4\cdot ebr_1 +2 \cdot ebr_2 -2\cdot ebr_4) = \frac{k_2}{2} ebr_3$  ). The necessary (but not sufficient) condition is 
	\[\boxed{y_1 - y_3 < 0, \;\;\; (y_1 - y_3) \mod 2 = 1} \]  With $k_2 = 2n+1$ we write
	\[B= (y_2+1)ebr_2 + (2+ y_1) ebr_1 - ebr_4 + n \cdot ebr_3 \] Since we have $ebr_3 +   ebr_4 = 4 \cdot ebr_1 + 2 \cdot ebr_2 - ebr_4 $ we see that the fragile phases only happen for the following combinations  of $y_2, y_1$:
	\[ y_1 = 0,1 \;\text{and} \;  y_2 \ge 0; \;\;\; \textbf{or} \;\;\; y_2 = 0  \; \text{and}\; y_1 \ge 0 \] Hence the indices for a fragile state (necessary and sufficient conditions) can be written as:
	
	\begin{eqnarray}
	&y_2+ y_3  \ge y_1 \ge 0, \; y_2 \ge 0, \; y_3 \ge 0 \nonumber \\ 
	& \boxed{y_1 - y_3 < 0, \;\;\; (y_1 - y_3) \mod 2 = 1;\;\;  2y_3 - y_1  \ge 0 } \nonumber \\ 
	& \boxed{  y_1 = 0,1 \;\text{and} \;  y_2 \ge 0;  \;\;\; \textbf{or} \;\;\; y_2 = 0  \; \text{and}\; y_1 \ge 0  }      \label{indices157}\end{eqnarray}
	Where the first line is just the $B\ge 0$ condition while the other lines represent the fragile condition.  There is one fragile root
	
	\[\boxed{V_{1,EBR_{157}}^{\text{fragile, root}}   = 2 \cdot ebr_1+ ebr_2 - ebr_4}   \]
	
	The decomposition of any fragile phase into EBRs and roots is
	\[ B= y_2 ebr_2 +y_1 ebr_1+ (y_3- y_1)V_{1,EBR_{157}}^{\text{fragile, root}}\] Notice that due to of Eq. (\ref{indices157}) all coefficients are positive.

	\item Case \textbf{C}:  $2y_3 - y_1 =-  k_1, \;\; y_1 - y_3 = k_2$  with $k_1 > 0$ $k_2 \ge 0$,  which means $y_3 =- k_1 + k_2, \;\; y_1=- k_1 +  2k_2$. The condition $y_3 \ge 0$ gives $k_2 \ge  k_1$. $y_2+ y_3 \ge y_1$ gives $y_2 \ge k_2$, and we write $y_2 = k_2+ r$. $B$ then becomes
	\[ B= r \cdot ebr_2 - k_1 \cdot ebr_1 + k_2 \cdot ebr_4 \] With $k_2 = y_3 + k_1$ we have
	\[ B= r \cdot ebr_2 + y_3 \cdot ebr_4 +k_2 \cdot (ebr_4 - ebr_1)  \]
	Hence the  other root is 
	
	\[\boxed{V_{2, EBR_{157}}^{\text{fragile, root}}   =  ebr_4 - ebr_1}   \]
	
	The root decomposition of any fragile state in SG 157 is:
	
	\[ B= (y_2+ y_3- y_1) \cdot ebr_2 + y_3 \cdot ebr_4 +(y_1 - y_3)  V_{2, EBR_{157}}^{\text{fragile, root}}   \]
	
	And the topological index is:
	\[ \boxed{2y_3 - y_1 <0, \;\; y_1 - y_3 \ge 0}\]
	
\end{itemize}

\[ \textbf{SG }159 : \text{EBR}_{159}= \left(
\begin{array}{cccc}
2 & 0 & 2 & 0 \\
0 & 2 & 0 & 2 \\
1 & 0 & 1 & 0 \\
1 & 0 & 1 & 0 \\
0 & 1 & 0 & 1 \\
1 & 1 & 1 & 1 \\
1 & 1 & 1 & 1 \\
2 & 2 & 2 & 2 \\
2 & 0 & 0 & 1 \\
2 & 0 & 0 & 1 \\
0 & 2 & 2 & 1 \\
2 & 0 & 0 & 1 \\
2 & 0 & 0 & 1 \\
0 & 2 & 2 & 1 \\
\end{array}
\right),\Delta= \left(
\begin{array}{cccc}
1 & 0 & 0 & 0 \\
0 & 1 & 0 & 0 \\
0 & 0 & 1 & 0 \\
0 & 0 & 0 & 0 \\
0 & 0 & 0 & 0 \\
0 & 0 & 0 & 0 \\
0 & 0 & 0 & 0 \\
0 & 0 & 0 & 0 \\
0 & 0 & 0 & 0 \\
0 & 0 & 0 & 0 \\
0 & 0 & 0 & 0 \\
0 & 0 & 0 & 0 \\
0 & 0 & 0 & 0 \\
0 & 0 & 0 & 0 \\
\end{array}
\right),\text{B=} \left(
\begin{array}{c}
-2 y_1+2 y_2+2 y_3 \\
2 y_1 \\
-y_1+y_2+y_3 \\
-y_1+y_2+y_3 \\
y_1 \\
y_2+y_3 \\
y_2+y_3 \\
2 y_2+2 y_3 \\
y_2 \\
y_2 \\
y_2+2 y_3 \\
y_2 \\
y_2 \\
y_2+2 y_3 \\
\end{array}
\right),V_{\text{EBR}_{159}}= \left(
\begin{array}{c}
y_1-y_3 \\
3 y_1-y_2-2 y_3 \\
-2 y_1+y_2+2 y_3 \\
-2 y_1+y_2+2 y_3 \\
\end{array}
\right)\]
By making the transformation $y_1 \rightarrow y_1$, $y_2 \rightarrow 2 y_3- y_2$, $y_3\rightarrow y_2 - y_3$, the proof of fragile phases in SG 159 is the same as the proof of fragile phases in SG 173 below.

\[ \textbf{SG }173 : \text{EBR}_{173}= \left(
\begin{array}{cccc}
2 & 0 & 2 & 0 \\
0 & 1 & 0 & 1 \\
0 & 1 & 0 & 1 \\
1 & 0 & 1 & 0 \\
1 & 0 & 1 & 0 \\
0 & 1 & 0 & 1 \\
0 & 1 & 0 & 1 \\
2 & 0 & 0 & 1 \\
0 & 2 & 2 & 1 \\
4 & 0 & 0 & 2 \\
0 & 2 & 2 & 1 \\
1 & 1 & 1 & 1 \\
1 & 1 & 1 & 1 \\
2 & 2 & 2 & 2 \\
\end{array}
\right),\Delta= \left(
\begin{array}{cccc}
1 & 0 & 0 & 0 \\
0 & 1 & 0 & 0 \\
0 & 0 & 1 & 0 \\
0 & 0 & 0 & 0 \\
0 & 0 & 0 & 0 \\
0 & 0 & 0 & 0 \\
0 & 0 & 0 & 0 \\
0 & 0 & 0 & 0 \\
0 & 0 & 0 & 0 \\
0 & 0 & 0 & 0 \\
0 & 0 & 0 & 0 \\
0 & 0 & 0 & 0 \\
0 & 0 & 0 & 0 \\
0 & 0 & 0 & 0 \\
\end{array}
\right),\text{B=} \left(
\begin{array}{c}
2 y_3-2 y_1 \\
y_1 \\
y_1 \\
y_3-y_1 \\
y_3-y_1 \\
y_1 \\
y_1 \\
2 y_3-y_2 \\
y_2 \\
4 y_3-2 y_2 \\
y_2 \\
y_3 \\
y_3 \\
2 y_3 \\
\end{array}
\right),V_{\text{EBR}_{173}}= \left(
\begin{array}{c}
y_1-y_2+y_3 \\
3 y_1-y_2 \\
y_2-2 y_1 \\
y_2-2 y_1 \\
\end{array}
\right)\]
The condition $B\ge 0$ gives
\[ y_3 \ge y_1\ge 0,\;\;\; 2 y_3 \ge y_2 \ge 0 \] The EBR linear dependence relations are:
\[ ebr_1+ 2 ebr_2 = ebr_3+ 2 ebr_4 \] We then have the decomposition:
\begin{eqnarray}
&B = (y_1- y_2+ y_3) \cdot ebr_1 + (3 y_1 - y_2) \cdot ebr_2 + (y_2 - 2 y_1) \cdot (ebr_3+ ebr_4) = \nonumber \\ &= (y_1- y_2+ y_3 + y_2- 2 y_1) \cdot ebr_1+ (3 y_1 - y_2 + 2 y_2 - 4 y_1) ebr_2 + (2y_1- y_2) \cdot ebr_4 =\nonumber \\ & =  ( y_3 -  y_1) \cdot ebr_1+ ( y_2 -  y_1) ebr_2 + (2y_1- y_2) \cdot ebr_4
\end{eqnarray} The first term coefficient is positive. In order to have a fragile phase, we can have 3 cases
\begin{itemize}
	\item Case \textbf{A}: $y_2 - y_1 = -k_1 ;\; 2 y_1 - y_2 = -k_2$ with $k_1, k_2 >0$. This gives $y_1 = -(k_1+ k_2)$ $y_2=-(2 k_1+ k_2)$, which is impossible since $y_1\ge 0 $ and $y_2 \ge 0$, and there are no fragile phases in case \textbf{A}.
	
	\item Case \textbf{B}:  $y_2 - y_1 = k_1 ;\; 2 y_1 - y_2 = -k_2$ with $k_1\ge 0 , k_2 >0$. This gives $y_1 = k_1- k_2$ $y_2=2 k_1- k_2$. From $y_1\ge 0$ we find $k_1\ge k_2$, which is obvious as $k_1 = k_2+ y_1$. We replace this in $B$ to get:
	\[ B=   ( y_3 -  y_1) \cdot ebr_1+ k_2( ebr_2 - ebr_4)+ y_1 \cdot ebr_2 \]
	
	We also have the inequality $2 y_3 \ge  y_2$ which gives $ 2 y_3 \ge 2 k_1- k_2$. We now differentiate two cases: 
	\begin{itemize}
		\item $k_2= 2 n, \; n\ge 0$ then $y_3= k-n + \delta$ with $\delta \ge 0$ and by direct substitution $B$ becomes
		\[ B = \delta \cdot ebr_1+ y_1 ebr_2 + n \cdot ebr_3 \] with all positive coefficients, and hence no fragile phase is possible. Hence the necessary condition for a fragile phase is: 
		\[\boxed{2y_1 - y_2 <0, \;\;\;\; 2y_1- y_2 \mod 2 = 1}\]
		with this, we add
		\item $k_2 = 2 n+1,\;\; n\ge 0$. This gives $y_3\ge 2(k_1- n)-1$ or $y_3 = k_1 - n + \delta,\;\; \delta \ge 0$. Directly plugging into $B$ we obtain
		\[ B= n\cdot ebr_3 + \delta \cdot ebr_1 + y_1\cdot ebr_2 + (ebr_1 + ebr_2 - ebr_4) \]
		The Root is then:
		
		\[\boxed{V_{1, EBR_{173}}^{\text{fragile, root}}   =  ebr_1+ ebr_2 - ebr_4}   \]
		
		Since $ebr_1+ 2 ebr_2- ebr_4= ebr_3+ ebr_4$, in order that $ebr_1+x\cdot ebr_2-ebr_4$ cannot be written in terms of EBRs with positive coefficients, we must make sure that the coefficient of $ebr_2$ stays unity and does not increase to 2. We then find another condition $\boxed{y_1 =0}$ for Eigenvalue Fragile phases. We then have the full set of conditions for the phase to be fragile:
		
		\[y_3 \ge y_1\ge 0,\;\;\; 2 y_3 \ge y_2 \ge 0,\;\;\ \boxed{ y_1=0, \;\;\;\; \;\; y_2 \mod 2 = 1}\]
		Where the un-boxed conditions are just the $B\ge 0 $ conditions re-written. The root and EBR decomposition of any state reads:
		\begin{eqnarray}
		&B= \frac{k_2-1}{2}\cdot ebr_3 + (y_3- k_1+ \frac{k_2-1}{2} )ebr_1 + y_1 ebr_2 +V_{1, EBR_{173}}^{\text{fragile, root}}  = \nonumber \\ & =  \frac{y_2- 2 y_1-1}{2}\cdot ebr_3 + (y_3- y_2 + y_1+ \frac{y_2- 2 y_1-1}{2} )ebr_1 + y_1 ebr_2 +V_{1, EBR_{173}}^{\text{fragile, root}}= \nonumber \\ & =\boxed{  \frac{y_2- 2 y_1-1}{2}\cdot ebr_3 + (y_3- \frac{y_2+1}{2} )ebr_1 + y_1 ebr_2 +V_{1, EBR_{173}}^{\text{fragile, root}}  }\end{eqnarray}
		(of course any state that does not have $y_1=0$ would not be fragile but would be and EBR.
	\end{itemize}

	\item Case \textbf{C}:  $y_2 - y_1 =- k_1 ;\; 2 y_1 - y_2 = k_2$ with $k_2\ge 0 , k_1 >0$. This gives $y_1 = - k_1+ k_2$ $y_2=-2 k_1+ k_2$. We replace back $k_2= 2 k_1+ y_2$ to get:
	\[B= (y_3- y_1) \cdot ebr_1  + k_1( 2 \cdot ebr_4 - ebr_2) + y_2 \cdot ebr_4, \] 
	\noindent Of course $y_3 \ge - k_1+ k_2$, because $y_3\ge y_1$ and $ 2 y_3 \ge - 2 k_1+ k_2$ because $2y_3\ge y_2$. We then have another root:

	\[\boxed{V_{2, EBR_{173}}^{\text{fragile, root}}   = 2\cdot  ebr_4- ebr_2}   \]
	
	The indices are:
	\[ \boxed{y_2 - y_1 <0}\]
	The decomposition of any such fragile state  into roots and EBRs is
	\[ \boxed{B=  (y_3- y_1) \cdot ebr_1  + (y_1 - y_2) ( 2 \cdot ebr_4 - ebr_2) + y_2 \cdot ebr_4 }\]

\end{itemize}

\[ \textbf{SG }182 : \text{EBR}_{182}= \left(
\begin{array}{cccccccc}
0 & 2 & 0 & 2 & 0 & 2 & 0 & 2 \\
1 & 0 & 1 & 0 & 1 & 0 & 1 & 0 \\
1 & 0 & 1 & 0 & 1 & 0 & 1 & 0 \\
0 & 1 & 0 & 1 & 0 & 1 & 0 & 1 \\
0 & 1 & 0 & 1 & 0 & 1 & 0 & 1 \\
1 & 0 & 1 & 0 & 1 & 0 & 1 & 0 \\
1 & 0 & 1 & 0 & 1 & 0 & 1 & 0 \\
0 & 2 & 0 & 2 & 1 & 0 & 1 & 0 \\
2 & 0 & 2 & 0 & 1 & 2 & 1 & 2 \\
0 & 2 & 0 & 2 & 1 & 0 & 1 & 0 \\
0 & 2 & 0 & 2 & 1 & 0 & 1 & 0 \\
2 & 0 & 2 & 0 & 1 & 2 & 1 & 2 \\
1 & 1 & 1 & 1 & 1 & 1 & 1 & 1 \\
1 & 1 & 1 & 1 & 1 & 1 & 1 & 1 \\
2 & 2 & 2 & 2 & 2 & 2 & 2 & 2 \\
\end{array}
\right),\Delta= \left(
\begin{array}{cccccccc}
1 & 0 & 0 & 0 & 0 & 0 & 0 & 0 \\
0 & 1 & 0 & 0 & 0 & 0 & 0 & 0 \\
0 & 0 & 1 & 0 & 0 & 0 & 0 & 0 \\
0 & 0 & 0 & 0 & 0 & 0 & 0 & 0 \\
0 & 0 & 0 & 0 & 0 & 0 & 0 & 0 \\
0 & 0 & 0 & 0 & 0 & 0 & 0 & 0 \\
0 & 0 & 0 & 0 & 0 & 0 & 0 & 0 \\
0 & 0 & 0 & 0 & 0 & 0 & 0 & 0 \\
0 & 0 & 0 & 0 & 0 & 0 & 0 & 0 \\
0 & 0 & 0 & 0 & 0 & 0 & 0 & 0 \\
0 & 0 & 0 & 0 & 0 & 0 & 0 & 0 \\
0 & 0 & 0 & 0 & 0 & 0 & 0 & 0 \\
0 & 0 & 0 & 0 & 0 & 0 & 0 & 0 \\
0 & 0 & 0 & 0 & 0 & 0 & 0 & 0 \\
0 & 0 & 0 & 0 & 0 & 0 & 0 & 0 \\
\end{array}
\right),\text{B=} \left(
\begin{array}{c}
2 y_3-2 y_1 \\
y_1 \\
y_1 \\
y_3-y_1 \\
y_3-y_1 \\
y_1 \\
y_1 \\
2 y_3-y_2 \\
y_2 \\
2 y_3-y_2 \\
2 y_3-y_2 \\
y_2 \\
y_3 \\
y_3 \\
2 y_3 \\
\end{array}
\right),V_{\text{EBR}_{182}}= \left(
\begin{array}{c}
y_2-y_1 \\
y_3-y_1 \\
0 \\
0 \\
2 y_1-y_2 \\
0 \\
0 \\
0 \\
\end{array}
\right)\]
With the substitution $y_1 \rightarrow y_2$, $y_2 \rightarrow y_1$, $ ebr_5 \rightarrow ebr_3$, $ebr_6 \rightarrow ebr_4$, the fragile phases of SG 182 are mapped into the fragile phases of SG 186.

\[ \textbf{SG }185 : \text{EBR}_{185}= \left(
\begin{array}{cccc}
0 & 2 & 4 & 0 \\
1 & 0 & 0 & 2 \\
1 & 0 & 0 & 2 \\
0 & 1 & 2 & 0 \\
1 & 0 & 0 & 2 \\
0 & 1 & 0 & 1 \\
0 & 1 & 0 & 1 \\
1 & 0 & 2 & 1 \\
0 & 2 & 0 & 2 \\
0 & 2 & 0 & 2 \\
2 & 0 & 4 & 2 \\
1 & 1 & 2 & 2 \\
2 & 2 & 4 & 4 \\
\end{array}
\right),\Delta= \left(
\begin{array}{cccc}
1 & 0 & 0 & 0 \\
0 & 1 & 0 & 0 \\
0 & 0 & 1 & 0 \\
0 & 0 & 0 & 0 \\
0 & 0 & 0 & 0 \\
0 & 0 & 0 & 0 \\
0 & 0 & 0 & 0 \\
0 & 0 & 0 & 0 \\
0 & 0 & 0 & 0 \\
0 & 0 & 0 & 0 \\
0 & 0 & 0 & 0 \\
0 & 0 & 0 & 0 \\
0 & 0 & 0 & 0 \\
\end{array}
\right),\text{B=} \left(
\begin{array}{c}
2 y_3-2 y_2 \\
y_2 \\
y_2 \\
y_3-y_2 \\
y_2 \\
y_3-y_1 \\
y_3-y_1 \\
y_1 \\
2 y_3-2 y_1 \\
2 y_3-2 y_1 \\
2 y_1 \\
y_3 \\
2 y_3 \\
\end{array}
\right),V_{\text{EBR}_{185}}= \left(
\begin{array}{c}
3 y_2-2 y_1 \\
-2 y_1+y_2+y_3 \\
y_1-y_2 \\
y_1-y_2 \\
\end{array}
\right)\]
With the substitution $y_1 \rightarrow y_3,\; y_2 \rightarrow  y_1, \; y_3 \rightarrow y_2 + y_2$, the SG 185 becomes identical to SG 157. 

\[ \textbf{SG }186 : \text{EBR}_{186}= \left(
\begin{array}{cccc}
0 & 2 & 0 & 2 \\
1 & 0 & 1 & 0 \\
1 & 0 & 1 & 0 \\
0 & 1 & 0 & 1 \\
1 & 0 & 1 & 0 \\
0 & 2 & 1 & 0 \\
2 & 0 & 1 & 2 \\
0 & 2 & 1 & 0 \\
0 & 2 & 1 & 0 \\
2 & 0 & 1 & 2 \\
1 & 1 & 1 & 1 \\
2 & 2 & 2 & 2 \\
\end{array}
\right),\Delta= \left(
\begin{array}{cccc}
1 & 0 & 0 & 0 \\
0 & 1 & 0 & 0 \\
0 & 0 & 1 & 0 \\
0 & 0 & 0 & 0 \\
0 & 0 & 0 & 0 \\
0 & 0 & 0 & 0 \\
0 & 0 & 0 & 0 \\
0 & 0 & 0 & 0 \\
0 & 0 & 0 & 0 \\
0 & 0 & 0 & 0 \\
0 & 0 & 0 & 0 \\
0 & 0 & 0 & 0 \\
\end{array}
\right),\text{B=} \left(
\begin{array}{c}
2 y_3-2 y_2 \\
y_2 \\
y_2 \\
y_3-y_2 \\
y_2 \\
2 y_3-y_1 \\
y_1 \\
2 y_3-y_1 \\
2 y_3-y_1 \\
y_1 \\
y_3 \\
2 y_3 \\
\end{array}
\right),V_{\text{EBR}_{186}}= \left(
\begin{array}{c}
y_1-y_2 \\
y_3-y_2 \\
2 y_2-y_1 \\
0 \\
\end{array}
\right)\] With the changes $y_1 \rightarrow y_2-y_1,\;\;\; y_2 \rightarrow y_1,\;\;\; y_3 \rightarrow 2y_1-y_2+y_3,\;\; ebr_3 \rightarrow ebr_4, \;\;\; ebr_4 \rightarrow ebr_3, \;\;\; ebr_2 \rightarrow ebr_1, \;\;\; ebr_1 \rightarrow ebr_2 $, the fragile phases, conditions and roots (and of course, proofs) of SG 186 become identical to those of SG 173.

\[ \textbf{SG }209 : \text{EBR}_{209}= \left(
\begin{array}{ccccccccc}
1 & 0 & 0 & 1 & 0 & 0 & 1 & 0 & 1 \\
0 & 1 & 0 & 0 & 1 & 0 & 1 & 0 & 1 \\
0 & 0 & 1 & 0 & 0 & 1 & 0 & 2 & 2 \\
1 & 0 & 1 & 1 & 0 & 1 & 1 & 2 & 3 \\
0 & 1 & 1 & 0 & 1 & 1 & 1 & 2 & 3 \\
0 & 0 & 1 & 0 & 0 & 1 & 0 & 2 & 2 \\
1 & 1 & 1 & 1 & 1 & 1 & 2 & 2 & 4 \\
1 & 1 & 2 & 1 & 1 & 2 & 2 & 4 & 6 \\
\end{array}
\right),\Delta= \left(
\begin{array}{ccccccccc}
1 & 0 & 0 & 0 & 0 & 0 & 0 & 0 & 0 \\
0 & 1 & 0 & 0 & 0 & 0 & 0 & 0 & 0 \\
0 & 0 & 1 & 0 & 0 & 0 & 0 & 0 & 0 \\
0 & 0 & 0 & 0 & 0 & 0 & 0 & 0 & 0 \\
0 & 0 & 0 & 0 & 0 & 0 & 0 & 0 & 0 \\
0 & 0 & 0 & 0 & 0 & 0 & 0 & 0 & 0 \\
0 & 0 & 0 & 0 & 0 & 0 & 0 & 0 & 0 \\
0 & 0 & 0 & 0 & 0 & 0 & 0 & 0 & 0 \\
\end{array}
\right)\],\[\text{B=} \left(
\begin{array}{c}
y_2-y_1 \\
y_1+y_2-y_3 \\
y_3-y_2 \\
y_3-y_1 \\
y_1 \\
y_3-y_2 \\
y_2 \\
y_3 \\
\end{array}
\right),V_{\text{EBR}_{209}}= \left(
\begin{array}{c}
y_2-y_1 \\
y_1+y_2-y_3 \\
y_3-y_2 \\
0 \\
0 \\
0 \\
0 \\
0 \\
0 \\
\end{array}
\right)\]
$B \ge 0$ implies $V_{\text{EBR}}\ge0$ hence there are no fragile phases in this group.

\[ \textbf{SG }211 : \text{EBR}_{211}= \left(
\begin{array}{cccccccc}
1 & 0 & 0 & 0 & 1 & 1 & 0 & 1 \\
0 & 1 & 0 & 1 & 0 & 1 & 0 & 1 \\
0 & 0 & 1 & 1 & 1 & 1 & 2 & 2 \\
1 & 0 & 0 & 0 & 1 & 1 & 0 & 1 \\
0 & 1 & 0 & 1 & 0 & 1 & 0 & 1 \\
0 & 0 & 1 & 1 & 1 & 1 & 2 & 2 \\
1 & 1 & 0 & 1 & 1 & 2 & 0 & 2 \\
0 & 0 & 1 & 1 & 1 & 1 & 2 & 2 \\
0 & 0 & 1 & 1 & 1 & 1 & 2 & 2 \\
1 & 1 & 2 & 3 & 3 & 4 & 4 & 6 \\
\end{array}
\right),\Delta= \left(
\begin{array}{cccccccc}
1 & 0 & 0 & 0 & 0 & 0 & 0 & 0 \\
0 & 1 & 0 & 0 & 0 & 0 & 0 & 0 \\
0 & 0 & 1 & 0 & 0 & 0 & 0 & 0 \\
0 & 0 & 0 & 0 & 0 & 0 & 0 & 0 \\
0 & 0 & 0 & 0 & 0 & 0 & 0 & 0 \\
0 & 0 & 0 & 0 & 0 & 0 & 0 & 0 \\
0 & 0 & 0 & 0 & 0 & 0 & 0 & 0 \\
0 & 0 & 0 & 0 & 0 & 0 & 0 & 0 \\
0 & 0 & 0 & 0 & 0 & 0 & 0 & 0 \\
0 & 0 & 0 & 0 & 0 & 0 & 0 & 0 \\
\end{array}
\right),\]\[\text{B=} \left(
\begin{array}{c}
-y_1-2 y_2+y_3 \\
y_1 \\
y_2 \\
-y_1-2 y_2+y_3 \\
y_1 \\
y_2 \\
y_3-2 y_2 \\
y_2 \\
y_2 \\
y_3 \\
\end{array}
\right),V_{\text{EBR}_{211}}= \left(
\begin{array}{c}
-y_1-2 y_2+y_3 \\
y_1 \\
y_2 \\
0 \\
0 \\
0 \\
0 \\
0 \\
\end{array}
\right)\]

$B \ge 0$ implies $V_{\text{EBR}_{211}}\ge0$ hence there are no fragile phases in this group.

\section{Example: determination of equivalences between different bands in the space group P$\overline{1}$ with TR.}
\label{equivalences:all}
In section \ref{equivalences} of the main text we have introduced two types of operations that map the set of irreps of the little groups at maximal $\mathbf{k}$-vectors into itself. In a given space group $\mathcal{G}$, the first set of equivalences is derived from the operations of the affine normalizer of the group. These equivalences can be expressed as a permutation matrix of dimension $N_{irr}\times N_{irr}$, being $N_{irr}$ the number of irreps at maximal $\mathbf{k}$-vectors. In section \ref{sup:equivnormal} we give an example of the calculation of these matrices in the space group P$\overline{1}$. The complete set of matrices form the group $\mathcal{N}_G$ (see section \ref{main:normalizers} in the main text).

The second set of equivalences can be obtained through the direct product of 1-dimensional irreps of the little group of all the $\mathbf{k}$-vectors in the space group whose star contains just one vector (the $\mathbf{k}$-vector itself). These equivalences can also be expressed as a permutation matrix of dimension $N_{irr}\times N_{irr}$. The complete set of matrices form the group $\mathcal{K}_G$ (see section \ref{kronecker} in the main text). In section \ref{sec:kronecker} we show an example of the derivation of one of these matrices in the space group P$\overline{1}$.

\subsection{Equivalences between band representations due to the affine normalizer of the space group: example of application in the space group P$\overline{1}$ with TR.}
\label{sup:equivnormal}
First we choose as set of generators of the double space group P$\overline{1}$ three translations that generate the lattice, $\{1|1,0,0\}$, $\{1|0,1,0\}$, $\{1|0,0,1\}$, one of the two inversion operations with no translation, $\{\overline{1}|0,0,0\}$, and the operation $\{ ^d1|0,0,0\}$, which represents a $2\pi$ rotation (it does not change the physical magnitudes in the 3-d space but it reverts the spins in the spinor space). However, this symmetry operation commutes with every operation of the double space group and, as a consequence, its trace remains invariant under conjugation in all the irreps. It can thus be omitted in the analysis. The traces of the remaining generators in the irreducible representations at maximal $\mathbf{k}$-vectors are given in Table (\ref{tb:equivalencesSG2}).

\begin{table}[!h]
	\caption{Traces of the $1\times1$ matrices of the generators of the space group P$\overline{1}$ in the irreducible representations at the maximal $\mathbf{k}$-vectors (\cite{irrepsSG}). The labels of the irreps, $\irrg{2}\irrg{2}$, etc\ldots have been split into two rows to make a compact table.\label{tb:equivalencesSG2}}	
	\begin{tabular}{c|rr|rr|rr|rr|rr|rr|rr|rr}
		&\multicolumn{2}{c}{$\Gamma$}&\multicolumn{2}{c}{X}&\multicolumn{2}{c}{Y}&\multicolumn{2}{c}{V}&\multicolumn{2}{c}{Z}&\multicolumn{2}{c}{U}&\multicolumn{2}{c}{T}&\multicolumn{2}{c}{R}\\
		\hline
		&$\irrg{2}$&$\irrg{3}$&$\irr{X}{2}$&$\irr{X}{3}$&$\irr{Y}{2}$&$\irr{Y}{3}$&$\irr{V}{2}$&$\irr{V}{3}$&$\irr{Z}{2}$&$\irr{Z}{3}$&$\irr{U}{2}$&$\irr{U}{3}$&$\irr{T}{2}$&$\irr{T}{3}$&$\irr{R}{2}$&$\irr{R}{3}$\\
		&$\irrg{2}$&$\irrg{3}$&$\irr{X}{2}$&$\irr{X}{3}$&$\irr{Y}{2}$&$\irr{Y}{3}$&$\irr{V}{2}$&$\irr{V}{3}$&$\irr{Z}{2}$&$\irr{Z}{3}$&$\irr{U}{2}$&$\irr{U}{3}$&$\irr{T}{2}$&$\irr{T}{3}$&$\irr{R}{2}$&$\irr{R}{3}$\\
		\hline
		$\{1|1,0,0\}$&1&1  &-1&-1 &1&1   &-1&-1 &1&1   &-1&-1 &1&1   &-1&-1\\
		$\{1|0,1,0\}$&1&1  &1&1   &-1&-1 &-1&-1 &1&1   &1&1   &-1&-1 &-1&-1\\
		$\{1|0,0,1\}$&1&1  &1&1   &1&1   &1&1   &-1&-1 &-1&-1 &-1&-1 &-1&-1\\
		$\{\overline{1}|0,0,0\}$&-1&1 &-1&1  &-1&1  &-1&1  &-1&1  &-1&1  &-1&1  &-1&1
	\end{tabular}
\end{table}

The elements of the affine normalizer $\{N|\mathbf{n}\}$ of the space group P$\overline{1}$ are,
\begin{equation}
N=\left(\begin{array}{rrr}
n_{11}&n_{12}&n_{13}\\
n_{21}&n_{22}&n_{23}\\
n_{31}&n_{32}&n_{33}
\end{array}
\right)\hspace{1cm}\textrm{and}\hspace{1cm}\mathbf{n}=\left(\frac{n_1}{2},\frac{n_2}{2},\frac{n_3}{2}\right)
\end{equation}
where the $n_{ij}$ and $n_i$ components are integer numbers and det$(N)=\pm1$. The database NORMALIZER \cite{normalizers} in the BCS gives the elements of the normalizer for any space group.

First we take the origin shift $N=\mathds{1}$ and $\mathbf{n}=(1/2,0,0)$. This element of the normalizer transform the little group of every $\mathbf{k}$-vector into itself. The result of the conjugation $\{N|\mathbf{n}\}^{-1}\{R|\mathbf{t}\}\{N|\mathbf{n}\}$ of the chosen generators are shown in Table \ref{tb:conjtrasSG2}.

Then, through conjugation, the traces of Table 	(\ref{tb:equivalencesSG2}) are transformed into the set of traces in Table \ref{tb:equivalencesSG2transf}.

\begin{table}[!h]
	\caption{Conjugation of the chosen generators of the space group P$\overline{1}$ through the member $\{1|1/2,0,0\}$ of the affine normalizer.\label{tb:conjtrasSG2}}		
	\begin{tabular}{c|c}
		$\{R|\mathbf{t}\}$&$\{1|1/2,0,0\}^{-1}\{R|\mathbf{t}\}\{1|1/2,0,0\}$\\
		\hline
		$\{1|1,0,0\}$&$\{1|1,0,0\}$\\
		$\{1|0,1,0\}$&$\{1|0,1,0\}$\\
		$\{1|0,0,1\}$&$\{1|0,0,1\}$\\
		$\{\overline{1}|0,0,0\}$&$\{\overline{1}|-1,0,0\}$
	\end{tabular}												
\end{table}

\begin{table}[!h]
	\caption{Traces of the matrices of the symmetry operations of Table \ref{tb:conjtrasSG2} of the space group P$\overline{1}$ in the irreducible representations at the maximal $\mathbf{k}$-vectors (\cite{irrepsSG}). The labels of the irreps, $\irrg{2}\irrg{2}$, etc\ldots have been split into two rows to make a compact table.\label{tb:equivalencesSG2transf}}	
	\begin{tabular}{c|rr|rr|rr|rr|rr|rr|rr|rr}
		&\multicolumn{2}{c}{$\Gamma$}&\multicolumn{2}{c}{X}&\multicolumn{2}{c}{Y}&\multicolumn{2}{c}{V}&\multicolumn{2}{c}{Z}&\multicolumn{2}{c}{U}&\multicolumn{2}{c}{T}&\multicolumn{2}{c}{R}\\
		\hline
		&$\irrg{2}$&$\irrg{3}$&$\irr{X}{2}$&$\irr{X}{3}$&$\irr{Y}{2}$&$\irr{Y}{3}$&$\irr{V}{2}$&$\irr{V}{3}$&$\irr{Z}{2}$&$\irr{Z}{3}$&$\irr{U}{2}$&$\irr{U}{3}$&$\irr{T}{2}$&$\irr{T}{3}$&$\irr{R}{2}$&$\irr{R}{3}$\\
		&$\irrg{2}$&$\irrg{3}$&$\irr{X}{2}$&$\irr{X}{3}$&$\irr{Y}{2}$&$\irr{Y}{3}$&$\irr{V}{2}$&$\irr{V}{3}$&$\irr{Z}{2}$&$\irr{Z}{3}$&$\irr{U}{2}$&$\irr{U}{3}$&$\irr{T}{2}$&$\irr{T}{3}$&$\irr{R}{2}$&$\irr{R}{3}$\\
		\hline
		$\{1|1,0,0\}$  &1&1  &-1&-1 &1&1   &-1&-1 &1&1   &-1&-1 &1&1   &-1&-1\\
		$\{1|0,1,0\}$  &1&1  &1&1   &-1&-1 &-1&-1 &1&1   &1&1   &-1&-1 &-1&-1\\
		$\{1|0,0,1\}$  &1&1  &1&1   &1&1   &1&1   &-1&-1 &-1&-1 &-1&-1 &-1&-1\\
		$\{\overline{1}|-1,0,0\}$ &-1&1 &1&-1  &-1&1  &1&-1  &-1&1  &1&-1  &-1&1  &1&-1
	\end{tabular}
\end{table}
Therefore, this element of the normalizer interchanges the following pairs of irreps,
\begin{equation}
\begin{array}{ccc}
\irr{X}{2}\irr{X}{2}&\leftrightarrow&\irr{X}{3}\irr{X}{3}\\
\irr{V}{2}\irr{V}{2}&\leftrightarrow&\irr{V}{3}\irr{V}{3}\\
\irr{U}{2}\irr{U}{2}&\leftrightarrow&\irr{U}{3}\irr{U}{3}\\
\irr{R}{2}\irr{R}{2}&\leftrightarrow&\irr{R}{3}\irr{R}{3}
\end{array}
\end{equation}

Thus, the conjugation of the maximal irreps under the given operation of the normalizer interchanges the two irreps at the points X, V, U and R and keep the irreps invariant at the other points. Given the irreps at the maximal $\mathbf{k}$-vectors in the order of Eq. (\ref{eq:labelsSG2}), (the same order has been used in Tables (\ref{tb:equivalencesSG2}) and (\ref{tb:equivalencesSG2transf})), the non-zero elements of the permutation matrix $M_{\{1|1/2,0,0\}}$ are: (1,1), (2,2), (3,4), (4,3), (5,5), (6,6), (7,8), (8,7), (9,9), (10,10), (11,12), (12,11), (13,13), (14,14), (15,16) and (16,15).

Following an equivalent calculation, it is straightforward to check that the conjugation under the members of the normalizer $\{1|0,1/2,0\}$ and $\{1|0,0,1/2\}$ interchange the even and odd irreps at maximal $\mathbf{k}$-vectors Y, V, T and Z in the first case and at Z, U, T and R in the second one. In the rest of $\mathbf{k}$-vectors, this operation maps the irreps into themselves. From these relations, it is also easy to calculate the matrices $M_{\{1|0,1/2,0\}}$ and $M_{\{1|0,0,1/2\}}$.

Next we consider the set of elements of the normalizer whose rotational part $N$ belong to the point group 432 (or O in the Sch\"onflies notation) and have $\mathbf{n}=(0,0,0)$ as translational part. It is enough to calculate the $M_{\{N|0,0,0\}}$ matrices only for a subset of symmetry rotational whose rotational parts form a set of generators of the point group. First, let's consider the operation $\{4_x|0,0,0\}$. The rotational part of this element of the normalizer keeps invariant the maximal $\mathbf{k}$-vectors $\Gamma$, X, T and R and interchanges the $\mathbf{k}$-vectors Y$\leftrightarrow$Z and U$\leftrightarrow$V. Then, the little groups of these pairs of $\mathbf{k}$-vectors are conjugated. Moreover, this operation relates irreps with the same parity (the inversion at the origin transforms into itself through conjugation by any symmetry operation with no translation part, i.e., it commutes with all symmetry operations with no translation, and then the character of the inversion remains invariant.) The traces of the conjugated symmetry operations given in Table (\ref{tb:conj4xSG2}) are shown in Table (\ref{tb:equivalencesSG2transf4x}).

\begin{table}[h]
	\caption{Conjugation of the chosen generators of the space group P$\overline{1}$ through the member $\{4_x|0,0,0\}$ of the affine normalizer.\label{tb:conj4xSG2}}		
	\begin{tabular}{c|c}
		$\{R|\mathbf{t}\}$&$\{4_x|0,0,0\}^{-1}\{R|\mathbf{t}\}\{4_x|0,0,0\}$\\
		\hline
		$\{1|1,0,0\}$&$\{1|1,0,0\}$\\
		$\{1|0,1,0\}$&$\{1|0,0,-1\}$\\
		$\{1|0,0,1\}$&$\{1|0,1,0\}$\\
		$\{\overline{1}|0,0,0\}$&$\{\overline{1}|0,0,0\}$
	\end{tabular}												
\end{table}

\begin{table}[h]
	\caption{Traces of the matrices of the symmetry operations that result under conjugation of $\{4_x|0,0,0\}$ of the chosen generators of the space group P$\overline{1}$ in the irreducible representations at the maximal $\mathbf{k}$-vectors (\cite{irrepsSG}). The labels of the irreps, $\irrg{2}\irrg{2}$, etc\ldots have been split into two rows to make a compact table.\label{tb:equivalencesSG2transf4x}}
	\begin{tabular}{c|rr|rr|rr|rr|rr|rr|rr|rr}
		&\multicolumn{2}{c}{$\Gamma$}&\multicolumn{2}{c}{X}&\multicolumn{2}{c}{Y}&\multicolumn{2}{c}{V}&\multicolumn{2}{c}{Z}&\multicolumn{2}{c}{U}&\multicolumn{2}{c}{T}&\multicolumn{2}{c}{R}\\
		\hline
		&$\irrg{2}$&$\irrg{3}$&$\irr{X}{2}$&$\irr{X}{3}$&$\irr{Y}{2}$&$\irr{Y}{3}$&$\irr{V}{2}$&$\irr{V}{3}$&$\irr{Z}{2}$&$\irr{Z}{3}$&$\irr{U}{2}$&$\irr{U}{3}$&$\irr{T}{2}$&$\irr{T}{3}$&$\irr{R}{2}$&$\irr{R}{3}$\\
		&$\irrg{2}$&$\irrg{3}$&$\irr{X}{2}$&$\irr{X}{3}$&$\irr{Y}{2}$&$\irr{Y}{3}$&$\irr{V}{2}$&$\irr{V}{3}$&$\irr{Z}{2}$&$\irr{Z}{3}$&$\irr{U}{2}$&$\irr{U}{3}$&$\irr{T}{2}$&$\irr{T}{3}$&$\irr{R}{2}$&$\irr{R}{3}$\\
		\hline
		$\{1|1,0,0\}$  &1&1  &-1&-1 &1&1   &-1&-1 &1&1   &-1&-1 &1&1   &-1&-1\\
		$\{1|0,0,-1\}$ &1&1  &1&1   &1&1   &1&1   &-1&-1 &-1&-1 &-1&-1 &-1&-1\\
		$\{1|0,1,0\}$  &1&1  &1&1   &-1&-1 &-1&-1 &1&1   &1&1   &-1&-1 &-1&-1\\
		$\{\overline{1}|0,0,0\}$  &-1&1 &-1&1  &-1&1  &-1&1  &-1&1  &-1&1  &-1&1  &-1&1
	\end{tabular}
\end{table}
The relations between the irreps of conjugated little groups are,
\begin{equation}
\begin{array}{ccc}
\irr{Y}{2}\irr{Y}{2}&\leftrightarrow&\irr{Z}{2}\irr{Z}{2}\\
\irr{Y}{3}\irr{Y}{3}&\leftrightarrow&\irr{Z}{3}\irr{Z}{3}\\
\irr{U}{2}\irr{U}{2}&\leftrightarrow&\irr{V}{2}\irr{V}{2}\\
\irr{U}{3}\irr{U}{3}&\leftrightarrow&\irr{V}{3}\irr{V}{3}
\end{array}
\end{equation}
and the non-zero elements of the permutation matrix $M_{\{4_x|0,0,0\}}$ are: (1,1), (2,2), (3,3), (4,4), (5,9), (6,10), (7,11), (8,12), (9,5), (10,6), (11,7), (12,8), (13,13), (14,14), (15,15) and (16,16).

An equivalent calculation using as element of the normalizer the operation $\{4_z|0,0,0\}$, gives a permutation matrix $M_{\{4_y|0,0,0\}}$ with the non-zero components: (1,1), (2,2), (3,5), (4,6), (5,3), (6,4), (7,7), (8,8), (9,9), (10,10), (11,13), (12,14), (13,11), (14,12), (15,15) and (16,16).

The matrices $M_{\{N|\mathbf{n}\}}$ calculated so far (those that correspond to the translations, $M_{\{4_x|0,0,0\}}$ and $M_{\{4_z|0,0,0\}}$) generate a group of order 48 of 16$\times$16 permutation matrices. Now, we calculate the matrix that correspond to the element of the normalizer given by
\begin{equation}
N=\left(
\begin{array}{rrr}
1&1&0\\
0&1&0\\
0&0&1
\end{array}
\right)
\end{equation}
and no translation. The rotational part $N$ keeps the maximal $\mathbf{k}$-vectors $\Gamma$, Y, Z and T invariant and interchanges the following pairs of $\mathbf{k}$-vectors; X$\leftrightarrow$V and U$\leftrightarrow$R. It does not interchange even and odd irreps, then, the permutation matrix $M_{\{N|0,0,0\}}$ has non-zero elements at the positions: (1,1), (2,2), (3,7), (4,8), (5,5), (6,6), (7,3), (8,4), (9,9), (10,10), (11,15), (12,16), (13,13), (14,14), (15,11) and (16,12). This permutation matrix together with the previous 48 calculated up to now form the automorphism group $\mathcal{N}_G$ of order 1344 (see Table \ref{normalizers:data}). 

In principle we could use additional operations of the normalizer of the space group, but none of them gives extra elements of the automorphism group.

The order of the automorphis group $\mathcal{K}_G$ for all the 230 space groups is given in Table (\ref{normalizers:data}).

\subsection{Equivalences between band representations due to the Kronecker product of the maximal irreps by a 1-dimensional single-valued irrep: example of application in the space group P$\overline{1}$ with TR.}
\label{sec:kronecker}
We take as generators of the space group the same subset as in the preceding section: three translations that define the lattice $\{1|1,0,0\}$, $\{1|0,1,0\}$, $\{1|0,0,1\}$ and one of the two inversion operations with no translation $\{\overline{1}|0,0,0\}$ (we can also omit the symmetry operation $\{ ^d1|0,0,0\}$). The point T in the first Brillouin zone $\mathbf{k}_1=(0,1/2,1/2)$ has a single vector in the star, and its little group has two 1-dimensional single-valued irreps, T$_1^{+}$ and T$_1^{-}$. In our example we will take the second one, T$_1^{-}$. The matrices of this irrep of the symmetry operations chosen as generators are,
\begin{equation}
	\label{kronecker:first}
\begin{array}{lcr}
	\{1|1,0,0\}&:&1\\
	\{1|0,1,0\}&:&-1\\
	\{1|0,0,1\}&:&-1\\
	\{\overline{1}|0,0,0\}&:&-1
\end{array}
\end{equation}

Now we consider the two double-valued irreps at the V point $\mathbf{k}_1=(1/2,1/2,0)$ of the first Brillouin zone. The matrices of the two irreps $\irr{V}{2}\irr{V}2$ and $\irr{V}{3}\irr{V}3$ are,
\begin{equation}
	\label{kronecker:second}
		\begin{array}{lcrr}
		&&\irr{V}{2}\irr{V}2&\irr{V}{3}\irr{V}3\\
		\{1|1,0,0\}&:&-\mathds{1}_2&-\mathds{1}_2\\
		\{1|0,1,0\}&:&-\mathds{1}_2&-\mathds{1}_2\\
		\{1|0,0,1\}&:&\mathds{1}_2&\mathds{1}_2\\
		\{\overline{1}|0,0,0\}&:&-\mathds{1}_2&\mathds{1}_2
	\end{array}
\end{equation}
Where $\mathds{1}_2$ is the 2-dimensional identity matrix.

The matrices of the Kronecker products T$_1^{-}\otimes \irr{V}{2}\irr{V}2$ and T$_1^{-}\otimes \irr{V}{3}\irr{V}3$ are,
\begin{equation}
	\label{kronecker:third}
	\begin{array}{lcrr}
		&&\textrm{T}_1^{-}\otimes\irr{V}{2}\irr{V}2&\textrm{T}_1^{-}\otimes\irr{V}{3}\irr{V}3\\
		\{1|1,0,0\}&:&-\mathds{1}_2&-\mathds{1}_2\\
		\{1|0,1,0\}&:&\mathds{1}_2&\mathds{1}_2\\
		\{1|0,0,1\}&:&-\mathds{1}_2&-\mathds{1}_2\\
		\{\overline{1}|0,0,0\}&:&\mathds{1}_2&-\mathds{1}_2
	\end{array}
\end{equation}
These irreps should be equivalent to the two double-valued irreps of the little group of the U vector $\mathbf{k}_3=\mathbf{k}_1+\mathbf{k}_2=(1/2,0,1/2)$ (mod a vector of the reciprocal lattice). The matrices of the irreps of the two double-valued irreps at U are,
\begin{equation}
	\label{kronecker:fourth}
	\begin{array}{lcrr}
		&&\irr{U}{2}\irr{U}2&\irr{U}{3}\irr{U}3\\
		\{1|1,0,0\}&:&-\mathds{1}_2&-\mathds{1}_2\\
		\{1|0,1,0\}&:&\mathds{1}_2&\mathds{1}_2\\
		\{1|0,0,1\}&:&-\mathds{1}_2&-\mathds{1}_2\\
		\{\overline{1}|0,0,0\}&:&-\mathds{1}_2&\mathds{1}_2
	\end{array}
\end{equation}
The sets of matrices for the generators of the space group allows us to make the following mapping between the irreps:
\begin{equation}
\begin{array}{lcl}
	\irr{V}{2}\irr{V}{2}&\rightarrow&\irr{U}{3}\irr{U}{3}\\
	\irr{V}{3}\irr{V}{3}&\rightarrow&\irr{U}{2}\irr{U}{2}
\end{array}
\end{equation}
If we perform the calculation for all the irreps at maximal $\mathbf{k}$-vectors we get the following relations:
\begin{equation}
\begin{array}{lcl|lcl|lcl|lcl}
	\irr{V}{2}\irr{V}{2}&\leftrightarrow&\irr{U}{3}\irr{U}{3}&\irrg{2}\irrg{2}&\leftrightarrow&\irr{T}{3}\irr{T}{3}&	\irr{R}{2}\irr{R}{2}&\leftrightarrow&\irr{X}{3}\irr{X}{3}&\irr{Y}{2}\irr{Y}{2}&\leftrightarrow&\irr{Z}{3}\irr{Z}{3}\\
	\irr{V}{3}\irr{V}{3}&\leftrightarrow&\irr{U}{2}\irr{U}{2}&\irrg{3}\irrg{3}&\leftrightarrow&\irr{T}{2}\irr{T}{2}&	\irr{R}{3}\irr{R}{3}&\leftrightarrow&\irr{X}{2}\irr{X}{2}&\irr{Y}{3}\irr{Y}{3}&\leftrightarrow&\irr{Z}{2}\irr{Z}{2}
\end{array}
\end{equation}
In the basis of Table (\ref{eq:labelsSG2}) this automorphism can be expressed as a 16$\times$16 permutation matrix whose non-zero elements are, (1,14), (2,13), (3,16), (4,15), (5,10), (6,9), (7,12), (8,11), (9,6), (10,5), (11,8), (12,7), (13,2), (14,1), (15,4) and (16,3).

In the space group P$\overline{1}$ there are 16 single-valued irreps of dimension 1: two irreps, even and odd, at each maximal $\mathbf{k}$-vector. If an equivalent analysis is performed for all the 16 irreps, the resulting 16 permutation matrix form an automorphism group of order 16, $\mathcal{K}_G$.

We have calculated the automorphism group $\mathcal{K}_G$ formed by the Kronecker product of the irreps of the little groups at the maximal $\mathbf{k}$-vectors with all the 1-dimensional single-valued irreps in the 230 space groups. The order of $\mathcal{K}_G$ for all the space groups is given in Table (\ref{normalizers:data}).

Finally, the complete automorphism group of any space group is the join of $\mathcal{N}_G$ and $\mathcal{K}_G$, $\mathcal{J}_G=\langle\mathcal{N}_G,\mathcal{K}_G\rangle$. In our example, $\mathcal{J}_G$ is the semidirect product of $\mathcal{N}_G$ and $\mathcal{K}_G$, then the order of $\mathcal{J}_G$ is $|\mathcal{J}_G|=1344*16=21504$. However, this is not the general rule in all the space groups. In general,  $|\mathcal{J}_G|\le|\mathcal{N}_G|*|\mathcal{K}_G|$, as can be checked in Table (\ref{normalizers:data}), where the last column gives the order of $\mathcal{J}_G$.
\begin{table}
\caption{For each space group $\mathcal{G}$ (column 1), the table gives the order of the automorphism group $\mathcal{N}_G$ (column 2), the order of the automorphism group  $\mathcal{K}_G$ (column 3) and the order of the automorphism group of the space group that results from the join of both automorphims groups $\mathcal{J}_G=\langle\mathcal{N}_G,\mathcal{K}_G\rangle$ (column 4). \label{normalizers:data}}
\begin{tabular}{|r|rrr|r|rrr|r|rrr|r|rrr|r|rrr|}\hline SG&$\left|\mathcal{N}_G\right|$&$\left|\mathcal{K}_G\right|$&$\left|\mathcal{J}_G\right|$&SG&$\left|\mathcal{N}_G\right|$&$\left|\mathcal{K}_G\right|$&$\left|\mathcal{J}_G\right|$&SG&$\left|\mathcal{N}_G\right|$&$\left|\mathcal{K}_G\right|$&$\left|\mathcal{J}_G\right|$&SG&$\left|\mathcal{N}_G\right|$&$\left|\mathcal{K}_G\right|$&$\left|\mathcal{J}_G\right|$&SG&$\left|\mathcal{N}_G\right|$&$\left|\mathcal{K}_G\right|$&$\left|\mathcal{J}_G\right|$\\
\hline
1&168&8&1344&47&48&16&768&93&2&8&16&139&2&8&16&185&2&4&4\\
\hline
2&1344&16&21504&48&12&4&48&94&2&8&16&140&1&8&8&186&2&4&4\\
\hline
3&6&8&48&49&8&8&64&95&2&4&8&141&2&4&8&187&12&4&48\\
\hline
4&48&8&192&50&4&4&16&96&2&4&8&142&2&8&8&188&6&4&24\\
\hline
5&2&4&8&51&4&8&32&97&1&4&4&143&12&2&24&189&4&4&16\\
\hline
6&6&8&48&52&8&4&16&98&2&8&8&144&12&2&24&190&4&4&16\\
\hline
7&8&8&32&53&8&8&32&99&2&8&16&145&12&2&24&191&2&8&16\\
\hline
\end{tabular}
\end{table}

\begin{table}
\caption{Number of fragile roots and independent fragile roots by space group. Only those space groups (column 1) that allow fragile bands are listed. The second and third columns show the number of fragile roots and the number of independent fragile roots under the equivalence operations described in section \ref{equivalences:all} of the supplementary material, respectively.\label{frag:ind:roots}}
\begin{tabular}{|r|rr|r|rr|r|rr|r|rr|r|rr|r|rr|}\hline SG&all&indep.&SG&all&indep.&SG&all&indep.&SG&all&indep.&SG&all&indep.&SG&all&indep.\\
\hline
2&1136&3&70&10&3&126&8&2&150&2&2&184&3&2&208&1&1\\
\hline
10&1136&7&71&132&6&127&328&23&156&6&2&185&2&2&210&1&1\\
\hline
11&8&1&72&4&1&128&128&17&157&2&2&186&2&2&214&1&1\\
\hline
12&224&9&74&60&5&129&16&2&158&6&2&187&615&22&215&16&5\\
\hline
13&8&1&81&8&1&130&8&2&159&2&2&188&102&15&216&36&5\\
\hline
14&8&1&82&8&1&131&700&33&162&668&89&189&49&9&217&8&3\\
\hline
15&60&5&83&58840&954&132&92&8&163&68&35&190&51&18&218&6&2\\
\hline
47&1136&7&84&700&33&134&16&2&164&668&89&191&72598&4615&219&6&2\\
\hline
49&8&1&85&16&2&135&8&2&165&40&21&192&723&190&220&5&3\\
\hline
51&8&1&86&16&2&136&44&7&166&140&19&193&975&253&221&51308&3243\\
\hline
53&8&1&87&1188&92&137&8&2&167&10&5&194&3070&503&222&22&7\\
\hline
55&8&1&88&78&12&138&8&2&168&3&2&199&1&1&223&57&17\\
\hline
58&8&1&111&8&1&139&1188&92&173&2&2&200&64&9&224&90&10\\
\hline
63&4&1&115&8&1&140&220&32&174&615&22&201&8&2&225&3208&407\\
\hline
64&4&1&119&8&1&141&78&12&175&72598&4615&202&48&13&226&334&88\\
\hline
65&224&12&121&4&1&143&6&2&176&3070&503&203&84&23&227&464&70\\
\hline
66&60&8&123&58840&954&147&668&89&177&3&2&204&24&8&228&7&4\\
\hline
67&8&1&124&252&19&148&140&19&182&2&2&205&6&2&229&868&126\\
\hline
69&52&4&125&16&2&149&6&2&183&3&2&206&13&3&230&19&7\\
\hline
\end{tabular}
\end{table}

\begin{longtable}{|l|}
\caption{List of double-valued irreducible representations of the little groups at $\mathbf{k}$-vecs of maximal symmetry in the 230 space groups. \label{allmaximalirreps}}\\

\hline 
SG: 1 , $P1$\\
$\overline{\Gamma}_{2}\overline{\Gamma}_{2}$,$\overline{\textrm{R}}_{2}\overline{\textrm{R}}_{2}$,$\overline{\textrm{T}}_{2}\overline{\textrm{T}}_{2}$,$\overline{\textrm{U}}_{2}\overline{\textrm{U}}_{2}$,$\overline{\textrm{V}}_{2}\overline{\textrm{V}}_{2}$,$\overline{\textrm{X}}_{2}\overline{\textrm{X}}_{2}$,$\overline{\textrm{Y}}_{2}\overline{\textrm{Y}}_{2}$,$\overline{\textrm{Z}}_{2}\overline{\textrm{Z}}_{2}$\\
\hline
SG: 2 , $P\bar{1}$\\
$\overline{\Gamma}_{2}\overline{\Gamma}_{2}$,$\overline{\Gamma}_{3}\overline{\Gamma}_{3}$,$\overline{\textrm{R}}_{2}\overline{\textrm{R}}_{2}$,$\overline{\textrm{R}}_{3}\overline{\textrm{R}}_{3}$,$\overline{\textrm{T}}_{2}\overline{\textrm{T}}_{2}$,$\overline{\textrm{T}}_{3}\overline{\textrm{T}}_{3}$,$\overline{\textrm{U}}_{2}\overline{\textrm{U}}_{2}$,$\overline{\textrm{U}}_{3}\overline{\textrm{U}}_{3}$,$\overline{\textrm{V}}_{2}\overline{\textrm{V}}_{2}$,$\overline{\textrm{V}}_{3}\overline{\textrm{V}}_{3}$,$\overline{\textrm{X}}_{2}\overline{\textrm{X}}_{2}$,$\overline{\textrm{X}}_{3}\overline{\textrm{X}}_{3}$,$\overline{\textrm{Y}}_{2}\overline{\textrm{Y}}_{2}$,$\overline{\textrm{Y}}_{3}\overline{\textrm{Y}}_{3}$,$\overline{\textrm{Z}}_{2}\overline{\textrm{Z}}_{2}$,$\overline{\textrm{Z}}_{3}\overline{\textrm{Z}}_{3}$\\
\hline
SG: 3 , $P2$\\
$\overline{\Gamma}_{3}\overline{\Gamma}_{4}$,$\overline{\textrm{A}}_{3}\overline{\textrm{A}}_{4}$,$\overline{\textrm{B}}_{3}\overline{\textrm{B}}_{4}$,$\overline{\textrm{C}}_{3}\overline{\textrm{C}}_{4}$,$\overline{\textrm{D}}_{3}\overline{\textrm{D}}_{4}$,$\overline{\textrm{E}}_{3}\overline{\textrm{E}}_{4}$,$\overline{\textrm{Y}}_{3}\overline{\textrm{Y}}_{4}$,$\overline{\textrm{Z}}_{3}\overline{\textrm{Z}}_{4}$\\
\hline
SG: 4 , $P2_1$\\
\hline
\end{longtable}

\bibliography{references.bib}
\newpage

\twocolumngrid
\LTcapwidth=0.45\textwidth
\clearpage

\input{SM2_basicbands}

\end{document}